\documentclass[aps,twocolumn,amsmath,amssymb,preprintnumbers]{revtex4}
\usepackage{amsmath} \usepackage{amsfonts} \usepackage{amssymb}
\usepackage{bbm}
\usepackage{epsfig}
\usepackage{graphics}
\usepackage{graphicx}
\newcommand{\be}{\begin{equation}}
\newcommand{\ee}{\end{equation}}
\newcommand{\ba}{\begin{eqnarray}}
\newcommand{\ea}{\end{eqnarray}}
\newcommand{\nn}{\nonumber}
\newcommand{\kr}{\rangle}
\newcommand{\kl}{\langle}
\newcommand{\M}{^{(M)}}
\newcommand{\hmn}{^{mn}}
\newcommand{\tmn}{_{mn}}
\newcommand{\tr}{{\rm tr}}
\newcommand{\x}{\tilde x_}
\newcommand{\s}{\tilde s^}
\newcommand{\f}{{\cal F}}
\newcommand{\D}{{\cal D}}
\newcommand{\g}{{\cal G}}
\newcommand{\h}{{\cal H}}
\newcommand{\bl}{\bar{\cal L}}
\newcommand{\hl}{\hat{\cal L}}
\newcommand{\m}{_{\mu_1}}
\newcommand{\ma}{^{m,ab}}
\newcommand{\je}{{j_1}}
\newcommand{\jz}{{j_2}}
\newcommand{\mc}{{\mathbbm C}}
\newcommand{\y}{(\tilde y)}

\begin{document}

\title[ ]{Geometry and symmetries in lattice spinor gravity}

\author{C. Wetterich}
\affiliation{Institut  f\"ur Theoretische Physik\\
Universit\"at Heidelberg\\
Philosophenweg 16, D-69120 Heidelberg}

\begin{abstract}
Lattice spinor gravity is a proposal for regularized quantum gravity based on fermionic degrees of freedom. In our lattice model the local Lorentz symmetry is generalized to complex transformation parameters. The difference between space and time is not put in a priori, and the euclidean and Minkowski quantum field theory are unified in one functional integral. The metric and its signature arise as a result of the dynamics, corresponding to a given ground state or cosmological solution. Geometrical objects as the vierbein, spin connection or the metric are expectation values of collective fields built from an even number of fermions. The quantum effective action for the metric is invariant under general coordinate transformations in the continuum limit. The action of our  model is found to be also invariant under gauge transformations. We observe a ``geometrical entanglement'' of gauge- and Lorentz-transformations due to geometrical objects transforming non-trivially under both types of symmetry transformations. 
\end{abstract}

\maketitle

\section{Introduction}
\label{Introduction}

Lattice spinor gravity \cite{LSG} has been proposed as a regularized model for quantum gravity. It is based on a Grassmann functional integral for fermions which is mathematically well defined for a finite number of lattice sites. For a realistic model of quantum gravity the decisive feature is the invariance of the quantum effective action under general coordinate transformations (diffeomorphism symmetry). The basic degrees of freedom used for the formulation of the functional integral are less important. In our fermionic formulation the metric and vierbein, as well as other geometrical objects, arise as expectation values of suitable collective fields built from an even number of fermions. 

For a realistic lattice quantum field theory for quantum gravity we require the following six criteria:

\begin{itemize}
\item [(1)] The functional integral is well defined for a finite number of lattice points.
\item [(2)] The functional measure and lattice action are lattice diffeomorphism invariant.
\item [(3)] Lattice diffeomorphism invariance turns to diffeomorphism symmetry for the quantum effective action in the continuum limit. 
\item [(4)] For a model with fermions the functional measure and lattice action are invariant under local Lorentz transformations.
\item [(5)] The continuum limit includes massless (or very light) degrees of freedom with gravitational interactions.
\item [(6)] A derivative expansion gives a reasonable approximation for the quantum effective action for the metric at long wavelength. 
\end{itemize}

We will present a model that obeys the first four criteria. The fifth criterion is not yet shown, but likely to hold in view of the diffeomorphism symmetry of the quantum effective action for the metric. For an investigation of the sixth criterion new methods for a reliable computation of the quantum effective action need to be developed, as sketched briefly in the conclusions. We emphasize that the derivative expansion of a diffeomorphism symmetric effective action for the metric permits only few invariants with a low number of derivatives. The two leading ones are a ``cosmological constant term'' with zero derivatives, and an ``Einstein-Hilbert term'' proportional to the curvature scalar with two derivatives. The coefficients of both terms may depend on additional scalar fields. If the cosmological constant term is small enough the gravitational field equations are close to Einstein's equations of general relativity and therefore to a realistic theory of gravity.

The use of fermions as basic variables has several advantages: (i) Fermions transform as scalars with respect to diffeomorphisms. This facilitates the formulation of a lattice diffeomorphism invariant functional measure which would be much harder (and has never been achieved so far) for fundamental metric degrees of freedom. (ii) For a Grassmann functional integral there is no problem of ``boundedness'' of the action. Criterion (1) is obeyed automatically. (iii) Fermions need to be included anyhow in any realistic model of particle physics. Bosons as the graviton, photon, $W$- and $Z$-boson, gluons and Higgs scalar can arise as collective states. Thus an extension of the present model of lattice spinor gravity can be a candidate for a unified description of all interactions. For these reasons we stick here to a purely fermionic functional integral and do not introduce bosonic lattice variables as in ref. \cite{Di}. The fermionic formulation and the implementation of lattice diffeomorphism invariance distinguish our approach from other lattice proposals for quantum gravity \cite{LQG1,LQG2,LQG3}. 

Diffeomorphism symmetry of the continuum action can be achieved rather easily for a purely fermionic model. It is underlying earlier versions of spinor gravity \cite{HCW,CWSG,CWA} and has been pioneered very early \cite{Aka,Ama,Den}. The work in \cite{HCW,CWSG,CWA} and \cite{Aka,Ama,Den} does not implement local Lorentz symmetry, however. If the action is invariant only with respect to global Lorentz symmetry additional torsion-type massless degrees of freedom are present. Their phenomenology is discussed in ref. \cite{CWSG}. In order to avoid such complications we stick here to the criterion (4) and formulate an action that is invariant under local Lorentz transformations. Our formulation of lattice spinor gravity differs therefore from ref. \cite{HCW,CWSG,CWA,Aka,Ama,Den}. We investigate the lattice action proposed in ref. \cite{LSG} for which local Lorentz symmetry is manifest. This model resembles in several aspects the higher-dimensional continuum action with local Lorentz symmetry proposed in ref. \cite{CWTS}.

In ref. \cite{LDI} we have formulated the concept of lattice diffeomorphism invariance. In this paper it was shown that a lattice diffeomorphism invariant action and functional measure imlpy diffeomorphism symmetry for the quantum effective action, including the quantum effective action for the metric as expectation value of a collective field. Our lattice model is lattice diffeomorphism invariant and therefore also obeys criteria (2) and (3). 

Our approach has at the basic level neither a metric not a vierbein. We also do not employ any objects of lattice geometry that replace these fields in a discrete formulation. The absence of a metric contrasts with ``induced gravity'' \cite{IG} where a metric is present, while its kinetic term in the effective action is induced by matter fluctuations. In our model the metric and the vierbein arise as expectation values of suitable collective fields. In this respect there is some resemblance to the appearance of a vierbein or a metric as condensates or order parameters in certain condensed matter systems \cite{GV}, or other ideas that the vierbein may originate from a fermion condensate \cite{FC}.

The formulation of a functional integral for quantum gravity without basic metric degrees of freedom opens the door for the interesting possibility that the signature of the metric which distinguishes between space and time is not introduced a priori. The difference between time and space can therefore arise dynamically, as induced by a particular expectation value of a collective field. For fermions, such a setting requires that we do not fix the choice between a euclidean rotation group $SO(4)$ or the Lorentz group $SO(1,3)$ a priori. This can be realized \cite{CWTS} if the model is invariant under the complexified orthogonal group $SO(4,{\mathbbm C})$. This group obtains if the real transformation parameters of $SO(4)$ are generalized to arbitrary complex transformation parameters. The group $SO(4,{\mathbbm C})$ contains both $SO(4)$ and $SO(1,3)$ as subgroups. Which one is realized depends again on the ground state (or cosmological solution for an appropriate Lorentz-type signature). 

Models with $SO(4,{\mathbbm C})$ symmetry have the important advantage that both the euclidean and the Minkowski setting are realized by one and the same functional integral. On the level of the basic theory there are no longer two different functional integrals that are related by an operation of analytic continuation. Now ``analytic continuation'' appears within one given functional integral and relates possible expectation values of objects that correspond to the vierbein. Such a realization of analytic continuation has been discussed previously in ref. \cite{CWMS}. The requirement of local $SO(4,{\mathbbm C})$ symmetry restricts the possible form of the lattice action. In short, the signature tensor $\eta_{mn}$ can no longer be used as a basic object for the construction of invariants.

The lattice action \cite{LSG} realizing the criteria (1)-(4) involves two species of Dirac fermions. It is found to be invariant under global chiral $SU(2)_L\times SU(2)_R$ gauge transformations acting in ``flavor-space''. These symmetries are actually extended to their complexified versions $SU(2,{\mathbbm C})_L\times SU(2,{\mathbbm C})_R$. In the continuum limit the action exhibits even local gauge symmetry. The presence of additional gauge symmetries has the interesting consequence that some of the collective geometrical degrees of freedom transform non-trivially under both the generalized Lorentz transformations $SO(4,{\mathbbm C})$ and the gauge transformations. This new ``geometric entanglement'' between Lorentz- and gauge-transformations results in interesting aspects of ``gauge-gravity unification''.

In the present paper we work out several aspects of lattice spinor gravity that are crucial for progress towards a realistic theory of gravity in this setting. We investigate the fermion-bilinears that can play the role of a vierbein, both for the continuum limit and the lattice version. We discuss their behavior under symmetry transformations, including gauge symmetries. We further establish the connection between the vierbein bilinears and a collective field for the metric. We investigate various continuous and discrete symmetries of the action, including an extension of the model where the Lorentz- and gauge transformations are unified within a larger group $SO(8,{\mathbbm C})$. We give a detailed account of the lattice formulation.

One of the important aspects of this work concerns the observation that a given lattice action can be seen from different geometrical perspectives. This is related to the different possibilities to group fermions into collective bilinear fields. From one point of view the proposed action for lattice spinor gravity appears a type of kinetic term for scalar bilinears, involving four derivatives contracted by an $\epsilon$-tensor. These scalars are invariant under generalized local Lorentz transformation such that the symmetry of the action is manifest. From a different point of view the action involves vierbein bilinears which transform non-trivially under Lorentz and gauge transformations. Finally, the action can also be seen from a purely fermionic point of view where eight spinors without derivatives and four derivatives of spinors are grouped into invariants. 

This paper is organized as follows: In sect. \ref{Action and symmetries} we formulate the functional integral and discuss the symmetry transformations $SO(4,{\mathbbm C})$ as well as the continuum limit of the action. This section has substantial overlap with the shorter presentation in the letter of ref. \cite{LSG} and permits a systematic and self-consistent presentation for the present paper. In sect. \ref{Geometry} we introduce the geometrical objects as the vierbein bilinears and the collective metric field in a continuum version. They will later be related to corresponding lattice objects. Sect. \ref{Symmetries} discusses the symmetries of the action.

In sect. \ref{Discretization} we turn to the detailed lattice formulation. Basic building blocks are invariant hyperlinks that correspond to the plaquettes in lattice gauge theories. We introduce lattice derivatives and compute the continuum limit, showing that it is diffeomorphism symmetric. In sect. \ref{Links and vierbein} we introduce fermion bilinears that act as links, somewhat similar to the link variables in lattice gauge theories. We present an equivalent expression for the lattice action in terms of links. The lattice vierbeins are closely connected to these links. Their continuum limit yield the vierbein bilinears of sect. \ref{Geometry}. In sect. \ref{Lattice metric} is devoted to the lattice metric collective field. In sect. \ref{Lattice action and diffeomorphism symmetry} we finally establish lattice diffeomorphism invariance of the proposed functional integral. Our conclusions are drawn in sect. \ref{conclusions}. In order to keep the main lines of the presentation clear we display the more technical aspects of our arguments in various appendices.

\section{Action}
\label{Action and symmetries}

{\bf 1. Functional integral}

It is our aim to formulate a quantum field theory for gravity based on the standard functional integral formalism. Our basic degrees of freedom are fermions, and the functional integral will therefore be a Grassmann functional integral. We also want all operations for this functional to be mathematically well defined. We therefore implement spinor gravity with a lattice regularization. 

Let us explore a setting with $16$ Grassmann variables $\psi^a_\gamma$ at every spacetime point $x$, $\gamma=1\dots 8,~a=1,2$. The coordinates $x$ parametrize the discrete points of a four dimensional lattice, i.e. $x^\mu=(x^0,x^1,x^2,x^3)$. We will later associate $t=x^0$ with a time coordinate, and $x^k,~k=1,2,3$, with space coordinates. There is, however, a priori no difference between time and space coordinates. The spinor index $\gamma$ denotes the eight real Grassmann variables that correspond to a complex four-component Dirac spinor. We consider two flavors of fermions, labeled by $a$, similar to the electron and neutrino, or up- and down-quark (without color). The ``real'' Grassmann variables $\psi^a_\gamma$ can be combined to complex Grassmann variables $\varphi^a_\alpha,\alpha=1\dots 4$, 
\be\label{AA1}
\varphi^a_\alpha(x)=\psi^a_\alpha(x)+i\psi^a_{\alpha+4}(x),
\ee
with $\alpha$ the ``Dirac index''. 

We concentrate on an action which involves twelve Grassmann variables and realizes diffeomorphism invariance and extended local Lorentz symmetry of the group $SO(4, {\mathbbm C})$
\ba\label{A1}
S&=&\alpha\int d^4x\varphi^{a_1}_{\alpha_1}\dots\varphi^{a_8}_{\alpha_8}
\epsilon^{\mu_1\mu_2\mu_3\mu_4}\\
&&\times J^{a_1\dots a_8b_1\dots b_4}_{\alpha_1\dots\alpha_8\beta_1\dots\beta_4}
\partial_{\mu_1}\varphi^{b_1}_{\beta_1}\partial_{\mu_2}\varphi_{\beta_2}^{b_2}\partial_{\mu_3}
\varphi_{\beta_3}^{b_3}\partial_{\mu_4}\varphi^{b_4}_{\beta_4}+c.c..\nn
\ea
We sum over repeated indices. The complex conjugation $c.c.$ replaces $\alpha\to\alpha^*,J\to J^*$ and $\varphi_\alpha(x)\to\varphi^*_\alpha(x)=\psi_\alpha(x)-i\psi_{\alpha+4}(x)$, such that $S^*=S$.  (Occasionally we use a notation where the flavor index $a$ is not written explicitly, such that each $\varphi_\alpha$ should be interpreted as a two-component complex vector.) In terms of the Grassmann variables $\psi_\gamma(x)$ the action $S$ as well as $\exp(-S)$ are elements of a real Grassmann algebra. 

The precise meaning of the derivatives $\partial_\mu\varphi(x)$ in the lattice formulation will be explained in sect. \ref{Discretization}. There we also relate $\int d^4x$ to a sum over the lattice points. In the continuum limit $x$ labels points of a region of ${\mathbbm R}^4$, $\partial_\mu\varphi(x)$ becomes a partial derivative of a Grassmann field, and $\int d^4x$ denotes the integration over the region in ${\mathbbm R}^4$. 

In the continuum limit the invariance of the action under general coordinate transformations follows from the use of the totally antisymmetric product of four derivatives $\partial_\mu=\partial/\partial x^\mu$. Indeed, with respect to diffeomorphisms $\varphi(x)$ transforms as a scalar, and $\partial_\mu\varphi(x)$ as a vector. The particular contraction with the totally antisymmetric tensor $\epsilon^{\mu_1\mu_2\mu_3\mu_4},\epsilon^{0123}=1$, allows for a realization of diffeomorphism symmetry without the use of a metric. In sect. \ref{Lattice action and diffeomorphism symmetry} we introduce for the discrete lattice the concept of lattice diffeomorphism invariance: This property of the action \eqref{A1} guarantees diffeomorphism symmetry in the continuum limit. Finally, the object $J$ will be chosen such that the action is invariant under local Lorentz transformations and their generalization to $SO(4,{\mathbbm C})$. 

The partition function $Z$ is defined as
\ba\label{1A}
Z&=&\int {\cal D}\psi g_f\exp (-S)g_{in},\nn\\
\int {\cal D}\psi&=&\prod_x \prod^{2}_{a=1}
\big\{ \int d\psi^a_1(x)\dots \int d\psi^a_8(x)\big\}.
\ea
For a finite number of discrete spacetime points on a lattice the Grassmann functional integral \eqref{1A} is well defined mathematically. We assume that the time coordinates $x^0=t$ obey $t_{in}\leq t\leq t_f$. The boundary term $g_{in}$ is a Grassmann element constructed from $\psi_\gamma(t_{in},\vec x)$, while $g_f$ involves terms with powers of $\psi_\gamma(t_f,\vec x)$, were $\vec x=(x^1,x^2,x^3)$. The product $g_fg_{in}$ can be generalized to a ``boundary matrix'' $\rho_{fi}$ that depends on $\psi(t_f,\vec x)$ and $\psi(t_{in},\vec x)$. If $g_fg_{in}$ or $\rho_{fi}$ are elements of a real Grassmann algebra the partition function is real. We may restrict the range of the space coordinates or use periodic or antiperiodic boundary conditions for $\varphi(x)$ with $x^k$ discrete points on a torus $T^3$. The Grassmann integration involves then a finite number of Grassmann variables. (For suitable $\rho_{fi}$ also the time-coordinate can be put on a torus.) The boundary term $g_{in}$ specifies the initial values of a pure quantum state, while $\rho_{fi}$ can be associated with a type of density matrix. The particular form of the boundary terms play no role in the discussion of this paper and we may, for simplicity, put both $t$ and $\vec x$ on a torus $T^4$. 

Observables ${\cal A}$ will be represented as Grassmann elements constructed from $\psi_\gamma(x)$. We will consider only bosonic observables that involve an even number of Grassmann variables. Their expectation value is defined as
\be\label{2A}
\kl {\cal A}\kr=Z^{-1}\int{\cal D}\psi g_f{\cal A}\exp (-S)g_{in}.
\ee
``Real observables'' are elements of a real Grassmann algebra, i.e. they are sums of powers of $\psi_\gamma(x)$ with real coefficients. For real $g_{in}$ and $g_f$ all real observables have real expectation values. We will take the continuum limit of vanishing lattice distance at the end. Physical observables are those that have a finite continuum limit. 

In the remainder of this section and sections \ref{Geometry}-\ref{Symmetries} we will discuss the properties of the action \eqref{A1} in the continuum limit where $\partial_\mu$ denotes partial derivatives. Most properties can be directly extended to the discrete formulation. Sects. \ref{Discretization}-\ref{Lattice action and diffeomorphism symmetry} will then provide an explicit discussion of the discrete setting on a lattice. 

\medskip\noindent
{\bf 2. Generalized Lorentz transformations}

We do not want to introduce a difference between time and space from the beginning. In consequence, we do not want to fix the signature of the Lorentz rotations a priori. This is achieved by extending the euclidean $SO(4)$ rotations to complex transformations $SO(4, {\mathbbm C})$. Depending on the choice of parameters both the euclidean rotations $SO(4)$ and the Lorentz group $SO(1,3)$ are subgroups of $SO(4,{\mathbbm C})$. 

Our aim is the construction of an action \eqref{A1} that is invariant under local $SO(4,{\mathbbm C})$ transformations. It is therefore necessary that the tensor $J^{a_1\dots a_8b_1\dots b_4}_{\alpha_1\dots \alpha_8\beta_1\dots\beta_4}$ is invariant under global $SO(4,{\mathbbm C})$ transformations. We will often use double indices $\epsilon=(\alpha,a)$ or $\eta=(\beta,b)$, $\epsilon, \eta=1\dots 8$. The tensor $J_{\epsilon_1\dots \epsilon_8\eta_1\dots\eta_4}$ is totally antisymmetric in the first eight indices $\epsilon_1\dots \epsilon_8$, and totally symmetric in the last four indices $\eta_1\dots\eta_4$. This follows from the anticommuting properties of the Grassmann variables $\varphi_\epsilon\varphi_\eta=-\varphi_\eta\varphi_\epsilon$. We will see that for an invariant $J$ the action \eqref{A1} is also invariant under local $SO(4, {\mathbbm C})$ transformations.

Local $SO(4,{\mathbbm C})$ transformations act infinitesimally as
\be\label{A2}
\delta\varphi^a_\alpha(x)=-\frac12\epsilon_{mn}(x)(\Sigma^{mn}_E)_{\alpha\beta}\varphi^a_\beta(x),
\ee
with arbitrary complex parameters $\epsilon_{mn}(x)=-\epsilon_{nm}(x),m=0,1,2,3$. The complex $4\times 4$ matrices $\Sigma^{mn}_E$ are associated to the generators of $SO(4)$ in the (reducible) four-component spinor representation. They can be obtained from the euclidean Dirac matrices
\ba\label{A3}
\Sigma^{mn}_E=-\frac14[\gamma^m_E,\gamma^n_E]~,~\{\gamma^m_E,\gamma^n_E\}=2\delta^{mn}.
\ea
Subgroups of $SO(4,{\mathbbm C})$ with different signatures obtain by appropriate choices of $\epsilon_{mn}$. Real parameters  $\epsilon_{mn}$ correspond to euclidean rotations $SO(4)$. Taking $\epsilon_{kl}, k,l=1,2,3$ real, and $\epsilon_{0k}=-i\epsilon^{(M)}_{0k}$ with real $\epsilon^{(M)}_{0k}$, realizes the Lorentz transformations $SO(1,3)$. The Lorentz transformations can be written equivalently with six real transformation parameters $\epsilon^{(M)}_{mn}~,~\epsilon\M_{kl}=\epsilon_{kl}$, using Lorentz-generators $\Sigma\hmn_M$ and signature $\eta\hmn=diag(-1,1,1,1)$, 
\be\label{3A}
\delta\varphi=-\frac12\epsilon\M\tmn\Sigma\hmn_M\varphi ,
\ee
with
\be\label{A4}
\Sigma\hmn_M=-\frac14[\gamma^m_M,\gamma^n_M]~,~\{\gamma^m_M,\gamma^n_M\}=\eta^{mn}.
\ee
The euclidean and Minkowski Dirac matrices are related by $\gamma^0_M=-i\gamma^0_E,\gamma^k_M=\gamma^k_E$. 

The transformation of a derivative involves an inhomogeneous part 
\be\label{A5}
\delta\partial_\mu\varphi_\beta=-\frac12\epsilon\tmn(\Sigma\hmn\partial_\mu
\varphi)_\beta-\frac12\partial_\mu\epsilon\tmn(\Sigma\hmn\varphi)_\beta,
\ee
with $\Sigma\hmn=\Sigma\hmn_E~,~\gamma^m=\gamma^m_E$. The first ``homogeneous term'' $\sim \partial_\mu\varphi$ transforms as $\varphi_\beta$. Contributions of the second ``inhomogeneous term'' to the  variation of the action $\delta S$ involve at least nine spinors at the same position $x$, i.e. $(\Sigma\hmn\varphi)^b_\beta(x)\varphi^{a_1}_{\alpha_1}(x)\dots\varphi^{a_8}_{\alpha_8}(x)$. This inhomogeneous contribution to $\delta S$ vanishes due to the identity $\varphi_\alpha(x)\varphi_\alpha(x)=0$ (no sum here). This invariance of $S$ under global $SO(4,{\mathbbm C})$ transformations entails the invariance under local $SO(4,{\mathbbm C})$ transformations. We have constructed in ref. \cite{CWTS} a sixteen dimensional spinor gravity with local $SO(16,{\mathbbm C})$ symmetry. The present four-dimensional model shows analogies to this. 

It is important that all invariants appearing in the action \eqref{A1} involve either only factors of $\varphi_\alpha=\psi_\alpha+i\psi_{\alpha+4}$ or only factors of $\varphi^*_\alpha =\psi_\alpha-i\psi_{\alpha+4}$. It is possible to construct $SO(1,3)$ invariants which involve both $\varphi$ and $\varphi^*$. Those will not be invariant under $SO(4,{\mathbbm C})$, however. We can also construct invariants involving $\varphi$ and $\varphi^*$ which are invariant under euclidean $SO(4)$ rotations. They will not be invariant under $SO(1,3)$. The only types of invariants invariant under both $SO(4)$ {\em and} $SO(1,3)$, and more generally $SO(4,{\mathbbm C})$, are those constructed from $\varphi$ alone or $\varphi^*$ alone, or products of such invariants. (Invariants involving both $\varphi$ and $\varphi^*$ can be constructed as products of invariants involving only $\varphi$ with invariants involving only $\varphi^*$.) 

We conclude that for a suitable invariant tensor $J$ the action has the symmetries required for a realistic theory of gravity for fermions, namely diffeomorphism symmetry and local $SO(1,3)$ Lorentz symmetry. No signature and no metric are introduced at this stage, such that there is no difference between time and space \cite{CWTS}. As discussed extensively in ref. \cite{CWSG}, global Lorentz symmetry may be sufficient for a realistic theory of gravity. Nevertheless, models with local $SO(1,3)$-symmetry may be preferable, since they contain only the metric as massless composite bosonic degree of freedom. 

For an action of the type (2) local $SO(4,{\mathbbm C})$ symmetry is realized for every invariant tensor $J$. We define the $SO(4,{\mathbbm C})$ variation of arbitrary tensors with Dirac indices $\alpha_1\dots \alpha_N$ as
\be\label{SA}
\delta T_{\alpha_1\dots\alpha_N}=T_{\tilde\alpha\alpha_2\dots\alpha_N}\Sigma_{\tilde\alpha\alpha_1}+\dots
+T_{\alpha_1\dots\tilde\alpha}\Sigma_{\tilde\alpha\alpha_N},
\ee
with 
\be\label{SB}
\Sigma_{\alpha\beta}=-\frac12\epsilon_{mn}\Sigma^{mn}_{\alpha\beta}.
\ee
We can express global $SO(4,{\mathbbm C})$-transformations (with $\epsilon_{mn}$ independent of $x$) of the action equivalently by a transformation \eqref{A2} of the spinors $\varphi$ with fixed $J$, or by a transformation \eqref{SA} of $J$ with fixed $\varphi$. For $\delta J=0$ the action is invariant under global $SO(4,{\mathbbm C})$-transformations.

\medskip\noindent
{\bf 3. Action with local Lorentz symmetry}

We will compose the invariant $J$ from a totally symmetric four-index invariant $L_{\eta_1\dots \eta_4}$ and a totally antisymmetric eight-index invariant $A^{(8)}_{\epsilon_1\dots\epsilon_8}$. For two flavors we can construct two real symmetric $SO(4,{\mathbbm C})$ invariants
\be\label{11A1}
S^\pm_{\eta_1\eta_2}=-(\tau_2)_{\beta_1\beta_2}(\tau_2)^{b_1b_2},
\ee
where $(\beta_1,\beta_2)$ are restricted to the values $(1,2)$ for $S^+$, and $(3,4)$ for $S^-$, and $\tau_k$ are the Pauli matrices. These invariants and their relation to Weyl spinors are described in more detail in appendix A. We work here in a basis (cf. app. A) where $\bar\gamma=diag(1,1,-1,-1)$ such that the Weyl spinors $\varphi_\pm=(1\pm\bar\gamma)\varphi/2$ correspond to the upper or lower two components of $\varphi$. 

A totally symmetric four-index invariant can be constructed by symmetrizing a product of two-index invariants
\ba\label{W9}
&&L_{\eta_1\eta_2\eta_3\eta_4}=\frac16
(S^+_{\eta_1\eta_2}S^-_{\eta_3\eta_4}+S^+_{\eta_1\eta_3}
S^-_{\eta_2\eta_4}+S^+_{\eta_1\eta_4}S^-_{\eta_2\eta_3}\nn\\
&&\hspace{1.0cm} +S^+_{\eta_3\eta_4}S^-_{\eta_1\eta_2}+S^+_{\eta_2\eta_4}
S^-_{\eta_1\eta_3}+S^+_{\eta_2\eta_3}S^-_{\eta_1\eta_4}).
\ea
Multiplication of $L$ with four spinor derivatives $\partial_\mu\varphi_\eta$ yields an expression $D$ that is invariant under global $SO(4,{\mathbbm C})$ transformations
\be\label{30D}
D=\epsilon^{\mu_1\mu_2\mu_3\mu_4}
\partial_{\mu_1}\varphi_{\eta_1}\partial_{\mu_2}\varphi_{\eta_2}\partial_{\mu_3}\varphi_{\eta_3}\partial_{\mu_4}\varphi_{\eta_4}
L_{\eta_1\eta_2\eta_3\eta_4}.
\ee
This invariant involves two Weyl spinors $\varphi_+$ and two Weyl spinors $\varphi_-$. 

Furthermore, an invariant with eight factors of $\varphi$ (without derivatives) involves the totally antisymmetric tensor for the eight values of the double-index $\epsilon$
\ba\label{30E}
A^{(8)}&=&\frac{1}{8!}\epsilon_{\epsilon_1\epsilon_2\dots\epsilon_8}
\varphi_{\epsilon_1}\dots\varphi_{\epsilon_8}\nn\\
&=&\frac{1}{(24)^2}\epsilon_{\alpha_1\alpha_2\alpha_3\alpha_4}
\varphi^1_{\alpha_1}\dots\varphi^1_{\alpha_4}\epsilon_{\beta_1\beta_2\beta_3\beta_4}
\varphi^2_{\beta_1}\dots\varphi^2_{\beta_4}\nn\\
&=&\varphi^1_1\varphi^1_2\varphi^1_3\varphi^1_4
\varphi^2_1\varphi^2_2\varphi^2_3\varphi^2_4.
\ea
It is easy to verify that $\delta(\epsilon_{\epsilon_1\dots \epsilon_8})=0$ in the sense of eq. \eqref{SA}. An invariant $J$ in eq. \eqref{A1} can therefore be constructed by multiplying $L_{\eta_1\dots \eta_4}$ with $\epsilon_{\epsilon_1\dots\epsilon_8}$. 

In conclusion of this discussion we will consider an action with local $SO(4,{\mathbbm C})$ symmetry which takes the form 
\be\label{30F}
S=\alpha\int d^4xA^{(8)}D+c.c.
\ee
Indeed, the inhomogeneous contribution \eqref{A5} to the variation of $D(x)$ contains factors $(\Sigma^{mn}\varphi^b)_\beta(x)$. It vanishes when multiplied with $A^{(8)}(x)$, since the Pauli principle $\big(\varphi^a_\alpha(x)\big)^2=0$ admits at most eight factors $\varphi$ for a given $x$. In consequence, the inhomogeneous variation of the action \eqref{30F} vanishes and $S$ is invariant under {\em local } $SO(4,{\mathbbm C})$ transformations. In contrast to $\int d^4xD(x)$ the action $S$ is not a total derivative. 

The derivative-invariant $D$ can be written in the form 
\be\label{16A}
D=\epsilon^{\mu_1\mu_2\mu_3\mu_4}D^+_{\mu_1\mu_2}D^-_{\mu_3\mu_4},
\ee
where
\be\label{32A}
D^\pm_{\mu\nu}=\partial_\mu\varphi_{\eta_1}S^\pm_{\eta_1\eta_2}\partial_\nu\varphi_{\eta_2},
\ee
involves two Weyl spinors $\varphi_+$ or two Weyl spinors $\varphi_-$, respectively. This shows that $D$ is invariant under an exchange $\varphi_{+}\leftrightarrow\varphi_{-}$ of the Weyl spinors. The transformation $\varphi\to\gamma^0\varphi$ maps $S^+_{\eta_1\eta_2}\leftrightarrow S^-_{\eta_1\eta_2}$ and therefore $D^+_{\mu_1\mu_2}\leftrightarrow D^-_{\mu_1\mu_2}$, such that again $D$ is invariant. (For a suitable choice $\gamma^0=\tau_1\otimes 1$ the transformation $\varphi\to\gamma^0\varphi$ actually corresponds to $\varphi_{+,\eta}\leftrightarrow \varphi_{-,\eta}$, cf. app. A.) We can also decompose
\be\label{32C}
A^{(8)}=A^+A^-,
\ee
with 
\be\label{32D}
A^+=\varphi^1_{+1}\varphi^1_{+2}\varphi^2_{+1}\varphi^2_{+2}
\ee
involving four Weyl spinors $\varphi_+$, and similarly for $A^-$. 

The invariants $A^\pm$ can be expressed in terms of the Lorentz invariant bilinears $H^\pm_k$ 
\be\label{41A1}
H^\pm_k=\pm\varphi^a_{\pm,\alpha}(\tau_2)_{\alpha\beta}(\tau_2 \tau_k)^{ab}\varphi^b_{\pm,\beta},
\ee
(with $\varphi_{\pm,\alpha}$ two-component Weyl spinors). These scalar fields are simple building blocks for $SO(4,{\mathbbm C})$ invariant actions. One finds (cf. sect. \ref{Links and vierbein}),
\be\label{b2}
A^\pm=-\frac{1}{24}H^\pm_kH^\pm_k.
\ee
The combinations 
\be\label{32E}
F^\pm_{\mu_1\mu_2}=A^\pm D^\pm_{\mu_1\mu_2}
\ee
are therefore composed of six Weyl spinors $\varphi_+$ or six Weyl spinors $\varphi_-$, respectively. The action involves products of $F^+$ and $F^-$, 
\be\label{32F}
S=\alpha\int d^4x\epsilon^{\mu_1\mu_2\mu_3\mu_4}F^+_{\mu_1\mu_2}F^-_{\mu_3\mu_4}+c.c.
\ee

Since the tensor $S_{\eta_1\eta_2}$ involves $(\tau_2)^{ab}$ all contributions to $D^+_{\mu\nu}$ have one spinor of each  flavor, i.e. only terms $\partial_\mu\varphi^1_\alpha\partial_\nu\varphi^2_\beta$ appear. This implies that $D^+_{\mu\nu}$ is odd under the transformation $\varphi^2_\alpha\to-\varphi^2_\alpha$. On the other hand, $A^+$ is even under this transformation, such that $F^+_{\mu\nu}$ is odd. Since $F^-_{\mu\nu}$ is invariant, the action changes sign. Similar involutions that change the sign of $S$ can be found by noting that each term in $S$ contains exactly three spinors of each of the four sorts $\varphi^1_+,\varphi^2_+,\varphi^1_-$ and $\varphi^2_-$. 

The action \eqref{30F} is not the only possible action with twelve spinors and $SO(4,{\mathbbm C})$ symmetry. The tensor $J_{\epsilon_1\dots\epsilon_8\eta_1\dots\eta_4}$ in eq. \eqref{A1} must be totally antisymmetric in the first eight indices $\epsilon_1\dots\epsilon_8$, implying $J=\epsilon^{(8)}\tilde L$. Since the totally antisymmetric tensor with eight indices $\epsilon_1\dots\epsilon_8$ is a singlet with respect to $SO(4,{\mathbbm C})$, the remaining piece $\tilde L_{\eta_1\dots\eta_4}$ must be a $SO(4,{\mathbbm C})$-singlet which is totally symmetric in the four indices $\eta_1\dots\eta_4$. Besides $L$  in eq. \eqref{W9} we can also construct a tensor $L_+$ where $S^-_{\eta_1\eta_2}$ is replaced by $S^+_{\eta_1\eta_2}$, and similarly for a tensor $L_-$. The three possible terms in $\tilde L=\alpha L+\beta L_++\gamma L_-$ all lead to local $SU(2,{\mathbbm C})_F$ gauge symmetry of the action, cf. sect. \ref{Symmetries}. The action \eqref{A1} becomes unique, however, if we require an equal number of six Weyl spinors $\varphi_+$ and six Weyl spinors $\varphi_-$. One can also find contributions to an action with local $SO(4,{\mathbbm C})$ symmetry that involve only eight or ten spinors. They are discussed in a separate publication.

\medskip\noindent
{\bf 4. Minkowski action}

Defining the Minkowski action by
\be\label{A23}
S=-iS_M~,~e^{-S}=e^{iS_M},
\ee
one finds the usual ``phase factor'' for the functional integral written in terms of $S_M$. We emphasize that one and the same functional integral \eqref{1A} describes the euclidean and the Minkowski setting. The use of the euclidean action $S$ or the Minkowski action $S_M$ is purely a matter of convenience. There is no ``analytic continuation'' between the euclidean and the Minkowski setting. They are the same, and different signatures arise only from different expectation values for collective bosonic fields describing the metric or the vierbein. Our setting realizes a version of analytic continuation in terms of the continuation of the possible values of the vierbein \cite{CWMS}. 

We can define the operation of a transposition as a total reordering of all Grassmann variables. The result of transposition for a product of Grassmann variables depends only on the number of factors $N_\varphi$. For $N_\varphi=2,3$ mod $4$ the transposition results in a minus sign, while for $N_\varphi=4,5$ mod $4$ the product is invariant. In consequence, one finds for the action \eqref{A1} with $12$ spinors
\be\label{33B}
S^T=S.
\ee
The hermitean conjugation $hc$ is the combination of transposition with the complex conjugation $c.c$, such that 
\ba\label{33C}
hc(S)=S.
\ea
With respect to the complex conjugation $c.c.$ used in eq. \eqref{A1} the Minkowski action is therefore antihermitean
\be\label{33D}
hc(S_M)=-S_M.
\ee
There exists a different complex structure for which $S_M$ is hermitean. This is discussed in appendix B. One can use the complex structure \eqref{AA1} in order to show that we deal with a real Grassmann algebra, while the different complex structure with hermitean $S_M$ can be employed for establishing a unitary time evolution.

\section{Geometry}
\label{Geometry}

The action \eqref{A1}, \eqref{30F} or \eqref{32F} has all required symmetries for a quantum field theory of gravity. The geometric content is not very apparent, however, in this formulation. In this section we will discuss collective bosonic fields whose expectation values correspond to usual geometric objects as the vierbein or metric. We will express the action in terms of such fields. 

\medskip\noindent
{\bf 1. Vierbein bilinear}

The action of spinor gravity has the symmetries of a theory for gravity, namely invariance under diffeomorphisms and local Lorentz transformations. One may therefore expect that a geometrical formulation in terms of a vierbein and metric should be possible. Indeed, the vierbein may appear as the expectation value of a suitable fermion bilinear. Bilinears of the type 
\be\label{A11}
(\tilde E^m_{1,\mu})^{ab}=\varphi^a_\alpha(C_1\gamma^m_M)_{\alpha\beta}\partial_\mu\varphi^b_\beta=-\partial_\mu\varphi^b_\alpha 
(C_1\gamma^m_M)_{\alpha\beta}\varphi^a_\beta 
\ee
transform as a vector under general coordinate transformations and as a vector under global $SO(4,{\mathbbm C})$ variations \eqref{A2} of $\varphi$, i.e. for $\epsilon\tmn$ independent of $x$, 
\be\label{20A}
\delta(\tilde E^m_{1,\mu})^{ab}=-(\tilde E^n_{1,\mu})^{ab}\epsilon^{(M)}_{np}\eta^{pm}.
\ee
The same holds for $\tilde E^m_{2,\mu}$ for which we replace $C_1\to C_2$ in eq. \eqref{A11}. In this case the sign of the second term in eq. \eqref{A11} is positive. The antisymmetric invariant $4\times 4$-matrices $C_1$ and $C_2$ are displayed explicitly in app. A. Under local generalized Lorentz transformation the vierbein bilinear acquires an inhomogeneous piece that we discuss in appendix C. 

A candidate for a vierbein can be obtained by a suitable contraction with a $2\times 2$ matrix $V^{ab}$,
\be\label{BB1}
\tilde E^m_\mu=V^{ab}(\tilde E^m_{1,\mu})^{ab}.
\ee
We observe that for antisymmetric $V^{ab}=-V^{ba}$ the vierbein is the derivative of a vector 
\be\label{BB2}
\tilde E^m_\mu=\frac12\partial_\mu(\varphi^a C_1\gamma^m_M\varphi^b V_{ab}),
\ee
while for symmetric $V^{ab}=V^{ba}$ one has
\ba\label{BB3}
\tilde E^m_\mu&=&\varphi^aC_1\gamma^m_M\partial_\mu\varphi^b V^{ab}\nn\\
&=&-\partial_\mu\varphi^a C_1\gamma^m_M\varphi^b V^{ab}.
\ea
Further objects $V^{ab}(\tilde E^m_{2,\mu})^{ab}$ with the transformation property of a vierbein can be found by replacing $C_1\to C_2=C_1\bar\gamma$. In this case a symmetric $V$ leads to the derivative of a vector. 

We conclude that for real $\epsilon^{(M)}$ the vierbein bilinear $\tilde E^{m}_\mu$ has almost the transformation properties of the vierbein in general relativity with a Minkowski signature. The only difference concerns the inhomogeneous piece for local Lorentz transformations. It is also not clear at this stage which one of the possible candidates for $\tilde E^m_\mu$ should be selected.

\medskip\noindent
{\bf 2. Absence of cosmological constant invariant}

We next investigate if the action \eqref{30F} can be written in terms of a suitable vierbein bilinear $\tilde E^m_\mu$ and its derivatives, plus suitable collective bosonic fields that transform as scalars under diffeomorphisms and global Lorentz transformations. We will see that this is indeed the case.

One may first ask if an invariant action of the type \eqref{A1} can be written in the intuitive form 
\be\label{A18}
S=\alpha\int d^4 x W \det (\tilde E^m_\mu)+c.c.,
\ee
with $\tilde E^m_\mu$ given by eqs. \eqref{A11}, \eqref{BB1} for some particular choice of $V_{ab}$, or by similar expressions with $C_2$ instead of $C_1$. In eq. \eqref{A18} $\tilde E^m_\mu$ is interpreted as a matrix with first index $\mu$ and second index $m$. The invariant $W$ must involve two Weyl spinors $\varphi_+$ and two Weyl spinors $\varphi_-$.
The invariance of the action \eqref{A18} under diffeomorphisms and $SO(4{\mathbbm C})$ transformations would be particularly transparent in this language. The transformation \eqref{20A} implies the invariance of $S$ under global Lorentz transformations in a simple way. With respect to diffeomorphisms the determinant $\tilde E=\det(\tilde E^{m}_\mu)$ has the same transformation properties as the determinant of the vierbein in general relativity. The latter equals the usual volume factor $\sqrt{g}=|\det (g_{\mu\nu})|^{1/2}$, and we recover the general coordinate invariance of the action \eqref{A18}. One may call $W\det\tilde E^m_\mu$ a ``cosmological constant invariant'' due to its resemblance to a cosmological constant in standard general relativity for $\kl W\kr =$const. However, we show in appendix D that such a ``cosmological constant invariant'' is not possible in the present formulation of spinor gravity.

This observation may have interesting consequences for the issue of a cosmological constant. For $\Delta\kl \tilde E^m_\mu\kr=e^m_\mu$ a term in the effective action $\Delta^{-4}\kl W\kr\det e^m_\mu$ could be associated with a cosmological constant $\kl W\kr/\Delta^4$. Earlier proposals for spinor gravity \cite{HCW,CWSG} or similar theories \cite{Aka,Ama,Den} have based the action on a ``cosmological constant invariant'' (with $W=1$). The absence of such an invariant in the present formulation is a distinctive feature. An expression of the action \eqref{30F} in terms of the vierbein bilinears $\tilde E^m_\mu$ must involve derivatives of those bilinears.

\medskip\noindent
{\bf 3. Flavored vierbein}

Besides the inhomogeneous transformation property under local generalized Lorentz transformations a second important difference between the vierbein bilinears and the usual vierbein concerns the nontrivial transformation of $\tilde E^m_\mu$ with respect to gauge transformations. We will discuss in the next section that the action \eqref{30F} is invariant under chiral gauge transformations $SU(2,{\mathbb C})_L\times SU(2,{\mathbbm C})_R$. Here the first factor $SU(2,{\mathbbm C})_L$ acts on the indices of the Weyl spinor $\varphi_+$, while the second factor $SU(2,{\mathbbm C})_R$ acts on $\varphi_-$. Since the vierbein bilinear involves one Weyl spinor $\varphi_+$ and one Weyl spinor $\varphi_-$ it transforms in the $(2,2)$ representation of this gauge group. The non-trivial transformation of geometrical objects under gauge transformations is a novel feature of our approach. Possible interesting observational consequences of this new type of ``gauge-gravity unification'' will be postponed to future investigations. We only describe here some features that will be needed later in this work. We also discuss in appendix E some other collective fields which show this entanglement between geometrical and gauge aspects.

With respect to the vectorlike gauge transformations $SU(2,{\mathbbm C})_F$ which consist of the diagonal subgroup of $SU(2,{\mathbbm C})_L\times SU(2,{\mathbbm C})_R$ the vierbein bilinears transform as singlets and three component vectors. The singlet that is not a pure derivative is given by
\ba\label{31A1a}
\bar E^{m}_{2,\mu}&=&\varphi^a(\tau_2)^{ab}C_2\gamma^m_M\partial_\mu\varphi^b\nn\\
&=&-\partial_\mu\varphi^a(\tau_2)^{ab}C_2\gamma^m_M\varphi^b,
\ea
where we observe that the matrices $(C_2\gamma^m_M)_{\alpha\beta}$ are antisymmetric, cf. app. A. 

The vector with respect to $SU(2,{\mathbbm C})_F$ which is not a total derivative involves the matrix $C_1$, 
\be\label{57A}
\bar E^m_{1(k)\mu}=\varphi^a(\tau_2\tau_k)^{ab}C_1\gamma^m_M\partial_\mu\varphi_b.
\ee
Now the matrices $C_1\gamma^m_M$ are symmetric, cf. app. A, and also $(\tau_2\tau_k)$ are symmetric $2\times 2$ matrices. The three ``components'' of $E_1$ labeled by $(k)$ transform indeed as a vector with respect to the gauge symmetry $SU(2,{\mathbbm C})_F$.

Objects with the transformation properties of the vierbein under diffeomorphisms and Lorentz rotations, but also transforming non-trivially with respect to gauge symmetries acting on flavor, may be called ``flavored vierbeins''. Such objects are not common in usual formulations of general relativity. They give a first glance on a more intrinsic unification of gravity and gauge symmetries that may be realized in our scenario. Any nonzero expectation value $\kl \bar E^m_{1(k)\mu}\kr$ would lead to spontaneous breaking of the gauge symmetry. 

\medskip\noindent
{\bf 4. Dimension of vierbein}

A third difference between $\tilde E^m_\mu$ and the usual vierbein does not concern the transformation property, but rather  the dimension. In fact, the spinors $\varphi$ are dimensionless, such that the presence of a derivative in eq. \eqref{A11} implies that $\tilde E^m_\mu$ has dimension of mass or inverse length. The discrete formulation in sect. \ref{Discretization} will introduce the lattice distance $\Delta$ with dimension of length. One may therefore consider the dimensionless vierbein bilinears
\be\label{57B}
\tilde e^m_{1(k)\mu}=\Delta\bar E^m_{1(k)\mu}~,~\tilde e^m_{2,\mu}=\Delta\bar E^m_{2,\mu}.
\ee

Other dimensionless fields transforming as scalars with respect to general coordinate transformations and vectors with respect to global generalized Lorentz transformations are
\be\label{57C}
\bar A^m=\varphi^a(\tau_2)^{ab}C_1\gamma^m_M\varphi^b
\ee
and 
\be\label{57D}
\bar S^m_{(k)}=\varphi^a(\tau_2\tau_k)^{ab}C_2\gamma^m_M\varphi^b.
\ee
These objects transform as scalars or vectors with respect to the vectorlike gauge transformation of $SU(2,{\mathbbm C})_F$. With respect to $SU(2,{\mathbbm C})_L\times SU(2,{\mathbbm C})_R$ they belong again to the representation $(2,2)$, similar to $\tilde e^m_{2,\mu}$ and $\tilde e^m_{1(k)\mu}$. 

The inhomogeneous part of the Lorentz transformation \eqref{A5} mixes the bilinears $\tilde e^m_{1(k)\mu},\bar S^m_{(k)}$ and $\tilde e^m_{2,\mu},\bar A^m$, cf. app. C,
\ba\label{57E}
\delta_{inh}\tilde e^m_{2,\mu}=\frac12\Delta\partial_\mu\tilde\epsilon^{(M)m}_n\bar A^n,\nn\\
\delta_{inh}\tilde e^m_{1(k)\mu}=\frac 12\Delta\partial_\mu \tilde\epsilon^{(M)m}_n\bar S^n_{(k)}.
\ea
One observes that this inhomogeneous part vanishes in the limit $\Delta\to 0$. We will discuss this important property in more detail later. If the inhomogeneous part can be neglected the expectation values $\kl \tilde e^m_{2,\mu}\kr$ and $\kl \tilde e^m_{1(k)\mu}\kr$ have precisely the local Lorentz-transformation properties of the vierbein in Cartan's formulation \cite{Ca} of general relativity.

\medskip\noindent
{\bf 5. Connection bilinear}

We will now proceed to an expression of the action \eqref{30F} in terms of the vierbein bilinears $\tilde e^m_{2,\mu}$ and $\tilde e^m_{1(k)_\mu}$. For this purpose we will introduce a collective field related to the spin connection. More generally, the expectation values of suitable bosonic collective fields can be used to define geometrical objects transforming as vierbein, metric, spin connection, curvature, tensor etc.. If we can find geometric fields with the standard transformation properties, they can be used to construct diffeomorphism and Lorentz invariant objects in the standard way. One only has to verify that such objects do not vanish identically due to the Pauli principle for spinors. We will see, however, that not all standard geometrical objects can be implemented in this way. In particular, the inverse vierbein cannot be obtained as a polynomial of spinors. 

We define ``spin connection bilinears'' by
\ba\label{82AB}
\tilde\Omega^m_{2,\mu\nu}&=&-\frac12(\partial_\mu\tilde e^m_{2,\nu}-\partial_\nu\tilde e^m_{2,\mu}),\nn\\
\tilde\Omega^m_{1(j)\mu\nu}&=&-\frac12(\partial_\mu\tilde e^m_{1(j)\nu}-\partial_\nu\tilde e^m_{1(j)\mu}).
\ea
The transformations properties of these objects and their relation to the usual spin connection are discussed in appendix F. There we also show that in terms of those bilinears the action \eqref{30F} can be written as
\ba\label{82AA}
S=\frac{\alpha}{16\cdot 24^2\Delta^2}\int d^4x\epsilon^{\mu_1\mu_2\mu_3\mu_4}
\eta_{mn}H^+_kH^+_kH^-_lH^-_l\nn\\
\{\tilde\Omega^m_{2,\mu_1\mu_2}\tilde \Omega^n_{2,\mu_3\mu_4}-
\tilde\Omega^m_{1(j)\mu_1\mu_2}\tilde\Omega^n_{1(j)\mu_3\mu_4}\}+c.c..
\ea
This is obtained by a suitable reordering of the Grassmann variables.

The expression \eqref{82AA} involves first derivatives of the vierbein, with the structure
\be\label{82AC}
\bar D=\frac{1}{16}\epsilon^{\mu\nu\rho\sigma}\Omega_{\mu\nu}{^p}\Omega_{\rho\sigma p}.
\ee
From the transformation properties in app. F, eq. \eqref{b10}, one finds that $\bar D$ is not invariant under local Lorentz transformations
\be\label{82AD}
\delta\bar D=\frac{1}{16}\epsilon^{\mu\nu\rho\sigma}\partial_\mu(e_\nu{^m}e_\rho{^n})\partial_\sigma\epsilon^{(M)}_{mn}.
\ee
Nevertheless, the action is invariant under local $SO(4,{\mathbbm C})$ transformations due to the Pauli principle.  (Recall that also $D$ in eq. \eqref{30D} transforms inhomogeneously under local $SO(4,{\mathbbm C})$.) The structure $\bar D$ can be written as a total derivative 
\be\label{82AE}
\bar D=\frac{1}{16}\partial_\mu(e_\nu{^p}\partial_\rho e_\sigma{^q})\eta_{pq}\epsilon^{\mu\nu\rho\sigma},
\ee
and the action can therefore also be written with a derivative $\partial_\mu$ acting on $H$.

Other forms of the action can be obtained by further reordering of the Grassmann variables. For example, we may use 
\be\label{AW1}
A^{(8)}=\frac{1}{192}H^+_kH^-_kH^+_lH^-_l,
\ee
which follows from squaring the relation
\ba\label{AW2}
H^+_kH^-_k&=&8\{\varphi^1_{+1}\varphi^1_{+2}\varphi^2_{-1}\varphi^2_{-2}+
\varphi^2_{+1}\varphi^2_{+2}\varphi^1_{-1}\varphi^1_{-2}\}\nn\\
&-&4\{\varphi^1_{+1}\varphi^2_{+2}\varphi^1_{-1}\varphi^2_{-2}+
\varphi^1_{+1}\varphi^2_{+2}\varphi^2_{-1}\varphi^1_{-2}\nn\\
&&+\varphi^2_{+1}\varphi^1_{+2}\varphi^1_{-1}\varphi^2_{-2}+\varphi^2_{+1}\varphi^1_{+2}\varphi^2_{-1}\varphi^1_{-2}\}.
\ea
A reordering can now be performed in the factor $H^+_kH^-_kD^+_{\mu\nu}D^-_{\rho\sigma}$.

\medskip\noindent
{\bf 6. Inverse vierbein}

What is not available on the level of multi-fermion fields is the inverse vierbein. Any given choice of the vierbein bilinear $\tilde E^m_\mu$ (given choice of $V_{ab}$) is an element of the Grassmann algebra. Inverse elements are not defined, however, for a Grassmann algebra. Nevertheless, with
\ba\label{b12}
\tilde E&=&\det (\tilde E^m_\mu)\\
&=& \frac{1}{24}\epsilon^{\mu_1\mu_2\mu_3\mu_4}\epsilon_{m_1m_2m_3m_4}
\tilde E^{m_1}_{\mu_1}\tilde E^{m_2}_{\mu_2}\tilde E^{m_3}_{\mu_3}\tilde E^{m_4}_{\mu_4},\nn
\ea
we can define an object that transforms as the product of the inverse vierbein with the determinant of the vierbein 
\ba\label{b13}
\tilde I^\mu_m&=&\frac16\epsilon^{\mu_1\mu_2\mu_3\mu_4}\epsilon_{m_1m_2m_3m_4}
\tilde E^{m_2}_{\mu_2}\tilde E^{m_3}_{\mu_3}\tilde E^{m_4}_{\mu_4},\\
&\widehat{=}&\tilde E\tilde E_m{^\mu}.\nn
\ea
It obeys
\be\label{b14}
\tilde I^\mu_m\tilde E^n_\mu=\tilde E\delta^n_m~,~\tilde I^\mu_m\tilde E^m_\nu=\tilde E\delta^\mu_\nu,
\ee
where we recall that $\tilde E^{-1}$ is not defined. 

Similarly, the antisymmetrized product of two inverse vierbeins, multiplied by $\tilde E$, can be defined as
\ba\label{b15}
\tilde I_{mn}^{\mu\nu}&=&\frac12\epsilon^{\mu\nu\mu_3\mu_4}
\epsilon_{mnm_3m_4}
\tilde E^{m_3}_{\mu_3}\tilde E^{m_4}_{\mu_4}\nn\\
&\widehat{=}&\tilde E(\tilde E_m{^\mu}\tilde E_n{^\nu}-\tilde E_m{^\nu}\tilde E_n{^\mu}).
\ea
It obeys
\ba\label{b16}
\tilde I^{\mu\nu}_{mn}\tilde E^p_\nu&=&\tilde I^\mu_m\delta^p_n-\tilde I^\mu_n\delta^p_m,\nn\\
\tilde I^{\mu\nu}_{mn}\tilde E^n_\rho&=&\tilde I^\mu_m\delta^\nu_\rho-\tilde I^\nu_m\delta^\mu_\rho.
\ea
One also has
\be\label{84}
\frac16\epsilon^{\mu\nu\rho\sigma}\epsilon_{mnpq}\tilde E^q_\sigma~\widehat{=}~\tilde E\hat A
\{\tilde E_m{^\mu}\tilde E_n{^\nu}\tilde E_p{^\rho}\},
\ee
where $\hat A$ stands for total antisymmetrization in the indices $(mnp)$, or equivalently, in $(\mu\nu\rho)$.

\newpage\noindent
{\bf 7. Metric collective field}

On the level of the metric we can define an invariant under local Lorentz transformations by use of the scalars $H^\pm_k$ in eq. \eqref{41A1},
\be\label{G1}
\tilde g_{\mu\nu}=\frac23\Delta^2(\partial_\mu H^+_k\partial_\nu  H^-_k+\partial_\mu H^-_k\partial_\nu H^+_k).
\ee
This object involves four spinors and transform as a second rank symmetric tensor under general coordinate transformations. We may identify its expectation value with the metric
\be\label{83AA}
g_{\mu\nu}=\frac12(\kl\tilde g_{\mu\nu}\kr+\kl \tilde g_{\mu\nu}\kr^*).
\ee
We note, however, the particularity that $\tilde g_{\mu\nu}$ is a singlet with respect to global vectorlike gauge transformations $SU(2,{\mathbbm C})_F$, while $\delta\tilde g_{\mu\nu}$ acquires an inhomogeneous term for local gauge transformations. Furthermore, for the chiral gauge group $SU(2,{\mathbbm C})_L\times SU(2,{\mathbbm C})_R$ the metric collective field $\tilde g_{\mu\nu}$ is not a representation. (It is an element of the $(3,3)$ representation.) Any nonvanishing metric therefore breaks this gauge symmetry. 

One can express the collective metric field \eqref{83AA} in terms of vierbein bilinears as
\ba\label{83AB}
\tilde g_{\mu\nu}&=&\eta_{mn}\{\tilde e^m_{2,\mu}\tilde e^n_{2,\nu}+\frac13\tilde e^m_{1(k)\mu}\tilde e^n_{1(k)\nu}\nn\\
&&\quad -\frac{\Delta^2}{4}\partial_\mu\bar A^m\partial_\nu\bar A^n-\frac{\Delta^2}{12}\partial_\mu
\bar S^m_{(k)}\partial_\nu\bar S^n_{(k)}\}.
\ea
This can be verified by a reordering of Grassmann variables and recombination to bilinears in the expression
\ba\label{83AC}
\tilde g_{\mu\nu} &=&-\frac83\Delta^2\{\partial_\mu\varphi^a_{+\alpha}\varphi^b_{+\beta}\partial_\nu\varphi^c_{-\gamma}\varphi^d_{-\delta}\nn\\
&&\times (\tau_2)_{\alpha\beta}(\tau_2)_{\gamma\delta}(\tau_2\tau_k)^{ab}(\tau_2\tau_k)^{cd}+\mu\leftrightarrow \nu\}.
\ea
For $\Delta \to 0$ the last two terms in eq. \eqref{83AB} can be neglected. If we associate, for example, the vierbein with $\tilde e^m_{2,\mu}$ we observe a relation between the collective metric and the vierbein bilinear similar to the usual one between metric and vierbein. 

\medskip\noindent
{\bf 8. Emergent geometry}

The task of determining the geometry for our model of spinor gravity consists in evaluating the metric as the expectation value \eqref{83AA}. We should do so in the presence of appropriate sources for the collective field, in order to account for the response of the metric to an energy momentum tensor. The formalism of this program involves the quantum effective action $\Gamma[g_{\mu\nu}]$ for the metric. In the regularized proposal for quantum gravity that we present in sect. \ref{Discretization} all steps for the definition of the effective action are mathematically well defined. We do not aim in this paper for a computation of $\Gamma$, but rather present here shortly its definition in a continuum language. 

We first introduce sources $\tilde T^{\mu\nu}(x)$ for the collective field $\tilde g_{\mu\nu}(x)$. The partition function \eqref{1A} becomes then a functional of the sources
\be\label{83AD}
Z[\tilde T]=\int {\cal D}\exp \{-S+\frac12\int_x\big(\tilde g_{\mu\nu}(x)+\tilde g^*_{\mu\nu}(x)\big)\tilde T^{\mu\nu}\}.
\ee
The metric obtains then by a functional derivative with respect to the sources
\be\label{83AE}
g_{\mu\nu}(x)=\frac{\delta W[\tilde T]}{\delta\tilde T^{\mu\nu}(x)}~,~W(\tilde T)=\ln Z[\tilde T].
\ee
The quantum effective action is defined by a Legendre transform
\be\label{83AF}
\Gamma[g_{\mu\nu}]=-W+\int_xg_{\mu\nu}(x)\tilde T^{\mu\nu}(x).
\ee
The metric obeys the exact field equation 
\be\label{83AG}
\frac{\delta\Gamma}{\delta g_{\mu\nu}(x)}=\tilde T^{\mu\nu}(x).
\ee
We recognize the relation between the source $\tilde T^{\mu\nu}$ and the energy momentum tensor $T^{\mu\nu}$
\be\label{83AH}
\tilde T^{\mu\nu}=\frac12\sqrt{g}T^{\mu\nu}~,~g=|\det g_{\mu\nu}|.
\ee

The effective action is diffeomorphism symmetric. This is a consequence of lattice diffeomorphism invariance of the lattice action, as discussed in ref. \cite{LDI} and briefly in sect. \ref{Lattice action and diffeomorphism symmetry}. Diffeomorphism symmetry constitutes a strong restriction for the possible form of the effective action. Realistic gravity can be obtained if $\Gamma$ admits a derivative expansion for metrics with a long wavelength, for example compared to $\Delta$. In this case the leading terms are a cosmological constant (no derivatives) and an Einstein-Hilbert term involving the curvature scalar (two derivatives). (The coefficients of both terms may depend on other fields as, for example, scalar fields.) If the cosmological constant term is small enough one would find the usual geometric setting for a massless graviton. It remains to be seen if the effective action for the metric can be computed in a satisfactory approximation by using suitable methods, for example the Schwinger-Dyson equation employed in ref. \cite{CWA}.

\section{Symmetries}
\label{Symmetries}

Symmetries consist in transformations of the Grassmann variables $\psi^a_\gamma(x)\to\psi'^a_\gamma(x)$ that leave the action and the functional measure invariant. We note that symmetry transformations do not involve a complex conjugation of parameters or other coefficients in the action, in contrast to hermitean conjugation or complex conjugation. Not all symmetries must be compatible with a given complex structure. 

Besides the generalized Lorentz transformations $SO(4,{\mathbbm C})$ the action \eqref{30F} is also invariant under continuous gauge transformations. By the same argument as for local $SO(4,{\mathbbm C})$ symmetry, any global continuous symmetry of the action is also a local symmetry due to the Pauli principle. 

\newpage\noindent
{\bf 1. Vectorlike gauge symmetry and electric charge}

The vectorlike gauge symmetry $SU(2,{\mathbbm C})_F$ transforms
\be\label{S1}
\delta\varphi^a_\alpha(x)=\frac i2\tilde\alpha_k(x)(\tau_k)^{ab}\varphi^b_\alpha(x),
\ee
with three complex parameters $\tilde\alpha_k$. For real $\tilde\alpha_k$ these are standard gauge transformations with compact gauge group $SU(2)$. The basic spinors $\varphi$ transform as a doublet, and the left- and right-handed Weyl spinors $\varphi_+$ and $\varphi_-$ have the same transformation property with respect to this gauge group. The spinors are therefore in a vectorlike representation, similar to quarks with respect to the color group $SU(3)_C$, but different from the chiral representation of quarks and leptons in the standard model of electroweak interaction. We do not aim in this paper for realistic gauge symmetries of the standard model and are rather interested in a consistent theory of gravity which is as simple as possible.

As one possibility one may identify the third component of the isospin with electric charge
\be\label{S2}
Q=2I_3=\tau_3,
\ee
where $\tau_3$ acts in flavor space. Then our model describes one Dirac spinor $\varphi^1_\alpha$ with charge $Q=1$, and another Dirac spinor $\varphi^2_\alpha$ with charge $Q=-1$. At the present stage these are distinguished fermions. We will discuss elsewhere the possibility to associate $\varphi^2$ with the antiparticle of $\varphi^1$. In this case the two Dirac spinors are no longer independent and our model describes an electron, and its antiparticle, the positron. Here we concentrate, however, on the setting where the antiparticle of $\varphi^1$ differs from $\varphi^2$. 

\medskip\noindent
{\bf 2. Axial $U(1)$-symmetry}

Let us next turn to further global continuous symmetries that leave the action invariant. A global phase rotation of $\varphi$ is not a symmetry. We may, however, decompose $\varphi$ into irreducible representations of $SO(4,{\mathbbm C})$ and use different phase rotations for the different representations. Since the action contains an equal number of Weyl spinors $\varphi_+$ and $\varphi_-$ it is invariant under global chiral $U(1)_A$ transformations
\be\label{A28}
\varphi_+\to e^{i\alpha}\varphi_+~,~\varphi_-\to e^{-i\alpha}\varphi_-.
\ee

We can express $\tilde E^m_\mu$ in terms of the Weyl spinors $\varphi_\pm$ as
\be\label{A26}
\tilde E^m_\mu=\varphi_+VC_+\gamma^m_M\partial_\mu\varphi_-\pm \varphi_- V C_-\gamma^m_M\partial_\mu\varphi_+,
\ee
where the $+$ sign applies for $C=C_1$, and the $-$ sign for $C=C_2,C_2=C_1\bar\gamma$. We employ here the matrices
\be\label{107A1}
C_\pm=\frac12(1\pm\bar\gamma)C_1
\ee
defined in app. A. We observe that the relations
\ba\label{A27}
\{\bar\gamma,\gamma^m_M\}=0~,~C\bar\gamma=\bar\gamma^TC
\ea
hold both for $C_1$ and $C_2$ independently of the particular representation of the Dirac matrices. (For details cf. ref. \cite{CWMS}.) The global chiral $U(1)$ symmetry \eqref{A28} leaves $\tilde E^m_\mu$ invariant. 

A transformation which is compatible with the complex structure has to obey (for real $\alpha$)
\be\label{A29}
\varphi\to\exp (i\alpha\bar\gamma)\varphi~,~\varphi^*\to\exp(-i\alpha\bar\gamma^*)\varphi^*.
\ee
Defining
\be\label{A30}
\varphi^*_\pm=\frac12(1\pm \bar\gamma^*)\varphi^*
\ee
yields
\be\label{A31}
\varphi^*_+\to e^{-i\alpha}\varphi^*_+~,~\varphi^*_-=e^{i\alpha}\varphi^*_-,
\ee
and 
\be\label{A32}
(\tilde E^m_\mu)^*=\varphi^*_+V^*C^*_+\gamma^{m*}_M\partial_\mu\varphi^*_-\pm \varphi^*_-V^*C^*_-\gamma^{m*}_M\partial_\mu
\varphi^*_+
\ee
is invariant. We can extend this axial symmetry to $U(1,{\mathbbm C})_A$ by using complex parameters $\alpha$. Invariance under this symmetry holds for all expressions containing an equal number of factors $\varphi_+$ and $\varphi_-$. 

\medskip\noindent
{\bf 3. Chiral $SU(2)_L\times SU(2)_R$ gauge symmetry}

The gauge symmetries $SU(2)_F\times U(1)_A$ leave the vierbein bilinear \eqref{31A1a} invariant. They are, however, not the only gauge symmetries of the action \eqref{30F}. We rather can extend the symmetry $SU(2,{\mathbbm C})_F$ to a chiral gauge symmetry $SU(2,{\mathbbm C})_L\times SU(2,{\mathbbm C})_R$, where the first factor acts only on the Weyl spinors $\varphi_+$, while the second acts only on $\varphi_-$. Altogether, we have four $SU(2,{\mathbbm C})$ factors, and with respect to $G=SU(2,{\mathbbm C})_+\times SU(2,{\mathbbm C})_-\times SU(2,{\mathbbm C})_L\times SU(2,{\mathbbm C})_R$ the Weyl spinors $\varphi_+$ and $\varphi_-$ transform as $(2,1,2,1)$ and $(1,2,1,2)$, respectively. 

In order to establish the extended chiral $SU(2,{\mathbbm C})_L\times SU(2,{\mathbbm C})_R$ symmetry we write the invariant $D$ in eq. \eqref{30D} in the form \eqref{32A}. Here
\be\label{S4a}
D^+_{\mu_1\mu_2}=-\partial_{\mu_1}\varphi^{b_1}_{\beta_1}(C_+)_{\beta_1\beta_2}(\tau_2)^{b_1b_2}
\partial_{\mu_2}\varphi^{b_2}_{\beta_2}
\ee
involves only the Weyl spinor $\varphi_+$ and is invariant under the transformation $SU(2,{\mathbbm C})_L$, 
\be\label{S5a}
\delta_L\varphi^a_\alpha=\frac i4\tilde\alpha_{L,k}(\tau_k)^{ab}(1+\bar\gamma)_{\alpha\beta}\varphi^b_\beta.
\ee
Similarly
\be\label{S6a}
D^-_{\mu_1\mu_2}=\partial_{\mu_1}\varphi^{b_1}_{\beta_1}(C_-)_{\beta_1\beta_2}
(\tau_2)^{b_1b_2}
\partial_{\mu_2}\varphi^{b_2}_{\beta_2}
\ee
involves only $\varphi_-$ and is invariant under $SU(2,{\mathbbm C})_R$, with 
\be\label{S7}
\delta_R\varphi^a_\alpha=\frac i4\tilde\alpha_{R,k}(\tau_k)^{ab}(1-\bar\gamma)_{\alpha\beta}\varphi^b_\beta.
\ee
Thus $D$ is invariant under $SU(2,{\mathbbm C})_L\times SU(2,{\mathbbm C})_R$. (The vectorlike subgroup $SU(2,{\mathbbm C})_F$ obtains for $\tilde\alpha_{L,k}=\tilde\alpha_{R,k}=\tilde\alpha_k$. We note that the total symmetrization in the indices $(\eta_1\dots \eta_4)$ in eqs. \eqref{30D}, \eqref{W9} results automatically from the permutation properties of Grassmann variables.) Similarly, we write the invariant $A^{(8)}$ in eq. \eqref{30E} as
\ba\label{S8a}
A^{(8)}&=&A^+A^-,\nn\\
A^+&=&\frac14\big (\varphi^a_1(\tau_2)^{ab}\varphi^b_1\big)
\big(\varphi^c_2(\tau_2)^{cd}\varphi^d_2\big),\nn\\
A^-&=&\frac14\big(\varphi^a_3(\tau_2)^{ab}\varphi^b_3\big)
\big(\varphi^c_4(\tau_2)^{cd}\varphi^d_4\big).
\ea
The factor $A^+$ involves only the Weyl spinor $\varphi_+$ and is invariant under $SU(2,{\mathbbm C})_L$, while $A^-$ involves $\varphi_-$ and is a singlet of $SU(2,{\mathbbm C})_R$. Thus $A^{(8)}$ is invariant under global $SU(2,{\mathbbm C})_L\times SU(2,{\mathbbm C})_R$ transformations. Since all inhomogeneous pieces of local symmetry transformations vanish due to the Pauli principle, the action \eqref{30F} exhibits a local $SU(2,{\mathbbm C})_L\times SU(2,{\mathbbm C})_R$ gauge symmetry. The compact part for real $\tilde \alpha_{L,R}$ corresponds to the chiral gauge group $SU(2)_L\times SU(2)_R$. 

The vierbein bilinear \eqref{31A1a} is not invariant under chiral $SU(2,{\mathbbm C})_L\times SU(2,{\mathbbm C})_R$ transformations. We may consider 
\be\label{S9a}
(\tilde E^m_{\pm,\mu})^{ab}=\varphi^a C_\pm\gamma^m_M\partial_\mu\varphi^b
\ee
which transforms in the $(2,2)$ representation of $SU(2,{\mathbbm C})_L\times SU(2,{\mathbbm C})_R$. Any expectation value $\kl \tilde E^m_\mu\kr$ will lead to spontaneous symmetry breaking of the chiral gauge symmetry.

\medskip\noindent
{\bf 4. Extended generalized Lorentz symmetry

 ~$SO(8,{\mathbbm C})$}

The generalized Lorentz transformations $SO(4,{\mathbbm C})$ and the chiral gauge transformations $SU(2,{\mathbbm C})_L\times SU(2,{\mathbbm C})_R$ are subgroups of an extended group $SO(8,{\mathbbm C})$ which unifies the Lorentz transformations and the gauge transformations into a common group. The eight complex spinors $\varphi^a_\alpha$ transform as a vector with respect to $SO(8,{\mathbbm C})$. With respect to the subgroup $SO(4,{\mathbbm C})_+\times SO(4,{\mathbbm C})_-$ it decomposes as
\be\label{SK1}
8\to (4,1)+(1,4)
\ee
with $(4,1)$ and $(1,4)$ decomposing as  $(2,2,1,1)$ and $(1,1,2,2)$ under $SU(2,{\mathbbm C})_+\times SU(2,{\mathbbm C})_L\times SU(2,{\mathbbm C})_-\times SU(2,\mc)_R$. The totally antisymmetric combination of eight spinors $A^{(8)}$ is invariant under $SO(8,\mc)$. Thus the action is invariant under local $SO(8,\mc)$-transformations if we can replace $D$ in eq. \eqref{30F} by an invariant $D'$ which is invariant under global $SO(8,\mc)$-transformations. 

Indeed, the totally symmetric product of four vectors contains a singlet. The corresponding invariant tensor reads, with $\eta=1\dots 8$,
\be\label{SK2}
T_{\eta_1\eta_2\eta_3\eta_4}=\hat S\{\delta_{\eta_1\eta_2}\delta_{\eta_3\eta_4}\},
\ee
where $\hat S\{\}$ denotes total symmetrization over all four indices. Replacing $L$ by $T$ in eq. \eqref{30D} yields a global invariant $D'$ with all required properties. The corresponding action, with $D$ replaced by $D'$ in eq. \eqref{30F} is manifestly $SO(8,\mc)$ invariant and therefore also invariant under the subgroup $SO(4,\mc)\times SU(2,\mc)_L\times SU(2,\mc)_R$. In the present paper we will not consider this interesting possible extension of our model and stick to the action \eqref{30F} which is not $SO(8,{\mathbbm C})$ symmetric.

\medskip\noindent
{\bf 5. Discrete symmetries}

Discrete symmetries are a useful tool to characterize the properties of the model. Simple symmetries of the action \eqref{30F} are $Z_{12}$ phase-transformations or multiplications with $\bar\gamma$, or $\gamma^0$ e.g.
\be\label{51}
\varphi\to \exp (2\pi in/12)\varphi~,~\varphi\to\bar\gamma\varphi~,~\varphi\to\gamma^0\varphi.
\ee
We observe that the transformations $\varphi\to i\varphi$ and $\varphi\to \bar\gamma\varphi$ change the sign of all $\tilde E^m_\mu$. On the other hand, all transformations $\varphi\to A\varphi$ with $A^TVC\gamma^mA=VC\gamma^m$ leave $\tilde E^m_\mu$ invariant. If $\{A,\bar\gamma\}=0$ the role of $\varphi_+$ and $\varphi_-$ in eq. \eqref{A26} are interchanged. In our conventions the transformation 
\be\label{53}
\varphi(x)\to\gamma^0\varphi(x).
\ee
corresponds to the exchange of the two Weyl spinors $\varphi_+\leftrightarrow \varphi_-$. With $(\gamma^0)^TC_2\gamma^m\gamma^0=C_2\gamma^0\gamma^m\gamma^0$ this transformation changes the sign of $\tilde E^k_\mu$ for $k\neq 0$, while $\tilde E^0_\mu$ is left invariant. (This holds for $\tilde E^m_\mu$ constructed with $C_2$ as in eq. \eqref{31A1a}, while using $C_1$ results in an overall additional minus sign.) Thus $\tilde E=\det(\tilde E^m_\mu)$ changes sign under this transformation, and any invariant action in the form \eqref{A18} would have to involve a scalar $W$ that is odd under $\varphi\to \gamma^0\varphi$, as discussed in sect. \ref{Geometry}.

Discrete flavor symmetries include
\be\label{79A}
\varphi^a\to(\tau_3)^{ab}\varphi^b.
\ee
This amounts to $\varphi^2\to-\varphi^2$, with $\varphi^1$ invariant. Since $S$ contains six factors $\varphi^2$ it is invariant under this flavor reflection. With respect to this transformation the vierbeins obey $\bar E^m_{2,\mu}\to-\bar E^m_{2,\mu}~,~\bar E^m_{1(3)\mu}\to-\bar E^m_{1(3)\mu}$, while $\bar E^m_{1(1)\mu}$ and $\bar E^m_{1(2)\mu}$ are invariant. Identifying any one of those with the vierbein $\tilde E^m_\mu$ one finds that $\tilde E=\det(\tilde E^m_\mu)$ is invariant. Another flavor reflection is 
\ba\label{79B}
\varphi^a\to(\tau_1)^{ab}\varphi^b,
\ea
which amounts to an exchange $\varphi^1\leftrightarrow\varphi^2$. The factors $A^+$ and $A^-$ in eq. \eqref{32D} are invariant. Since $\tau_1\tau_2\tau_1=-\tau_1$ the invariants $S^\pm$ in eq. \eqref{W3a} both change sign, such that $L$ in eq. \eqref{W9} is invariant. Correspondingly, this transformation maps $D^\pm_{\mu\nu}\to-D^\pm_{\mu\nu},D\to D$ in eqs. \eqref{32B}, \eqref{32A}. The action is invariant. Under the transformation \eqref{79B} the possible vierbeins transform as $\bar E^m_{2,\mu}\to-\bar E^m_{2,\mu}~,~\bar E^m_{1(1)\mu}\to-\bar E^m_{1(1)\mu}$, while $\bar E^m_{1(2)\mu}$ and $\bar E^m_{1(3)\mu}$ are invariant. Again $\tilde E$ is invariant. 

The reflection of the three space coordinates
\be\label{52}
\psi^a_\gamma(x)\to \psi^a_\gamma(Px)~,~P(x^0,x^1,x^2,x^3)=(x^0,-x^1,-x^2,-x^3),
\ee
changes the sign of the action. If this transformation is accompanied by any other discrete transformation which inverts the sign of $S$ the combined transformation amounts to a type of parity symmetry. As an example, we may consider the transformation,
\be\label{81A}
\varphi^1(x)\to\gamma^0\varphi^1(x)~,~\varphi^2(x)\to \gamma^0\bar\gamma\varphi^2(x).
\ee
This results in $S\to -S$. Many other discrete transformations changing the sign of $S$ exist, as $\varphi^2\to\bar\gamma\varphi^2$ which corresponds to $\varphi^2_-\to-\varphi^2_-$, with $\varphi^1_\pm,\varphi^2_+$ invariant. Parity transformations can be constructed by combining the transformations \eqref{52} and \eqref{81A}, together with some transformation that leaves $S$ invariant. For example, the transformation 
\be\label{54}
\varphi^1(x)\to\gamma^0_M\bar\gamma\varphi^1(Px)~,~\varphi_2(x)\to\gamma^0_M\varphi^2(Px)
\ee
leaves the action invariant. 

Time reflection symmetry can be obtained in a similar way by combining $\psi^a_\gamma(x)\to\psi^a_\gamma(-Px)$ with a suitable transformation that changes the sign of $S$, as for eq. \eqref{53}. Reflections of an even number of coordinates, including the simultaneous space and time reflections, $\psi^a_\gamma(x)\to\psi^a_\gamma(-x)$, leave the action invariant.

\section{Discretization}
\label{Discretization}
In this section we formulate a regularized version of the functional integral \eqref{1A}. For this purpose we will use a lattice of space-time points. We recall that the action \eqref{30F} is invariant under $SO(4)$ and $SO(1,3)$ transformations and does not involve any metric. The regularization will therefore be valid simultaneously for a Minkowski and a euclidean theory.

\medskip\noindent
{\bf 1. Spacetime lattice}

Let us consider a four-dimensional hypercubic lattice with lattice distance $\Delta$. We distinguish between the ``even sublattice'' of points $y^\mu=\tilde y^\mu\Delta$, $\tilde y^\mu$ integer, $\Sigma_\mu\tilde y^\mu$ even, and the ``odd sublattice'' $z^\mu=\tilde z^\mu\Delta~,~\tilde z^\mu$ integer, $\Sigma_\mu\tilde z^\mu$ odd. The odd sublattice is considered as the fundamental lattice, and we associate to each position $z^\mu$ the $16$ (``real'') Grassmann variables $\psi^a_\gamma(z)$, or their complex counterpart $\varphi^a_\alpha(z)$. (We use here $z$ instead of $x$ in sect. \ref{Action and symmetries} in order to make the distinction to the continuum limit more visible.) For a finite number of lattice points the number of Grassmann variables is finite and the regularized functional integral is mathematically well defined. For example, this can be realized by a periodic lattice with $L$ lattice points on a torus in each ``direction'' $\mu$, such that the total number of lattice points is $N_L=L^4/2$. Alternatively, we could take some finite number of lattice points $L_t$ in some direction, without imposing a periodicity constraint. The continuum limit corresponds to $N_L\to \infty $ and is realized by keeping fixed $z^\mu$ with $\Delta\to 0$.

\begin{figure}[htb]
\begin{center}
\includegraphics[width=0.45\textwidth]{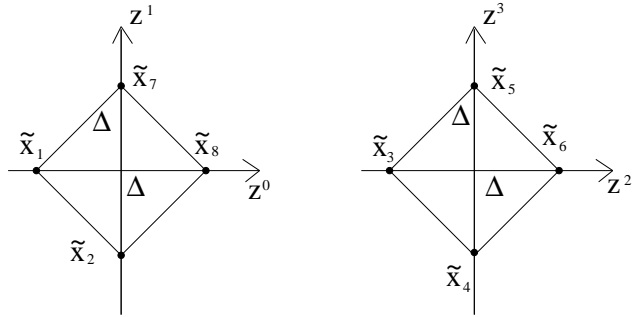}
\caption{Positions of $\tilde x_j$. The center of the squares corresponds to $\tilde y$.}
\label{fig2}
\end{center}
\end{figure}

We write the action as a sum over local terms or Lagrangians ${\cal L}(\tilde y)$, 
\be\label{L1-1}
S=\tilde\alpha\sum_{\tilde y}{\cal L}(\tilde y)+c.c.
\ee
Here $\tilde y^\mu$ denotes a position on the even sublattice or ``dual lattice''. It has eight nearest neighbors on the fundamental lattice, with unit distance from $\tilde y$. To each point $\tilde y$ we associate a ``cell'' of those eight points $\tilde x_j(\tilde y),j=1\dots 8$, with $\tilde z$-coordinates given by 
\be\label{V1}
\tilde z^\mu\big(\tilde x_j(\tilde y)\big)=\tilde y^\mu+V^\mu_j.
\ee
The eight vectors $V_j$ obey
\ba\label{V2}
V_1=(-1,0,0,0)&,&V_5=(0,0,0,1)\nn\\
V_2=(0,-1,0,0)&,&V_6=(0,0,1,0)\nn\\
V_3=(0,0,-1,0)&,&V_7=(0,1,0,0)\nn\\
V_4=(0,0,0,-1)&,&V_8=(1,0,0,0),
\ea
and we indicate the positions $\tilde x_j$ in the $z^0-z^1$ plane and the $z^2-z^3$ plane, respectively in Fig. \ref{fig2}. 
The distance between two neighboring $\tilde x_j$ is $\sqrt{2}$, and each point in the cell has six nearest neighbors. There is further an ``opposite point'' at distance $2$, with pairs of opposite points given by $(\tilde x_1,\tilde x_8),(\tilde x_2,\tilde x_7),(\tilde x_3,\tilde x_6),(\tilde x_4,\tilde x_5)$.

The Lagrangian ${\cal L}(\tilde y)$ is given by a sum of ``hyperloops''. A hyperloop is a product of an even number of Grassmann variables located at positions $\tilde x_j(\tilde y)$ within the cell at $\tilde y$. In accordance with eq. \eqref{A1} we will consider hyperloops with twelve spinors. The reason for the name ``hyperloop'' and the number twelve will become more apparent below. In a certain sense the hyperloops are a four-dimensional generalization of the plaquettes in lattice gauge theories. 

\medskip\noindent
{\bf 2. Local $SO(4,{\mathbbm C}$ symmetry}

We want to preserve the local $SO(4,{\mathbbm C})$-symmetry for the lattice regularization of spinor gravity. We therefore employ hyperloops that are invariant under local $SO(4,{\mathbbm C})$ transformations. 

Local $SO(4,{\mathbbm C})$ symmetry can be implemented by constructing the hyperloops as products of invariant bilinears involving two spinors located at the same position $\tilde x_j(\tilde y)$,
\be\label{L4a}
\tilde \h^k_\pm (\tilde x)=\varphi^a_\alpha(\tilde x)(C_\pm)_{\alpha\beta}(\tilde \tau_k)^{ab}\varphi^b_\beta(\tilde x).
\ee
Since the local $SO(4,{\mathbbm C})$ transformations \eqref{A2} involve the same $\epsilon_{mn}(\tilde x)$ for both spinors the six bilinears $\tilde \h^k_\pm$ are all invariant. The three matrices 
\be\label{123A}
\tilde\tau_k=\tau_2\tau_k
\ee
are symmetric, such that $C_\pm\otimes \tilde \tau_k$ is antisymmetric, as required by the Pauli principle.

An $SO(4,{\mathbbm C})$ invariant hyperloop can be written as a product of six factors $\tilde \h(\tilde x_j\big(\tilde y)\big)$, with $\tilde x_j$ belonging to the hypercube $\tilde y$ and obeying eq. \eqref{V1}. We will take all six positions $\tilde x_{j_1}\dots \tilde x_{j_6}$ to be different. Furthermore, we will take three factors $\tilde \h_+$ and three factors $\tilde \h_-$ in order to realize the global symmetries of the continuum limit discussed in sect. \ref{Action and symmetries}. The values of $k$ for the three factors $\tilde \h_+$ will be taken all different, and similar for the three factors $\tilde \h_-$. An invariant hyperloop is therefore fully specified by three positions $\{j_+\}=(j_1,j_2,j_3)$ for the bilinears $\tilde \h^1_+,\tilde \h^2_+$ and $\tilde \h^3_+$, and three positions $\{j_-\}=(j_4,j_5,j_6)$ for the bilinears $\tilde \h^1_-,\tilde \h^2_-$ and $\tilde \h^3_-$, such that
\ba\label{L5}
&&{\cal L}(\tilde y)=\sum_{\{j_+\},\{j_-\}}h_{\{j_+\}\{j_-\}}\\
&\times&\tilde \h^1_+(\tilde x_{j_1})\tilde \h^2_+(\tilde x_{j_2}) \tilde \h^3_+(\tilde x_{j_3})\tilde \h^1_-(\tilde x_{j_4})\tilde \h^2_-(\tilde x_{j_5})\tilde \h^3_-(\tilde x_{j_6}).\nn
\ea
We recall that all $j_m$ can take the values $1\dots 8$ according to eqs. \eqref{V1}, \eqref{V2}, and they are all different. This means that for six out of the possible eight sites $\tilde x_j(\tilde y)$ a bilinear $\tilde \h^k_+$ or $\tilde \h^k_-$ is present. For a given term the bilinears are absent only for two of the possible sites within a cell.

\medskip\noindent
{\bf 3. Lattice action}

The lattice action is a sum of local terms ${\cal L}(\tilde y)$ for all hypercubes $\tilde y$, where each ${\cal L}(\tilde y)$ is a combination of hyperloops. We consider a Lagrangian of the form 
\be\label{145A1}
{\cal L}\y=\frac{1}{24}\epsilon^{\mu\nu\rho\sigma}\f^+_{\mu\nu}\y\f^-_{\rho\sigma}\y.
\ee
Here we use the basic building blocks (no sums over $\mu,\nu$ here)
\be\label{145A2}
\f^\pm_{\mu\nu}\y=\frac{1}{12}\epsilon^{klm}\h^k_{\pm\mu\nu}\y p^\pm_{l,\mu}\y p^\pm_{m,\nu}\y,
\ee
with 
\ba\label{145A3}
\h^k_{\pm\mu\nu}\y&=&\frac14\{\tilde\h^k_\pm(\tilde y+v_\mu)+\tilde\h^k_\pm(\tilde y-v_\mu)\nn\\
&&+\tilde\h^k_{k\pm}(\tilde y+v_\nu)+\tilde \h^k_\pm(\tilde y-v_\nu)\}
\ea
and 
\be\label{145A4}
p^\pm_{k,\mu}\y=\tilde \h^k_\pm(\tilde y+v_\mu)-\tilde\h^k_\pm(\tilde y-v_\mu).
\ee
We employ lattice vectors $v_\mu$ with components $(v_\mu)^\nu=\delta^\nu_\mu$. We show in appendix G that this action is invariant under $\pi/2$-rotations in all lattice planes, and odd under the reflection of a single coordinate, as well as under diagonal reflections $\tilde z^\mu\leftrightarrow \tilde z^\nu$. Further details of the lattice geometry can be found in appendix H. 

Since all terms in ${\cal L}\y$ are products of $SO(4,{\mathbbm C})$-singlets the lattice action is invariant under local generalized Lorentz transformations. Furthermore, the contraction with the invariant $\epsilon^{klm}$ in eq. \eqref{145A2} guarantees that $\f^\pm_{\mu\nu}$ and therefore the action are invariant under global chiral $SU(2,{\mathbbm C})_L\times SU(2,{\mathbbm C})_R$ gauge transformations. The local gauge symmetry that is present in the continuum limit is not realized on the lattice level since the factors $\tilde\h^k_\pm$ are placed at different points of the cell. We discuss this issue in app. G.

In the continuum limit we can associate
\be\label{145A5}
\h^k_{\pm\mu\nu}\y\to H^\pm_k(y)~,~
p^\pm_{k,\mu}\y\to 2\Delta\partial_\mu H^\pm_k(y).
\ee
From eqs. \eqref{145A1}, \eqref{145A2} we conclude that ${\cal L}(\tilde y)$ is proportional to $\Delta^4$. On the other hand, for a computation of the action we have to convert the sum over cells $\Sigma_y$ into an integral 
\be\label{VV37}
\Delta^4\sum_{\tilde y}=\frac12\int_y,
\ee
where the factor $1/2$ accounts for the fact that the positions $\tilde y$ of the cells are placed only on the even sublattice of a hypercubic lattice. The continuum action is independent of $\Delta$, 
\ba\label{151C}
S&=&\frac{\tilde\alpha}{432}\epsilon^{klm}\epsilon^{k'l'm'}\epsilon^{\mu\nu\rho\sigma}\\
&&\int_yH^+_k\partial_\mu H^+_l\partial_\nu H^+_mH^-_{k'}\partial_\rho H^-_{l'}\partial_\sigma H^-_{m'}+c.c..\nn
\ea
In this form the continuum action is a type of unusual kinetic term for the scalar bilinears $H^\pm_k$ which involves four derivatives. The diffeomorphism symmetry is manifest by the contraction of the derivatives with the $\epsilon$-tensor.

In app. G we establish for the continuum limit the relation between $\f^\pm_{\mu\nu}$ in eq. \eqref{145A2} and $F^\pm_{\mu\nu}$ in eq. \eqref{32E}, 
\be\label{145A6}
\f^\pm_{\mu\nu}\y\to\pm 16i\Delta^2 F^\pm_{\mu\nu}(y),
\ee
such that
\be\label{VV38}
S=\frac{16}{3}\tilde\alpha\int_y\epsilon^{\mu_1\mu_2\mu_3\mu_4}F^+_{\mu_1\mu_2}
F^-_{\mu_3\mu_4}+c.c
\ee
This coincides with the continuum action \eqref{32F}, provided we choose $\tilde\alpha=3\alpha/16$. At this stage we have established three equivalent forms of the continuum action, given by eqs. \eqref{32F}, \eqref{82AA} and \eqref{151C}. The form \eqref{151C} makes the local $SO(4,{\mathbbm C})$ symmetry manifest since $H^\pm$ is invariant.

\section{Links and vierbein on the lattice}
\label{Links and vierbein}

In order to understand the appearance of geometrical objects as the vierbein bilinear in the lattice formulation it is useful to consider the building blocks ${\cal L}(\tilde y)$ of the action from a different perspective. The local $SO(4,{\mathbbm C})$ symmetry is manifest by the formulation in terms of invariant spinor bilinears at every lattice site. Alternatively, one can group the spinors into bilinears with the two spinors at neighboring lattice sites. These bilinears will show similarities to the link variables in lattice gauge theories and will therefore be called links. In a geometrical view the links have two endpoints. Under local transformations they transform according to the spinors ``sitting'' at their endpoints. We show in this section that the vierbein bilinear is closely connected to appropriate links. 

\newpage\noindent
{\bf 1. Links}

Links are bilinears involving two spinors at two different sites $\tilde x_{j_1}(\tilde y)$ and $\tilde x_{j_2}(\tilde y)$. Since both $\tilde x_{j_1}$ and $\tilde x_{j_2}$ belong to the fundamental lattice, the distance between two nearest neighbors is $\sqrt{2}\Delta$. We recall that not all pairs $(j_1,j_2)$ within a cell  are nearest neighbors, for example $\tilde x_1$ and $\tilde x_8$ specify positions with distance $2\Delta$. According to the presence of Weyl spinors $\varphi_+$ and $\varphi_-$ we distinguish four types of links labeled by $(s_1,s_2)$, 
\ba\label{L6}
&&L^{(s_1,s_2)ab}_{(k)\alpha\beta}(\tilde x_{j_1},\tilde x_{j_2})=\left(\frac{1+s_1\bar\gamma}{2}\varphi^a(\tilde x_{j_1})\right)_\alpha
\nn\\
&&\hspace{1.5cm}\times\left(\varphi^c(\tilde x_{j_2})C_1\frac{1+s_2\bar\gamma}{2}\right)_\beta(\tilde \tau_k)^{cb},
\ea
with $s_1,s_2=\pm 1$ or $+,-$. (The link $L^{(++)}$ involves two Weyl spinors $\varphi_+$, with $C_1(1\pm\bar\gamma)/2=C_\pm$). For each type $(s_1,s_2)$ we employ three links with different flavor structure, labeled by $k$. 

We will consider hyperloops that can be written as closed loops of six nearest neighbor links within a cell, with indices contracted when two links join. Indeed, one has the identity
\ba\label{L7}
&&L^{(s_1,s_2){ab}}_{(k)\alpha\beta}(\tilde x_{j_1},\tilde x_{j_2}) L^{(s_2,s_3)bc}_{(l)\beta\gamma}(\tilde x_{j_2},\tilde x_{j_3})=\nn\\
&&\hspace{1.5cm}\left(\frac{1+s_1\bar\gamma}{2}\varphi^a(\tilde x_{j_1})\right)_\alpha\tilde \h^k_{s_2}(\tilde x_{j_2})\nn\\
&&\hspace{1.5cm}\times\left (\varphi^d(\tilde x_{j_3})C_1\frac{1+s_3\bar\gamma}{2}\right)_\gamma(\tilde\tau_l)^{dc},
\ea
while $L^{(s_1,s_2)}L^{(-s_2,s_3)}=0$. This can be continued for chains of links, such that a closed loop obeys 
\ba\label{L8a}
&&L^{(s_1,s_2)a_1a_2}_{(k_2)\alpha_1\alpha_2}(\tilde x_{j_1},\tilde x_{j_2})
L^{(s_2,s_3)a_2a_3}_{(k_3)\alpha_2\alpha_3}
(\tilde x_{j_2},\tilde x_{j_3})\dots\nn\\
&&\qquad\quad\times\dots L^{(s_6,s_1)a_6a_1}_{(k_1)\alpha_6\alpha_1}(\tilde x_{j_6},\tilde x_{j_1})\nn\\
&&\qquad\quad=\tilde \h^{k_1}_{s_1}(\tilde x_{j_1})\tilde \h^{k_2}_{s_2}(\tilde x_{j_2})\dots\tilde \h^{k_6}_{s_6}(\tilde x_{j_6}).
\ea
In the sum \eqref{L5} we only include hyperloops which can be written in the form \eqref{L8} - this explains the name. We observe that hyperloops follow a similar construction principle as the plaquettes in lattice gauge theory. 

The lattice action \eqref{L1-1}, \eqref{145A1}-\eqref{145A4} can be written in the form 
\be\label{L8A}
{\cal L}(\y)=\hat s\{\tilde {\cal  C}(\tilde x_1,\tilde x_3, \tilde x_2,\tilde x_4,\tilde x_8,\tilde x_6)\},
\ee
where 
\ba\label{L8Aa}
&&\tilde {\cal  C}(\tilde x_1,\tilde x_3,\tilde x_2,\tilde x_4,\tilde x_8,\tilde x_6)=\frac{1}{36}\epsilon^{klm}
\epsilon^{pqr}\nn\\
&&\times L^{+-}_k(\tilde x_1,\tilde x_3)L^{-+}_p(\tilde x_3,\tilde x_2)L^{+-}_l
(\tilde x_2,\tilde x_4)\nn\\
&&\times L^{-+}_q(\tilde x_4,\tilde x_8) L^{+-}_m(\tilde x_8,\tilde x_6)L^{-+}_r(\tilde x_6,\tilde x_1)
\ea
denotes a combination of hyperloops with positions given by $\{j_+\}=(1,2,8)$ or $(\tilde x_1,\tilde x_2,\tilde x_8)$ and $\{j_-\}=(3,4,6)$ or $(\tilde x_3,\tilde x_4,\tilde x_6)$. In eq. \eqref{L8Aa} we have suppressed the contracted Dirac and flavor indices. The symmetrization operation $\hat s$ guarantees the proper behavior under rotations and discrete symmetries and will be specified below. We observe that ${\cal L}(\tilde y)$ indeed involves six Weyl spinors $\varphi_+$ and six Weyl spinors $\varphi_-$. If we rewrite the loop according to eq. \eqref{L8a} all invariant bilinears $\tilde \h^k_\pm(\tilde x_j)$ occur for different positions $\x j$ within the cell. (We have omitted the specification $\tilde x_j(\tilde y)$.) The invariance under local $SO(4,{\mathbbm C})$ transformations is guaranteed by construction.

The three locations $\x1,\x2,\x8$ for the Weyl spinors $\varphi_+$ define a rectangular triangle, with vectors
\ba\label{LLA1}
\x2-\x1&=&(1,-1,0,0)\nn\\
\x8-\x2&=&(1,1,0,0).
\ea
(The vector $\x8-\x1$ has length $2$ such that $\x1$ and $\x8$ are not nearest neighbors.) The same holds for the positions $\x3,\x4,\x6$ of the Weyl spinors $\varphi_-$, with orthogonal vectors
\ba\label{LLA2}
\x4-\x3&=&(0,0,1,-1),\nn\\
\x6-\x4&=&(0,0,1,1).
\ea
All four vectors $\x2-\x1,\x8-\x2,\x4-\x3$ and $\x6-\x4$ are mutually orthogonal. They may be considered as the basis vectors of a four dimensional space. The hyperloops in eq. \eqref{L8Aa} can therefore be considered as a four-dimensional objects.

We can write the expression \eqref{L8Aa} as a product of two terms
\ba\label{LLA3}
\tilde {\cal C}(\x1,\x3,\x2,\x4,\x8,\x6)=\tilde {\cal C}_+(\x1,\x2,\x8)\tilde {\cal C}_-(\x3,\x4,\x6),\nn\\
\ea
with 
\be\label{LLA4}
\tilde {\cal C}_\pm (\x {j_1},\x{j_2},\x{j_3})=\frac16\epsilon^{klm}
\tilde \h^k_\pm(\x{j_1})\tilde \h^l_\pm (\x{j_2})\tilde \h^m_\pm (\x{j_3}).
\ee
Both $\tilde {\cal C}_+(\x1,\x2,\x8)$ and $\tilde {\cal C}_-(\x3,\x4,\x6)$ can be considered as two-dimensional objects, defining each a two-dimensional plane. The two planes are orthogonal to each other. In terms of links we can write
\ba\label{LLA5}
&&\tilde {\cal C}_+(\x1,\x2,\x8)\nn\\
&&\qquad =\frac16 \epsilon^{klm}L^{++}_k(\x1,\x2)L^{++}_l(\x2,\x8)L^{++}_m(\x8,\x1),\nn\\
&&\tilde {\cal C}_-(\x3,\x4,\x6)\nn\\
&&\qquad =\frac16 \epsilon^{klm}L_k^{--}(\x3,\x4)L^{--}_l(\x4,\x6)L^{--}_m(\x6,\x3).\nn\\
\ea
We note that the expression \eqref{LLA5} does not only contain next-neighbor links. 

The symmetrization in eq. \eqref{L8A} can be understood most easily in terms of the quantities employed in app. G. From eqs. \eqref{LLA3} and \eqref{V4} it is apparent that $\tilde {\cal C}(\x1,\x3,\x2,\x4,\x8,\x6)$ has the structure of a product of the first term in $\f^{1,2,8,7}_+$ with the first term in $\f^{3,4,6,5}_-$. The symmetrization $\hat s$ in eq. \eqref{L8A} can therefore be performed in two steps,
\be\label{160A}
\hat s\{\tilde {\cal C}\}=s\big \{\bar s\{\tilde {\cal C}\}\big\}.
\ee
The first part averages over the sixteen terms that are obtained by $\pi/2$-rotations in the $z^0-z^1$ plane and $z^2-z^3$-plane, e.g. exchanging $(1,2,8)$ by $(2,8,7)$ etc.. This yields, cf. app. G,
\be\label{160B}
\bar s\big\{\tilde {\cal C}(\x1,\x3,\x2,\x4,\x8,\x6)\big\}=\f^+_{01}\f^-_{23}=\f^{1,2,8,7}_+\f^{3,4,6,5}_-.
\ee
The second step averages over the remaining rotations and is the same as in eq. \eqref{V3}, leading to eq. \eqref{VV3}, and finally to the lattice action \eqref{145A1}.

Comparing the construction of ${\cal L}(y)$ in terms of links with lattice gauge theories it becomes apparent why our discretized version does not preserve the local gauge symmetries. Joining two links produces factors $\h^k_\pm$, which are $SO(4, {\mathbbm C})$-invariant, but not invariant under the gauge transformations. Local gauge symmetry of the lattice formulation can be achieved if $\h^k_\pm$ is replaced by an object that is invariant under gauge transformations. Within our setting this is not compatible with local $SO(4,{\mathbbm C})$ transformations. Indeed, one would need to replace $\tilde \tau_k$ in eqs. \eqref{L4a} and \eqref{L6} by the invariant tensor $\tilde \tau_0=\tau_2$. However, $\tilde \tau_0$ is antisymmetric such that $C_\pm\otimes\tilde\tau_0$ is symmetric. Due to the Pauli principle such a modification of $\tilde \h$ vanishes, and therefore the product of two links vanishes. 

On the other hand, we could realize local $SU(2,{\mathbbm C})_L\times SU(2,{\mathbbm C})_R$ gauge symmetries by abandoning the generalized local Lorentz symmetry, for example replacing $(C_\pm)_{\alpha\beta}$ by the symmetric matrices $(C_\pm\tau_k)_{\alpha\beta}$. Indeed, replacing in eq. \eqref{L4a} $C_+\otimes\tau_2\tau_k$ by $C_+\tau_k\otimes\tau_2$ our model would be invariant under local $SU(2, {\mathbbm C})_L \times SU(2,{\mathbbm C})_R$ gauge transformations and global $SO(4,{\mathbbm C})$-Lorentz rotations. To some extent this issue resembles the setting of anomalies in gauge theories. It is rather obvious that a simultaneous realization of gauge symmetries and local Lorentz symmetry becomes possible for an extended flavor structure or in other dimensions. It is sufficient that an anti-symmetric tensor $A_{\eta_1\eta_2}$ exists which is invariant under both Lorentz and gauge transformations. We emphasize that our construction can also be used for a formulation of local gauge theories only in terms of spinors, without introducing gauge fields. The issue of local Lorentz symmetry is not important for this purpose, global Lorentz symmetry is sufficient. 

\medskip\noindent
{\bf 2. Lattice vierbein bilinears}

We define ``lattice vierbein bilinears'' by $(\tilde\tau_0=\tau_2,\tilde\tau_k=\tau_2\tau_k)$
\ba\label{207A1}
\bar E^m_{1(k)}(\x{j_1},\x{j_2})&=&\varphi(\x{j_1})C_1\gamma^m_M\otimes\tilde\tau_k\varphi(\x{j_2}),\nn\\
\bar E^m_2(\x{j_1},\x{j_2})&=&\varphi(\x{j_1})C_2\gamma^m_M\otimes\tau_0\varphi(\tilde x_{j_2}).
\ea
They involve spinors at two different points $\x{j_1}$ and $\x{j_2}$ of a cell $\tilde y$. The matrices $C_1\gamma^m_M\otimes\tilde\tau_k$ and $C_2\gamma^m_M\otimes\tilde\tau_0$ are symmetric, cf. app. A, such that in the continuum limit the leading contribution is a derivative. Using the definitions \eqref{31A1a}-\eqref{57B} and \eqref{V2} one obtains
\ba\label{207A2}
\bar E^m_{1(k)}(\x{j_1},\x{j_2})&\to& (V^\mu_{j_2}-V^\mu_{j_1})\tilde e^m_{1(k)\mu}(y),\nn\\
\bar E^m_2(\x{j_1},\x{j_2})&\to&(V^\mu_{j_2}-V^\mu_{j_1})\tilde e^m_{2,\mu}(y).
\ea
This establishes the direct connection between the lattice vierbeins and the continuum vierbeins. 

The lattice vierbein bilinears are linear combinations of links. This algebraic relation is discussed in appendix I. We show here only the expression of the links in terms of the vierbein bilinears and the two scalars
\ba\label{A21a}
\bar S^m_{(k)}(\x{j_1},\x{j_2})&=&\varphi(\x{j_1})C_2\gamma^m_M\otimes \tilde\tau_k\varphi(\x{j_2},\nn\\
\bar A^m(\x{j_1},\x{j_2}&=&\varphi(\x{j_1})C_1\gamma^m_M\otimes\tilde\tau_0\varphi(\x{j_2}).
\ea
(Since $C_2\gamma^m_M\otimes\tilde\tau_k$ and $C_1\gamma^m_M\otimes\tilde\tau_0$ are antisymmetric matrices, the continuum limit is given in leading order by the scalars \eqref{57C}, \eqref{57D}, $\bar S^m_k(\x{j_1},\x{j_2})\to\bar S^m_k(y),\bar A^m(\x{j_1},\x{j_2})\to\bar A^m\y$.) One finds the relation 
\ba\label{ED}
L^{(+-)ab}_{(k)\alpha\beta}(\x{j_1},\x{j_2})&=&\frac i4(\tau_m)_{\alpha\beta}
(V^m_{+k})^{ba}(\x{j_2},\x{j_1}),\\
L^{(-+)ab}_{(k)\alpha\beta}(\x{j_1},\x{j_2})&=&-\frac i4(\hat\tau_m)_{\alpha\beta}
(V^m_{-k})^{ba}(\x{j_2},\x{j_1}),\nn
\ea
with $\hat\tau_m=(-\tau_0,\tau_k)=(-1,\tau_k)$, and 
\ba\label{EP}
(V^m_{\pm k})^{ab}&=&\frac12(\bar E^m_{1(k)}\mp\bar S^m_{(k)})\delta^{ab}\nn\\
&&+\frac i2\epsilon_{klj}(\bar E^m_{1(l)}\mp\bar S^m_{(l)})(\tau_2\tau_j\tau_2)^{ab}\nn\\
&&\pm\frac12(\bar E^m_2\mp\bar A^m)(\tau_2\tau_k\tau_2)^{ab}.
\ea
This algebraic relation between the links and the lattice vierbein bilinears can be used in order to express the lattice action in terms of the lattice vierbein bilinears. The procedure is sketched in appendix J. 

\medskip\noindent
{\bf 3. Local Lorentz transformation of lattice vierbein}

The 32 bilinears $\bar E^m_{1(k)},\bar E^m_2,\bar S^m_{(k)}$ and $\bar A^m$ are a complete basis for bilinears within a cell which are formed from two different Weyl spinors $\varphi_+$ and $\varphi_-$. Local Lorentz transformations act within the space of these bilinears. Let us consider the possibility of different transformation parameters $\epsilon_{pq}(\x{j_1})$ and $\epsilon_{pq}(\x{j_2})$ and define $\bar\epsilon^{(M)}_{pq}$ and $\delta^{(M)}_{pq}$ by 
\ba\label{LL1}
\epsilon^{(M)}_{pq}(\x{j_1})&=&\bar\epsilon^{(M)}_{pq}(\x{j_1},\x{j_2})+\delta^{(M)}_{pq}(\x{j_1},\x{j_2}),\nn\\
\epsilon^{(M)}_{pq}(\x{j_2})&=&\bar \epsilon^{(M)}_{pq}(\x{j_1},\x{j_2})-\delta^{(M)}_{pq}(\x{j_1},\x{j_2}).
\ea
The generalized Lorentz transformation of the bilinear $\bar E^m_2$ becomes
\ba\label{LL2}
&&\delta\bar E^m_2(\x\je,\x\jz) =-\bar E^n_2(\x\je,\x\jz)\bar\epsilon^{(M)m}_n(\x\je,\x\jz)\nn\\
&&\qquad-\frac12\bar A^n(\x\je,\x\jz)\epsilon_n{^{mpq}}\delta^{(M)}_{pq}
(\x\je,\x\jz),
\ea
and similarly
\ba\label{LL3}
&&\delta\bar A^m(\x\je,\x\jz)=-\bar A^n(\x\je,\x\jz)\bar\epsilon^{(M)m}_n(\x\je,\x\jz)\nn\\
&&\qquad\qquad-\frac12 \bar E^n_2(\x\je,\x\jz)
\epsilon_n{^{mpq}}\delta^{(M)}_{pq}(\x\je,\x\jz).
\ea
Here the ``Lorentz-indices'' $m,n,p,q$ are raised and lowered with $\eta^{mn}$ or $\eta_{mn}$. The same transformation law applies for the pair $\bar E^M_{1(k)},\bar S^m_{(k)}$ if we replace in eqs. \eqref{LL2}, \eqref{LL3} $\bar E_2\to\bar E_{1(k)},\bar A\to\bar S_{(k)}$. We may treat the transformations with $\bar \epsilon^{(M)}_{mn}$ and $\bar\delta^{(M)}_{mn}=\frac12\epsilon_{mn}{^{pq}}\delta^{(M)}_{pq}$ as independent symmetry transformations. The first acts on the bilinears which are formed from spinors at $\x\je$ and $\x\jz$ as a global Lorentz-rotation. The second with parameter $\bar\delta^{(M)}_{mn}$ rotates, in addition, between $\bar E_2$ and $\bar A$ and between $\bar E_{1(k)}$ and $\bar S_{(k)}$. Thus the combinations $(\bar E^m_2\bar E^n_2+\bar A^m\bar A^n)\eta_{mn}$ and $(\bar E^m_{1(k)}\bar E^n_{1(k)}+\bar S^m_{(k)}\bar S^n_{(k)})\eta_{mn}$ are invariant under local Lorentz transformations. 

In the continuum limit we write
\ba\label{LL4}
\epsilon^{(M)}_{pq}(\x j)=\epsilon\M_{pq}(y)+\Delta V^\mu_j\partial_\mu\epsilon\M_{pq}(y).
\ea
The transformation \eqref{LL2} reads, with $\tilde \epsilon\M_{mn}=\frac12\epsilon_{mn}{^{pq}}\epsilon\M_{pq}$,
\ba\label{LL5}
&&\delta\bar E^{(m)}_2(\x\je,\x\jz)=-\bar E^n_2(\x\je,\x\jz)\epsilon_n^{(M)m}(y)\nn\\
&&-\frac12\Delta(V^\mu_\je-V^\mu_\jz)\bar A^n(\x\je,\x\jz)\partial_\mu\tilde\epsilon^{(M)m}_n(y),
\ea
and similar for the other transformations as eq. \eqref{LL3}. For fixed $\bar E_2$ and $\bar A$ the inhomogeneous term $\sim\partial_\mu\tilde\epsilon^{(M)m}_n$ vanishes for $\Delta\to 0$. This feature  persists if we express $\bar E^m_2(\x\je,\x\jz)$ by its continuum limit in terms of the dimensionless vierbein bilinear $\tilde e^m_{2,\mu}(y)$ according to eq. \eqref{207A2},
and similar $\bar A^m(\x\je,\x\jz)\to A^m(y)$. Again, the inhomogeneous term vanishes for $\Delta\to 0$,
\be\label{LL8}
\delta\tilde e^m_{2,\mu}=-\tilde e^n_{2,\mu}\epsilon^{(M)m}_n+\frac12\Delta A^n\partial_\mu\tilde\epsilon^{(M)m}_n.
\ee

This situation is remarkable. While the inhomogeneous term in the local Lorentz transformation of the bilinears $\tilde E^{m,ab}_{2,\mu}$ or $\bar E^m_{2,\mu}$ survives in the continuum limit and involves the bilinear $A^m$, it vanishes for $\tilde e^m_{2,\mu}$ due to the additional factor $\Delta$ in its definition \eqref{57B}. In turn, the inhomogeneous term in the local Lorentz transformation of $A^m$ involves $\Delta\tilde e^m_{2,\mu}=\Delta^2\bar E^m_{2,\mu}$ and vanishes again in the continuum limit. This suggests that the relevant vierbein bilinear for the formulation of a geometry is actually $\tilde e^m_{2,\mu}$ or $\tilde e^m_{1(k)\mu}$. This  quantity is dimensionless and transforms homogeneously under Lorentz transformations in the continuum limit. We observe that for the discrete formulation $\tilde e^m_{2,\mu}$ corresponds to bilinears of the type $\bar E^m_2$ and does not involve any factors of $\Delta$. A nonzero and finite expectation value for $\Delta\to 0$,
\be\label{LL9}
\kl \tilde e^m_{2,\mu}\kr=e^m_\mu,
\ee
or similar for $\tilde e^m_{1(k)\mu}$, seems quite reasonable. For an action formulated in terms of $\tilde e^m_\mu$ and $A^m,S^m$ we observe the presence of a factor $\Delta^{-4}$ which arises from $\Sigma_y=\frac12\Delta^{-4}\int_y$. It is an interesting question if this is canceled by a sufficient number of derivatives acting of $\tilde e^m_\mu,A^m,S^m$. For dimensionless reason such derivatives always appear in the combination $\Delta\partial_\mu$. 

\section{Lattice metric}
\label{Lattice metric}
In this section we construct a lattice metric. This is an object whose continuum limit is a symmetric second rank tensor with respect to general coordinate transformations. It is a singlet under generalized Lorentz transformations $SO(4,{\mc})$. 

\medskip\noindent
{\bf 1. Lorentz-invariant lattice metric}

Consider the $SO(4,{\mathbbm C})$-invariant object
\ba\label{L1}
&&G^{kl}_{01}(\tilde y)=\frac12
\Big\{\big(\tilde \h^k_+(\x8)-\tilde\h^k_+(\x1)\big)
\big(\h^l_-(\x7)-\tilde\h^l_-(\x2)\big)\nn\\
&&\qquad +\big(\tilde\h^k_+(\x7)-\tilde\h^k_+(\x2)\big)\big (\tilde\h^l_-(\x8)-\tilde\h^l_-(\x1)\big)\Big\}.
\ea
All spinors are placed in the $z^0-z^1$-plane of the cell, and we indicate this plane as an index $(01)$ of $G$. The reflection $z^0\leftrightarrow z^1$ leaves $G$ invariant, $G^{kl}_{10}=G^{kl}_{01}$, and the same holds for the other diagonal reflection $z^0\leftrightarrow-z^1$. On the other hand, $G^{kl}_{01}$ is odd under the reflections $z^0\to - z^0$ or $z^1\to- z^1$. Furthermore, $G^{kl}_{01}$ changes sign under a $\pi/2$-rotation in the $z^0-z^1$-plane. These are the transformation properties of the $0,1$-component of a metric tensor. We can easily generalize eq. \eqref{L1} to the other planes in the cell, defining the off-diagonal components $G_{\mu\nu},\mu\neq \nu$. For the diagonal components as $G^{kl}_{00}$ we replace in eq. \eqref{L1} $\x7\to\x8,\x2\to\x1$. We can summarize this by writing 
\ba\label{L2}
G^{kl}_{\mu\nu}&=&\frac12\Big\{\big(\tilde\h^k_+(\tilde y+v_\mu)-\tilde\h^k_+(\tilde y-v_\mu)\big)\\
&&\times\big(\tilde\h^l_-(\tilde y+v_\nu)-\tilde\h^l_-(\tilde y-v_\nu)\big)
+(\mu\leftrightarrow\nu)\big\},\nn
\ea
with $(v_\mu)^\nu=\delta^\nu_\mu$, and we note
\be\label{L6b}
G^{kl}_{\mu\nu}=G^{kl}_{\nu\mu}=G^{lk}_{\mu\nu}.
\ee

In terms of lattice derivatives and the derivative-bilinear $\D^k_{\pm\mu}$ in eq. \eqref{W8} we can express $G$ as 
\be\label{L3}
G^{kl}_{\mu\nu}=8\Delta^2\big \{ \D^k_{+\mu}\D^l_{-\nu}+(\mu\leftrightarrow\nu)\big\}.
\ee
In the continuum limit \eqref{VV22}  $G^{kl}_{\mu\nu}$ becomes indeed a second rank tensor with respect to general coordinate transformations. Writing the continuum limit in the form 
\be\label{L4}
G^{kl}_{\mu\nu}\to 2\Delta^2 (\partial_\mu H^k_+\partial_\nu H^l_-+\partial_\nu H^k_+\partial_\mu H^l_-),
\ee
the invariance under generalized Lorentz transformations is particularly transparent. With respect to $SU(2,{\mathbbm C})_F$ flavor transformations $G^{kl}_{\mu\nu}$ transforms as a singlet and a five-component symmetric tensor, and we may concentrate on the singlet 
\be\label{L6c}
\bar G_{\mu\nu}=\frac13\sum_k G^{kk}_{\mu\nu}.
\ee
(With respect to $SU(2,{\mathbbm C})_L\times SU(2,{\mathbbm C})_R$-transformations $\bar G_{\mu\nu}$ is not invariant, however. Further Lorentz-singlets transforming as second rank symmetric tensors under diffeomorphisms can be constructed by using one type of Weyl spinors $\varphi_+$ or $\varphi_-$, e.g. replacing $\h_-\to\h_+$ in eq. \eqref{L1}. Then the contraction of the type \eqref{L6c} yields $SU(2,{\mathbbm C})_L\times SU(2,{\mathbbm C})_R$ singlets.) We may identify the expectation value of $\bar G_{\mu\nu}$ with the lattice metric,
\ba\label{240A}
g_{\mu\nu}(\tilde y)&=&\frac12
\kl\bar G_{\mu\nu}(\tilde y)+\bar G^*_{\mu\nu}(\tilde y)\kr.
\ea
The continuum limit of $\bar G_{\mu\nu}$ 
\ba\label{L7}
\bar G_{\mu\nu}(\tilde y)&=&\frac16\Big\{\big(\tilde\h^k_+(\tilde y+v_\mu)-\tilde \h^k_+(\tilde y-v_\mu)\big)\nn\\
&&\times \big(\tilde \h^k_-(\tilde y+v_\nu)-\tilde \h^k_-(\tilde y-v_\nu)\big)\nn\\
&&+(\mu\leftrightarrow\nu)\Big\}.
\ea
coincides with $\tilde g_{\mu\nu}$ in eq. \eqref{G1}.

\medskip\noindent
{\bf 2. Lattice metric and vierbein}

In general relativity the vierbein and the metric are related. A corresponding, but somewhat more complicated, relation exists for the collective field $\bar G_{\mu\nu}$ and the vierbein bilinears. In order to establish this relation we reorder the four spinors in the definition of $\bar G_{\mu\nu}$. Using identities of the type \eqref{VV24} and 
\be\label{L8}
(\tilde\tau_k)^{a_1b_1}(\tilde\tau_k)^{a_2b_2}=-\frac32(\tilde\tau_0)^{a_1a_2}(\tilde\tau_0)^{b_1b_2}-\frac12(\tilde\tau_k)^{a_1a_2}
(\tilde\tau_k)^{b_1b_2},
\ee
one obtains
\ba\label{L9}
&&\bar G_{\mu\nu}=-\frac{1}{24}\Big(3(\tilde\tau_0)^{a_1a_2}(\tilde\tau_0)^{b_1b_2}+(\tilde\tau_k)^{a_1a_2}(\tilde\tau_k)^{b_1b_2}\Big)\nn\\
&&\quad \Big\{ \Big\{ \big[\varphi^{a_1}(\tilde y+v_\mu)C_+\gamma^m\varphi^{a_2}(\tilde y+v_\nu)\big]\nn\\
&&\quad \times \big[\varphi^{b_1}(\tilde y+v_\mu) C_+\gamma^n\varphi^{b_2}(\tilde y+v_\nu)\big]\delta_{mn}\nn\\
&&\quad -(v_\mu\to-v_\mu)-(v_\nu\to-v_\nu)\nn\\
&&\quad +(v_\mu\to-v_\mu~,~v_\nu\to-v_\nu)\Big\}+(\mu\leftrightarrow\nu)\Big\}.
\ea
This can be expressed in terms of the bilinears \eqref{L10a} as 
\ba\label{L11}
\bar G_{\mu\nu}&=&\frac12-\frac{1}{48}\eta_{mn}\Big\{ \big[\bar E^m_{1(k)}\bar E^n_{1(k)}
+3\bar E^m_2\bar E^n_2+\bar S^{m}_{(k)}\bar S^n_{(k)}\nn\\
&&+3\bar A^m\bar A^n\big](\tilde y+v_\mu~,~\tilde y+v_\nu)\\
&&\hspace{-1.2cm}-(v_\mu\to-v_\mu)-(v_\nu\to-v_\nu)+(v_\mu\to-v_\mu,v_\nu\to-v_\nu)\Big\}.\nn
\ea
The bilinears $\bar E^m_i,\bar S^m$ and $\bar A^m$ all transform as vectors under global $SO(4,{\mathbbm C})$-transformations. The inhomogeneous part of the local $SO(4,{\mathbbm C})$-transformations transforms these objects into each other in a way such that $\bar G_{\mu\nu}$ remains invariant, cf. eqs. \eqref{LL2}, \eqref{LL3}.

It is interesting to consider the continuum limit 
\ba\label{L12}
&&\bar E^m_i(\tilde y+v_\mu,\tilde y+v_\nu)\to\Delta(\bar E^m_{i,\nu}\y-\bar E^m_{i,\mu}\y\big),\nn\\
&&\bar S^m_{(k)}(\tilde y+v_\mu,\tilde y+v_\nu)\to\bar S^m_{(k)}\y+\frac12\Delta(\partial_\mu+\partial_\nu)
\bar S^m_{(k)}\y\nn\\
&&\qquad +\Delta^2\tilde S^m_{(k)\mu\nu}\y,\nn\\
&&\bar A^m(\tilde y+v_\mu,\tilde y+v_\nu)\to\bar A^m(y)+\frac12\Delta(\partial_\mu+\partial_\nu)\bar A^m\y\nn\\
&&\qquad +\Delta^2\tilde A^m_{\mu\nu}\y,
\ea
with
\ba\label{L13}
\tilde S^m_{(k)\mu\nu}&=&\partial_\mu\varphi C_2\gamma^m_M\otimes \tilde \tau_k\partial_\nu\varphi,\nn\\
\tilde A^m_{\mu\nu}&=&\partial_\mu\varphi C_1\gamma^m_M\otimes\tilde \tau_0\partial_\nu\varphi.
\ea
Expanding $\bar G_{\mu\nu}$ in terms of $\Delta$ we note that the lowest order terms $\sim \bar S^2,\bar A^2$ drop out. Since $v_\mu\to-v_\mu$ corresponds to $\partial_\mu\to -\partial_\mu$ there is also no term linear in $\Delta$. The leading term is quadratic in $\Delta$,
\ba\label{L14}
\bar G_{\mu\nu}&=&\frac16\Delta^2\eta_{mn}\Big\{\bar E^m_{1(k)\mu}\bar E^n_{1(k)\nu}+3\bar E^m_{2,\mu}\bar E^n_{2,\nu}\nn\\
&&-\frac14\partial_\mu\bar S^m_{(k)}\partial_\nu\bar S^n_{(k)}-\frac34\partial_\mu\bar A^m\partial_\nu\bar A^n\nn\\
&&-\bar S^m_{(k)}\tilde S^n_{(k)\mu\nu}-3\bar A^m\tilde A^m_{\mu\nu}\Big\}. 
\ea
The last two terms containing $\tilde S$ and $\tilde A$ can again be reordered and one obtains
\ba\label{L15}
\bar G_{\mu\nu}&=&\Delta^2\eta_{mn}\left\{\frac13\bar E^m_{1(k)\mu}\bar E^n_{1(k)\nu}+\bar E^m_{2,\mu}\bar E^n_{2,\nu}\right.\nn\\
&&\left.-\frac{1}{12}\partial_\mu\bar S^m_{(k)}\partial_\nu\bar S^n_{(k)}-\frac14\partial_\mu\bar A^m\partial_\nu\bar A^n\right \}.
\ea
This can also be obtained by performing a suitable reordering of Grassmann variables directly in the continuum limit \eqref{L4}, cf. \eqref{83AB}.

We may consider the case where we can approximate $\kl \bar E^m_{2,\mu}\bar E^n_{2,\mu}\kr\approx \kl\bar E^m_{2,\mu}\kr
\kl\bar E^n_{2,\mu}\kr$ and this contribution dominates. Denoting 
\be\label{L16}
\Delta\kl \bar E^m_{2,\mu}\kr=e^m_\mu,
\ee
this yields the familiar relation between the metric and the vierbein
\be\label{L17}
\kl \bar G_{\mu\nu}\kr\approx\eta_{mn}e^m_\mu e^n_\nu.
\ee
The situation is similar if a different single vierbein dominates. (Note that $\partial_\mu\bar S^m,\partial_\mu\bar A^m$ also have the transformation property of a vierbein.) For real $e^m_\mu$ we obtain a metric with a Minkowski signature, whereas other signatures can arise for complex $e^m_\mu$. The metric defining the geometry may be chosen either as 
\be\label{L18}
g^{(1)}_{\mu\nu}=\frac12\eta_{mn}(e^m_\mu e^n_\nu+e^{m*}_\mu e^{n*}_\nu),
\ee
or as 
\be\label{L19}
g^{(2)}_{\mu\nu}=\frac12\kl\bar G_{\mu\nu}+\bar G^*_{\mu\nu}\kr. 
\ee
The two definitions only differ by higher correlation functions for the spinors. 

\section{Lattice action and diffeomorphism symmetry}
\label{Lattice action and diffeomorphism symmetry}
The diffeomorphism symmetry of the continuum action can be rooted in particular properties of the lattice action. In this section we discuss this lattice origin of the invariance under general coordinate transformations. We will describe ``lattice diffeomorphism invariance'' which becomes the usual diffeomorphism symmetry in the continuum limit. Lattice diffeomorphism invariance is not a symmetry of the type discussed in sect. \ref{Symmetries}, i.e. it is not a pure transformation among Grassmann variables. It is rather related to the redundancy that appears when a formulation of the lattice action in terms of a continuous manifold is chosen. 

\medskip\noindent
{\bf 1. Lattice diffeomorphism invariance}

Coordinates are used to parametrize some continuous manifold. For a suitable given region of this manifold we will choose coordinates $x^\mu$ parametrizing a region in ${\mathbbm R}^4$. We now place the lattice points, labeled by $\{\tilde z^\mu\}$, on this manifold. This means that we associate a position $x^\mu\big(\{\tilde z^\nu\}\big)$ or $x(\tilde z)$ to each lattice point. One particular possibility is the choice $x^\mu=z^\mu=\tilde z\Delta$, that we have employed in sects.  \ref{Discretization}-\ref{Lattice metric}. In the present section we will consider now much more general choices of positions. Instead of a regular lattice we will now consider rather arbitrary irregular lattices. (This is somewhat reminiscent of random lattices. We consider, however, one given lattice and do not take averages over different lattices.)

Lattice derivatives are defined in terms of the positions of the lattice points on the manifold $x(\tilde z)$. In turn, the lattice action can be expressed in terms of lattice derivatives and average quantities within cells $\tilde y$ whose position on the manifold is some conveniently defined $x(\tilde y)$. Then the action is given as
\be\label{DL1}
S=\sum_{\tilde y}\hat {\cal L}\y,
\ee
with $\hat{ \cal L}\y$ an element of the Grassmann algebra generated by variables $\varphi\y~,~\hat\partial_\mu\varphi\y~,~\hat\partial_\mu\hat\partial_\nu\varphi\y$, etc. (More generally, if $\varphi$ is not necessarily a spinor, $\hat {\cal L}\y$ is a function of $\varphi\y~,~\hat\partial_\mu\varphi\y$ etc.).

Furthermore, we associate to each cell $\tilde y$ a ``cell volume'' $V\y$ which depends on the positions $x^\mu\big(\tilde z(\x j\y\big)$ of the lattice points $\x j\y$ belonging to the cell. An integral over a region of the manifold is defined by $\int d^4x=\sum_{\tilde y}V\y$. We introduce 
\be\label{DL2}
\bar{\cal L}\y=\hat{\cal L}\y/V\y,
\ee
such an action takes the form 
\be\label{DL3}
S=\sum_{\tilde y}V\y\bar{\cal L}\y=\int d^4 x\bar{\cal L}(x),
\ee
where $\bar {\cal L}(x)$ equals $\bar{\cal L}\y$ at the position of the cell $x\y$. Here we recall that in the discrete lattice formulation $\bar{\cal L}(x)$ is only defined for a discrete set of points, corresponding to the cell positions. Thus the action remains regularized, with a finite number of Grassmann variables in a finite region of ${\mathbbm R}^4$. The number of variables within a given region in ${\mathbbm R}^4$ may vary, however, in dependence on the assignment of positions $x\y$. The regularized definition of the integral is given by eq. \eqref{DL1}.

We may now change the positions of the lattice points to $x'(\tilde z)$, for example by an infinitesimal change 
\be\label{DL4}
x'^\mu(\tilde z)=x^\mu(\tilde z)+\xi^\mu(\tilde z),
\ee
and ask how the action depends on the assignment of positions. More precisely, since we use a fixed manifold described by the coordinates $x^\mu$, we want to know how $\bar{\cal L}\y$, expressed in terms of average fields and lattice derivatives, depends on $\xi^\mu(\tilde z)$. If $\bar{\cal L}\y$ is found to be independent of $\xi^\mu(\tilde z)$ the action is lattice diffeomorphism invariant. In this case the particular positioning of lattice points plays no role for the form of the action in terms of average fields and lattice derivatives. Lattice diffeomorphism invariance is a property of a particular class of lattice actions and will not be realized for some arbitrary lattice action. For example, the standard Wilson action for lattice gauge theories is not lattice diffeomorphism invariant.

The continuum limit of a lattice diffeomorphism invariant action for spinor gravity is rather straightforward. Lattice derivatives are replaced by ordinary derivatives. The independence on the positioning of the variables $\varphi(x)$ results in the independence of the continuum action $S=\int_x\bar{\cal L}(x)$ on the variable change $\varphi(x)\to\varphi(x-\xi)=\varphi(x)-\xi^\mu\partial_\mu\varphi(x)$. This amounts to diffeomorphism symmetry of the continuum action. Indeed, in the continuum diffeomorphisms can be formulated as a genuine symmetry corresponding to a map in the space of variables. This is due to the fact that for any given point $x$ there will be some variable, originally located at $x-\xi$, which will be moved to this position by the change of the positioning \eqref{DL4}. This property is only realized in the continuum. For a lattice typically no variable will be positioned precisely at $x$ after the change of positions \eqref{DL4}, if before a variable was located there. More precisely, the connection between lattice diffeomorphism invariance and diffeomorphism symmetry of the continuum limit both for the action and the quantum effective action can be understood in terms of interpolating functions. This issue is discussed in detail in ref. \cite{LDI}. 

We emphasize that we employ here a fixed ``coordinate manifold'' for the coordinates $x^\mu$. No metric is introduced at this stage. Physical distances do not correspond to cartesian distances on the coordinate manifold. They will rather be induced by appropriate correlation functions, as described in general term by ``geometry from general statistics'' \cite{CWG}. The appropriate metric for the computation of infinitesimal physical distances will be associated to the expectation value of a suitable collective field, as discussed in sect. \ref{Geometry}.

\medskip\noindent
{\bf 2. Lattice diffeomorphism invariant action in two dimensions}

The lattice action \eqref{DL1} as given by eq. \eqref{145A1}, is lattice diffeomorphism invariant. The basic construction principle can be understood in two dimensions and is then easily generalized to four dimensions. The concept of lattice diffeomorphism invariance is not linked in a crucial way to a formulation of the model in terms of fermions. We will therefore first present a simple example with three real bosonic fields $\h_k(\tilde z)~,~k=1,2,3$ where we restrict the lattice points to the plane spanned by $\tilde z^0,\tilde z^1$. We keep the same notation as before, i.e. cell positions $\tilde y$ given by $(\tilde y^0,\tilde y^1)$, with $\tilde y^\mu$ integers with even $\tilde y^0+\tilde y^1$. To each cell $\tilde y$ we associate four lattice points $\x1\y,\x2\y,\x7\y,\x8\y$ as given by eqs. \eqref{V1}, \eqref{V2} and depicted in the left part of fig. 1. We specify the lattice action by
\ba\label{DL5}
&&{\cal L}\y=\frac18\big[\h_3(\x1)+\h_3(\x2)+\h_3(\x7)+\h_3(\x8)\big]\nn\\
&&\quad \Big[\big(\h_1(\x8)-\h_1(\x1)\big)\big(\h_2(\x7)-\h_2(\x2)\big)\nn\\
&&\quad -\big(\h_2(\x8)-\h_2(\x1)\big)\big(\h_1(\x7)-\h_1(\x2)\big)\Big].
\ea
(We may also consider complex fields $\h_k$ and add to ${\cal L}(\tilde y)$ in eq. \eqref{DL5} its complex conjugate. Actually, two fields $\h_{1,2}$ would be sufficient and $\h_3$ could be a linear combination of them. We keep here an independent $\h_3$ for the sake of analogy with our setting of spinor gravity.)

\begin{figure}[htb]
\begin{center}
\includegraphics[width=0.25\textwidth]{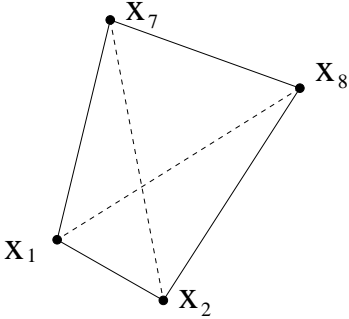}
\caption{Positions of variables in a cell}
\label{fig3}
\end{center}
\end{figure}

To each position $\x j\y$ within a cell a lattice point $\tilde z_j$ is associated by eq. \eqref{V1}, and to every $\tilde z_j$ we associate a position $x(\tilde z_j)$ on the manifold. In an intuitive notation we will denote these four positions by $x_1,x_2,x_7,x_8$. As depicted in fig. \ref{fig3} these positions are at (almost) arbitrary points in the plane. 

The volume $V\y$ associated to the cell corresponds to the surface enclosed by the solid lines in fig. \ref{fig3}. It is given by
\ba\label{DL6}
V\y&=&\frac12|(x^0_8-x^0_1)(x^1_7-x^1_2)-(x^1_8-x^1_1)(x^0_7-x^0_2)|\nn\\
&=&\frac12\epsilon_{\mu\nu}(x^\mu_8-x^\mu_1)(x^\nu_7-x^\nu_2).
\ea
For the particular case $x^\mu=\tilde z^\mu\Delta$ this yields 
\ba\label{DL6a}
x^0_8-x^0_1&=&2\Delta~,~x^1_8-x^1_1=0,\nn\\
x^1_7-x^1_2&=&2\Delta~,~x^0_7-x^0_2=0,
\ea
and therefore $V\y=2\Delta^2$. For simplicity we restrict the discussion to deformations from this simple case where $\epsilon_{\mu\nu}(x^\mu_8-x^\mu_1)(x^\nu_7-x^\nu_2)$ remains positive, i.e. $V\y>0$. We also assume that in the course of such deformations no position ever crosses any boundary line of the surface between two other positions.

The next step needs the definition of average fields and lattice derivatives. The average field takes the mean over all positions in the cell 
\be\label{DL7}
\h\y=\frac14\big\{\h(\x1)+\h(\x2)+\h(\x7)+\h(\x8)\big\}.
\ee
We also have to associate a position $x\y$ to the cell. The prescription for the location of $x\y$ is unimportant, however. We only require that $x\y$ is a point situated insider the cell volume. The definition of lattice derivatives is rather straightforward. We define them implicitly by
\ba\label{DL8}
\h(\x8)-\h(\x1)&=&(x^\mu_8-x^\mu_1)\hat\partial_\mu\h\y\nn\\
\h(\x7)-\h(\x2)&=&(x^\mu_7-x^\mu_2)\hat\partial_\mu\h\y,
\ea
assuming nonzero differences $x_8-x_1$ and $x_7-x_2$. The two equations \eqref{DL8} determine the two derivatives $\hat\partial_0\h\y$ and $\tilde\partial_1\h\y$ uniquely.

We can now compute $\hat{\cal L}\y$ from the lattice action \eqref{DL5} and find 
\ba\label{DL12}
\hat{\cal L}\y&=&\frac12\h_3\y\big[(x^0_8-x^0_1)(x^1_7-x^1_2)\nn\\
&&-(x^1_8-x^1_1)(x^0_7-x^0_2)\big]\\
&&\big[\hat\partial_0\h_1\y\hat\partial_1\h_2\y-\hat\partial_1\h_1\y\hat\partial_0\h_2\y\big].\nn
\ea
We recognize that the position dependent factor precisely corresponds to the volume $V\y$, such that
\be\label{DL13}
\bar{\cal L} =\epsilon^{\mu\nu}\h_3\y\hat\partial_\mu\h_1\y\hat\partial_\nu\h_2\y.
\ee
All dependence on the positions $x(\tilde z_j)$ has dropped out and $\bl\y$ is indeed independent of $\xi$ in eq. \eqref{DL4}. Thus the particular lattice action \eqref{DL5} is lattice diffeomorphism invariant. It is obvious that this property does not hold for an arbitrary lattice action. What is required is that the $x$-dependence of $\hl\y$ is exactly canceled by dividing out the volume $V\y$. For example, an omission of the second term in the last bracket of eq. \eqref{DL5} would not yield a lattice diffeomorphism invariant action. We also note that the lattice action \eqref{DL5} would remain lattice diffeomorphism invariant even if we leave out the factor containing $\h_3$. It would be, however, a total derivative. 

\medskip\noindent
{\bf 3. Lattice diffeomorphism invariance for four- 

~dimensional spinor gravity}

The generalization to higher dimensions is straightforward. For the four-dimensional cells discussed in sect. \ref{Discretization} the cell volume is 
\be\label{Dl14}
V\y=\frac18\epsilon_{\mu\nu\rho\sigma}(x^\mu_8-x^\mu_1)(x^\nu_7-x^\nu_2)
(x^\rho_6-x^\rho_3)(x^\sigma_5-x^\sigma_4).
\ee
The structure of a lattice diffeomorphism invariant action is now
\ba\label{DL15}
{\cal L}\y&=&{\cal M}\y A^{ABCD}\\
&&\big (\h_A(\x8)-\h_A(\x1)\big)\big(\h_B(\x7)-\h_B(\x2)\big)\nn\\
&\times&\big(\h_C(\x6)-\h_C(\x3)\big)
\big(\h_D(\x5)-\h_D(\x4)\big),\nn
\ea
with ${\cal M}\y$ containing only average fields in the cell. This structure requires at least four different bosonic fields $\h_{A,B,C,D}$ in order to permit a nonvanishing contraction with an object $A^{ABCD}$ that is totally antisymmetric in the four indices $A,B,C,D$. There may be more than four different bosons $\h_A,A=1\dots A_{{\rm max}}$, $A_{{\rm max}}\geq 4$. Expressed in terms of lattice derivatives this action yields
\ba\label{DL16}
\hl\y&=&\frac13 V\y {\cal M}\y  A^{ABCD}\epsilon^{\mu\nu\rho\sigma}\nn\\
&\times&\hat\partial_\mu\h_A\y\hat\partial_\nu\h_B\y\hat\partial_\rho\h_C\y\hat\partial_\sigma \h_D\y.
\ea
As it should be, no dependence on the positions $x_1\dots x_8$ remains after dividing out the cell volume in $\bl\y=\hl\y/V\y $. Up to terms that vanish in the continuum limit the lattice action \eqref{145A1} has the structure \eqref{DL15}. We establish this in appendix K.

\section{Conclusions and discussion}
\label{conclusions}

We have investigated a lattice regularized functional integral as a model for quantum gravity. It is based on fermion degrees of freedom, while geometrical objects as the vierbein and metric arise as expectation values of collective fields. The lattice action is invariant under local Lorentz transformations and their generalization to the complexified group $SO(4, {\mathbbm C})$. The functional integral exhibits lattice diffeomorphism invariance which implies diffeomorphism symmetry for the quantum effective action. Our model obeys the criteria (1)-(4) for a realistic lattice quantum field theory that we have mentioned in the introduction. 

One of the interesting consequences of our model is the appearance of geometrical quantities that carry flavor indices. As an example we cite the vierbein candidate $(e^m_{1,\mu})^{ab}=\Delta\kl (\tilde E^m_{1,\mu})^{ab}\kr$ in eq. \eqref{A11}. It has flavor indices $(a,b)$ and one expects different $(a,b)$ components to have different couplings to different fermion flavors or generations. It will be interesting to investigate if such a ``flavored geometry'', once implemented in a more realistic setting, is compatible with observation. We actually expect that only one linear combination of the different ``flavor vierbeins'' will remain massless, while the other components correspond to massive fields. The massless vierbein and the associated metric then still play a universal role. Possible different couplings to different fermion generations can be absorbed by a rescaling of the fermion fields. As long as only the massless vierbein is involved no observable deviation from standard gravity is expected.

This may change due to effects involving the heavy vierbein components. As long as Lorentz symmetry is preserved by all vierbeins. e.g. $(e^m_{1,\mu})^{ab}=Z^{ab}\delta^m_\mu$, it seems likely that effects from possible small expectation values of the heavy vierbeins can still be absorbed by a rescaling of the fermion fields. This would be different if the expectation values of some heavy fields violate Lorentz symmetry. Then the speed of massless fermions belonging to different generations or flavors would be different from each other, and differ from the speed of light. 

The next important step towards a realistic model for quantum gravity will be the computation of the quantum effective action for collective fields as the vierbeins $e^m_{1(k)\mu}$ and $e^m_{2,\mu}$, scalars $h^\pm_k=\kl H^\pm_k\kr$ etc.. The formal setting is straightforward by introducing sources for the various collective fields and performing a Legendre transform similar to eq. \eqref{83AF}. The general form of the continuum limit of the action is restricted by diffeomorphism symmetry. Still, many different invariants involving the various expectation values of collective bosonic fields are possible if one only uses symmetry considerations. 

A suitable method for a computation of the effective action for bilinears could be the bosonic effective action \cite{BEA} which is based on the two-particle-irreducible formalism \cite{LW}. This method has already been applied to six-fermion interactions (induced by instantons in QCD) in ref. \cite{JW}, and to a different setting of spinor gravity in ref. \cite{CWA}. The lowest order term is a ``classical action'' for the collective bilinears, which is supplemented by a fermion loop in the presence of background fields. The extremum of the bosonic effective action determines the expectation values $e^m_{2,\mu}$ etc.. The extremum condition amounts to gap equation as familiar from Schwinger-Dyson equations \cite{SD}.

The classical action replaces in the action \eqref{30F} the spinors by different combinations of bilinears, with computable coefficients. For example, it contains terms of type \eqref{82AA} or \eqref{151C} with $H^\pm_k$ replaced by $h^\pm_k$ and $\tilde e^m_{2,\mu}$ replaced by $e^m_{2,\mu}$, etc. (The coefficients differ, however, from the coefficients in eqs. \eqref{82AA} or \eqref{151C}.) There are many other similar terms, corresponding to different groupings of the fermions into bilinears. The absence of the cosmological constant invariant discussed in sect. \ref{Geometry} implies that the classical action does not contain a potential term for the scalars $h^\pm_k$ which would be of the form $W(h)\det (e^m_\mu)$. It will be interesting to see if the classical action contains an Einstein-Hilbert term of the form
\be\label{ZZA}
\int_xeF(h)R=\frac14\int_xR_{\mu\nu}{^{mn}} e_\rho{^r}e_\sigma{^s} \epsilon_{mnrs} \epsilon^{\mu\nu\rho\sigma}F(h),
\ee
with $R_{\mu\nu}{^{mn}}$ an object transforming as the curvature tensor and built from bilinears.

The classical bosonic action will directly inherit the symmetries from the action \eqref{30F}, including diffeomorphism symmetry. All coefficients of the various terms will be dimensionless if we use the bilinears $h$ and $E^m_\mu=e^m_\mu/\Delta$. In terms of these fields the lattice distance does not appear in the classical action which is therefore scale invariant. The $n$-point functions for the fields $h$ and $E^m_\mu$ derived from the classical action will depend on momentum as the only scale, without the appearance of $\Delta$. Their dependence on the momentum scale is therefore uniquely determined by appropriate dimensions. In turn, if we use the dimensionless vierbein $e_\mu^m$ instead of $E^m_\mu$ one has a factor $\Delta^{-1}$ for every factor of $e^m_\mu$ in the classical action. This situation may be changed by running couplings if the fermion loop term is included. 

This short discussion highlights that substantial work is still needed in order to see if the present model can describe realistic quantum gravity which also obeys the criteria \eqref{A2} and \eqref{A3} mentioned in the introduction. So far there is, however, no obvious obstacle in sight. The perspective of a classical bosonic action with at most four derivatives gives some hope that the loop correction does not destroy the validity of the derivative expansion for the bosonic effective action for a low number of derivatives. One should be ready, however, to explore some uncommon features as ``flavored geometry''. 

\section*{APPENDIX A: INVARIANT QUADRATIC FORMS FOR WEYL SPINORS}
\renewcommand{\theequation}{A.\arabic{equation}}
\setcounter{equation}{0}
\label{INVARIANT QUADRATIC FORMS}

Our model with two flavors allows us to construct symmetric invariants with two Dirac indices
\be\label{W3a}
S^\pm_{\eta_1\eta_2}=(S^\pm)^{b_1b_2}_{\beta_1\beta_2}
=\mp(C_\pm)_{\beta_1\beta_2}(\tau_2)^{b_1b_2},
\ee
with Pauli matrices $\tau_k$. The invariant tensors $C_\pm$ are antisymmetric
\be\label{SD}
(C_\pm)_{\beta_2\beta_1}=-(C_\pm)_{\beta_1\beta_2},
\ee
such that $S^\pm$ is symmetric under the exchange $(\beta_1,b_1)\leftrightarrow(\beta_2,b_2)$, or, in terms of the double index $\eta=(\beta,b)$, 
\be\label{SE}
S^\pm_{\eta_2\eta_1}=S^\pm_{\eta_1\eta_2}.
\ee
We will employ a representation of the Dirac matrices where $C_\pm$ are given by the $4\times 4$ matrices
\ba\label{SEa}
C_+=\left(\begin{array}{ccc}\tau_2&,&0\\0&,&0\end{array}\right)~,~
C_-=\left(\begin{array}{ccc}0&,&0\\0&,&-\tau_2\end{array}\right).
\ea

The $SO(4,{\mathbbm C})$-invariants $C_\pm$ can best be understood in terms of Weyl spinors. The matrix 
\be\label{37A}
\bar\gamma=-\gamma^0\gamma^1\gamma^2\gamma^3
\ee
commutes with $\Sigma\hmn$ such that the two doublets
\be\label{A25}
\varphi_+=\frac12(1+\bar\gamma)\varphi~,~\varphi_-=\frac12(1-\bar\gamma)\varphi
\ee
correspond to inequivalent two component complex spinor representations (Weyl spinors). We employ here a representation of the Dirac matrices $\gamma^m$ where $\bar\gamma=diag(1,1,-1,-1)$, namely
\be\label{16A}
\gamma^0=\tau_1\otimes 1~,~\gamma^k=\tau_2\otimes \tau_k.
\ee
(The general structure is independent of this choice. Our representation corresponds to the Weyl basis of ref. \cite{CWMS} where details of conventions can be found.) In this representation one has $\varphi_+=(\varphi_1,\varphi_2),\varphi_-=(\varphi_3,\varphi_4)$. We may order the double index $\eta$ or $\epsilon$ such that 
\ba\label{W1}
\varphi_{+,\eta}&=&(\varphi_1,\varphi_2,\varphi_3,\varphi_4)=(\varphi^1_1,\varphi^2_1,\varphi^1_2,\varphi^2_2)\nn\\
\varphi_{-,\eta}&=&(\varphi_5,\varphi_6,\varphi_7,\varphi_8)=(\varphi^1_3,\varphi^2_3,\varphi^1_4,\varphi^2_4),
\ea
i.e. $\beta=1,b=2$ corresponds to $\eta=2$.

According to eq. \eqref{SA} a general invariant matrix $C$ obeys
\be\label{A6H}
\Sigma^TC+C\Sigma=0.
\ee
In four dimensions, this condition implies that the matrix $C$ must be antisymmetric \cite{CWS}. There exist two matrices $C_1$ and $C_2$ which obey the condition \eqref{A6H}, which can also be written as
\be\label{A8}
C\Sigma\hmn C^{-1}=-(\Sigma\hmn)^T.
\ee
We can choose $C=C_1$ obeying 
\be\label{A9}
C_1\gamma^mC^{-1}_1=-(\gamma^m)^T~,~C^T_1=-C_1~,~C^\dagger_1C_1=1.
\ee
Then $C_1\gamma^m$  is a symmetric matrix
\be\label{A10}
(C_1\gamma^m)^T=C_1\gamma^m.
\ee
Another possible choice for $C$ obeying eq. \eqref{A8} is the antisymmetric matrix $C_2=C_1\bar\gamma$ which obeys
\be\label{22A}
C_2\gamma^mC^{-1}_2=(\gamma^m)^T~,~C^T_2=-C_2.
\ee
Now $C_2\gamma^m$ is antisymmetric
\be\label{A12}
(C_2\gamma^m)^T=-C_2\gamma^m.
\ee

The bilinears $\varphi C_1\varphi$ and $\varphi C_2\varphi$ correspond to the two singlets contained in the antisymmetric product of two Dirac spinors. In our basis one has $C_1=$diag($\tau_2,-\tau_2),~C_2=$diag$(\tau_2,\tau_2)$. The matrices $C_1$ and $C_2$ are related to $C_+$ and $C_-$ by 
\ba\label{W2}
C_+&=&\frac12(C_1+C_2)=\frac12 C_1(1+\bar\gamma),\nn\\
C_-&=&\frac12(C_1-C_2)=\frac12 C_1(1-\bar\gamma).
\ea
For the transposed spinors one has the identities $\varphi^T_\pm C_1=\varphi^T_\pm C_\pm=\varphi^TC_\pm$.

The invariant $S^+$ involves only the indices $\eta=1\dots 4$, while $S_-$ involves only $\eta=5\dots 8$. Both can be represented as $4\times 4$ matrices $\tilde S$,
\be\label{W4}
\tilde S=\left(\begin{array}{ccccccc}
0&,&0&,&0&,&1\\0&,&0&,&-1&,&0\\0&,&-1&,&0&,&0\\1&,&0&,&0&,&0
\end{array}\right),
\ee
such that the representation as $8\times 8$ matrices reads
\be\label{W5}
S^+=\left(\begin{array}{ccc}\tilde S&,&0\\0&,&0\end{array}\right)~,~
S^-=\left(\begin{array}{ccc}0&,&0\\0&,&\tilde S\end{array}\right).
\ee

It is straightforward to construct invariants only involving the two Weyl spinors $\varphi^1_+$ and $\varphi^2_+$. For this purpose we can restrict the index $\eta$ to the values $1\dots 4$. The action of $SO(4,{\mathbbm C})$ on $\varphi_+$ is given by the subgroup of complexified $SU(2,{\mathbbm C})_+$ transformations. In our basis the generators of $SU(2,{\mathbbm C})_+$ read
\be\label{W6}
\Sigma^{0k}=-\frac i2\tau_k~,~\Sigma^{kl}=\epsilon^{klm}\Sigma^{0m},
\ee
such that $\Sigma^{kl}$ is linearly dependent on $\Sigma^{0k}$. (For $SU(2,{\mathbbm C})_-$ the generators $\Sigma^{kl}$ are identical, while $\Sigma^{0k}=\frac i2\tau_k$. The subgroup of unitary transformations $SU(2)$ obtains for real transformation parameters, while we consider here arbitrary complex transformation parameters.) 

We observe that we can also consider a group $SU(2,{\mathbbm C})_L$ acting on the flavor indices. With respect to $SU(2,{\mathbbm C})_+\times SU(2,{\mathbbm C})_L$ the four component spinor $\varphi_{+,\eta}~(\eta=1\dots 4)$ transforms as the $(2,2)$ representation. Since the matrix $(\tau_2)^{ab}$ in eq. \eqref{W3a} is invariant under $SU(2,{\mathbbm C})_L$, the invariant $S^+$ is invariant under the group 
\be\label{W7}
SO(4,{\mathbbm C})_+\equiv SU (2,{\mathbbm C})_+\times SU(2,{\mathbbm C})_L. 
\ee
Here $SO(4,{\mathbbm C})_+$ should be distinguished from the generalized Lorentz transformation since it acts both in the space of Dirac and flavor indices. With respect to $SO(4,{\mathbbm C})_+$ the two-flavored spinor $\varphi_+$ transforms as a four component vector $4_v$. The classification of tensors, invariants and symmetries can be directly inferred from the analysis of four-dimensional vectors. 

Invariants only involving $\varphi_-$ can be constructed in a similar way. Now $S^-$ is left invariant under the group 
\be\label{29AA}
SO(4,{\mathbbm C})_-=SU(2,{\mathbbm C})_-\times SU(2,{\mathbbm C})_R,
\ee
which acts on the spinor and flavor indices $\varphi_-$, respectively. In consequence, the fermion bilinears
\be\label{32B}
D^\pm_{\mu_1\mu_2}=\partial_{\mu_1}\varphi_{\eta_1}S^\pm_{\eta_1\eta_2}\partial_{\mu_2}\varphi_{\eta_2}.
\ee
are invariant under generalized global Lorentz transformations
\be\label{29AB}
SO(4,{\mathbbm C})=SU(2,{\mathbbm C})_+\times SU(2,{\mathbbm C})_-,
\ee
as well as under global gauge transformations $SU(2,{\mathbbm C})_L\times SU(2,{\mathbbm C})_R$ acting on the flavor indices. 

We can also construct bilinears that do not involve derivatives and are invariant under local generalized Lorentz transformations,
\be\label{b1}
H^\pm_k=\varphi^a_\alpha(C_\pm)_{\alpha\beta}(\tau_2\tau_k)^{ab}\varphi^b_\beta.
\ee
They transform as vectors with respect to the flavor gauge symmetries $SU(2,{\mathbbm C})_L$ and $SU(2,{\mathbbm C})_R$, respectively. The bilinears $H^\pm_k$ are important building blocks for the construction of actions with local generalized Lorentz symmetry. We will employ this for our discretized setting in sect. \ref{Discretization}. We will also construct a metric collective field from derivatives of $H^\pm_k$.

\section*{APPENDIX B: COMPLEX STRUCTURES}
\renewcommand{\theequation}{B.\arabic{equation}}
\setcounter{equation}{0}
\label{Complex structure}

We can express the action \eqref{A1} or \eqref{30F} as an element of a real Grassmann algebra, based on the sixteen ``real'' Grassmann variables $\psi^a_\gamma$. On the level of $\psi$ a complex structure consists of an involution $\psi\to\psi^*=K\psi~,~K^2=1$, accompanied by a map $\psi\to I\psi~,~I^2=-1$, which anticommutes with $K$,
\be\label{C1}
K^2=1~,~I^2=-1~,~\{K,I\}=0.
\ee
The complex structure used in eqs. \eqref{AA1}, \eqref{A1} amounts to $K_1$: $\psi^a_\alpha\to\psi^a_\alpha~,~\psi^a_{\alpha+4}\to-\psi^a_{\alpha+4}$, for $\alpha=1\dots 4$, and $I_1:~\psi^a_\alpha\to-\psi^a_{\alpha+4}~,~\psi^a_{\alpha+4}\to\psi^a_\alpha$. The maps $K_1$ and $I_1$ are flavor blind and their action on the index $\gamma$ can be represented by the $8\times 8$-matrices
\be\label{C2}
K_1=\left(\begin{array}{ccc}1_4&,&0\\0&,&-1_4\end{array}\right)~,~
I_1=\left(\begin{array}{ccc}0&,&-1_4\\1_4&,&0\end{array}\right).
\ee
The action \eqref{A1} is invariant under this transformation and therefore ``real'' with respect to this complex structure.

There are many possible ways to define complex structures obeying eq. \eqref{C1}. For example, we can replace $K_1$ by $K_2:\psi_\alpha\leftrightarrow\psi_{\alpha+4}$,     
\be\label{C3}        
K_2=\left(\begin{array}{ccc}0&,&1_4\\1_4&,&0\end{array}\right),
\ee
while keeping $I=I_1$. If we keep the definition \eqref{AA1} this amounts to the map 
\be\label{C3}
K_2:~\varphi\to \varphi^{**}=i\varphi^*~,~\varphi^*\to(\varphi^*)^{**}=-i\varphi,
\ee
where $(**)$ denotes complex conjugation according to $K_2$. With respect to $K_2$ the meaning of ``real and imaginary parts'' is changed as compared to $K_1$. They are now defined as linear combinations that are even or odd with respect to $K_2,(\alpha=1\dots 4)$
\be\label{C4}
\check{\psi}_\alpha=\frac{1}{\sqrt{2}}(\psi_\alpha+\psi_{\alpha+4})~,~\check{\psi}_{\alpha+4}=\frac{1}{\sqrt{2}}
(\psi_\alpha-\psi_{\alpha+4}),
\ee
and we may replace eq. \eqref{AA1} by
\be\label{C5}
\check\varphi_\alpha=\check{\psi}_\alpha+i\check\psi_{\alpha+4}.
\ee
(In the basis $\check\psi$ the map $K_2$ is represented by $diag(1,-1)$, similar to the representation of $K_1$ in the basis $\psi$.) 

As another possibility we consider the involution $K_3:~\psi_{1,2,7,8}\to\psi_{1,2,7,8}~,~\psi_{3,4,5,6}\to-\psi_{3,4,5,6}$,
\be\label{C6} 
K_3=\left(\begin{array}{ccc}\bar\gamma&,&0\\0&,&-\bar\gamma\end{array}\right)~,~
\bar\gamma=\left(\begin{array}{ccc}
1_2&,&0\\0&,&-1_2
\end{array}\right),
\ee
again with $I=I_1$. With eq. \eqref{AA1} this amounts to $\varphi\to\bar\gamma\varphi^*~,~\varphi^*\to\bar\gamma\varphi$. Finally, we consider $K_4:~\psi_1\leftrightarrow\psi_3,~\psi_2\leftrightarrow\psi_4,~\psi_5\leftrightarrow-\psi_7,~\psi_6\leftrightarrow-\psi_8$ (again with $I=I_1$), which reads for our convention of $\gamma^0$
\be\label{C7}
K_4=\left(\begin{array}{ccc}\gamma^0&,&0\\0&,&-\gamma^0\end{array}\right)~,~
\gamma^0=\left(\begin{array}{ccc}0&,&1_2\\1_2&,&0\end{array}\right).
\ee
With eq. \eqref{AA1} one has $\varphi\to\gamma^0\varphi^*~,~\varphi^*\to\gamma^0\varphi$. Complex Grassmann variables corresponding to the complex conjugation $K_4$ are
\ba\label{AxA}
\hat\varphi_1&=&\frac{1}{\sqrt{2}}(\psi_1+\psi_3+i\psi_5+i\psi_7),\nn\\
\hat\varphi_2&=&\frac{1}{\sqrt{2}}(\psi_2+\psi_4+i\psi_6+i\psi_8),\nn\\
\hat\varphi_3&=&\frac{1}{\sqrt{2}}(\psi_5-\psi_7-i\psi_1+i\psi_3),\nn\\
\hat\varphi_4&=&\frac{1}{\sqrt{2}}(\psi_6-\psi_8-i\psi_2+i\psi_4).
\ea

The ``reality'' of the action depends on the choice of the complex structure. The action \eqref{30F} is invariant with respect to the complex conjugations $K_1$, $K_2$, $K_3$ and $K_4$. There exist other complex structures, however, where the action \eqref{30F} is odd under the complex conjugation $K$ and therefore ``imaginary''. As an example we consider the involution $K_5$ given by
\ba\label{91A}
K_5:\quad\psi^1&\to&
\left(
\begin{array}{ccccccc}
1_2&,&0&,&0&,&0\\
0&,&1_2&,&0&,&0\\
0&,&0&,&-1_2&,&0\\
0&,&0&,&0&,&-1_2
\end{array}
\right),\nn\\
\psi^2&\to &
\left(\begin{array}{ccccccc}
1_2&,&0&,&0&,&0\\
0&,&-1_2&,&0&,&0\\
0&,&0&,&-1_2&,&0\\
0&,&0&,&0&,&1_2
\end{array}
\right),
\ea
while we keep the same transformation $I_1$ given in eq. \eqref{C2}. Under the complex conjugation $K^5$ the Grassmann variables $\psi^a_\gamma$ that change sign are $\psi^1_5,\psi^1_6,\psi^1_7,\psi^1_8,\psi^2_3,\psi^2_4,\psi^2_5,\psi^2_6$. For $a=1$ we keep the same definition for complex Grassmann variables $\varphi^1_\alpha$ as in eq. \eqref{AA1}, and similar for the first two components of $\varphi^2$, i.e. $\varphi^2_1$ and $\varphi^2_2$. However, instead of $\varphi^2_3$ and $\varphi^2_4$ we use new complex variables
\be\label{91B}
\xi^2_3=\psi^2_7-i\psi^2_3~,~\xi^2_4=\psi^2_8-i\psi^2_4.
\ee

The complex conjugation is now realized by a standard complex conjugation of $\xi^2_3$ and $\xi^2_4$,
\be\label{91C}
(\xi^2_3)^*=\psi^2_7+i\psi^2_3~,~(\xi^2_4)^*=\psi^2_8+i\psi^2_4.
\ee
(Note that under the complex conjugation $K_5$ the combinations $\varphi^2_3$ and $\varphi^2_4$ transform as $K_5(\varphi^2_3)=K_5(\psi^2_3+i\psi^2_7)=-\psi^2_3+i\psi^2_7=-(\varphi^2_3)^*~,~K_5(\varphi^2_4)=K_5(\psi^2_4+i\psi^2_8)=-\psi^2_4+i\psi^2_8=-(\varphi^2_4)^*$, such that $\varphi^2_3$ and $\varphi^2_4$ are no longer mapped to $(\varphi^2_3)^*$ and $(\varphi^2_4)^*$.) The complex conjugation $K_5$ can be understood as a multiplication of the conjugation $K_1$ (corresponding to c.c. in eq. \eqref{A1}) with the transformation $\varphi^2_-\to-\varphi^2_-$ or $\varphi^2\to\bar\gamma\varphi^2$. The action changes sign under the involution $K_5$. 

If we write the euclidean action $S$ in eq. \eqref{30F} in terms of $\varphi^1_\pm,\varphi^2_+$ and $\xi^2_-=(\xi^2_3,\xi^2_4)$, it changes sign under the new complex conjugation $\varphi^1_\pm\to(\varphi^1_\pm)^*~,~\varphi^2_+\to(\varphi^2_+)^*~,~(\xi^2_-)\to(\xi^2_-)^*$, such that $S$ is purely imaginary. With respect to the complex conjugation $K_5$ the Minkowski action $S_M=iS$ is now real and symmetric, and therefore hermitean. We can use the involution $K_1$ in order to establish that $S$ is an element of a real Grassmann algebra (with Grassmann variables $\psi^a_\gamma(x)$). The complex structure based on $K_5$ can be employed to define hermiticity of $S_M$, which is related to a unitary time evolution. 

\section*{APPENDIX C: INHOMOGENEOUS LORENTZ TRANSFORMATION OF THE VIERBEIN BILINEARS}
\renewcommand{\theequation}{C.\arabic{equation}}
\setcounter{equation}{0}
\label{Inhohomgeneous Lorentz transformation}

With respect to diffeomorphisms $\tilde E^m_\mu$ transforms as a covariant vector. It also transforms as a vector with respect to global $SO(4,{\mathbbm C})$ transformations. This is seen most easily by defining the  ``euclidean vierbein bilinear''
\be\label{A19}
\tilde E^{(E)m}_\mu=\varphi V C\gamma^m\partial_\mu\varphi 
\ee
by multiplication with an appropriate factor $i$,
\be\label{A20}
\tilde E^{(E)0}_\mu=i\tilde E^0_\mu~,~\tilde E^{(E)k}_\mu=\tilde E^k_\mu.
\ee
One obtains the standard homogeneous transformation property of a vector 
\be\label{A21}
\delta\tilde E^{(E)m}_\mu=-\tilde E^{(E)n}_\mu
\epsilon_{np}\delta^{pm},
\ee
with arbitrary complex transformation parameters $\epsilon_{np}$ for $SO(4,{\mathbbm C})$. The transformation \eqref{A21} is equivalent to transformation \eqref{20A}. For real $\epsilon^{(M)}_{np}$ the ``vierbein bilinears'' $\tilde E^m_\mu$ are Lorentz-vectors with respect to global $SO(1,3)$ transformations. 

Thus the bilinears $\tilde E^m_\mu$ transform as the usual vierbein with respect to global Lorentz transformations. With respect to local $SO(4,{\mathbbm C})$ transformations, however, the vierbein bilinears do not transform homogeneously. The transformation of $\tilde E^m_\mu$ involves an inhomogeneous term proportional to the derivative of the transformation parameter. Indeed, for $x$-dependent $\epsilon_{mn}$ the variation of the euclidean vierbein bilinear $\tilde E^{(E)m}_\mu$ reads (with $\epsilon^m{_n}=\delta^{mp}\epsilon_{pn}$ and $C=C_1$)
\ba\label{31A}
\delta\tilde E^{(E)m}_\mu&=&\epsilon^m{_n}\tilde E^{(E)n}_\mu+\delta_{inh}\tilde E^{(E)m}_\mu,\nn\\
\delta_{inh}\tilde E^{(E)m}_\mu&=&-\frac14\partial_\mu\epsilon_{np}\varphi V C_1\{\gamma^m,\Sigma^{np}\}\varphi\nn\\
&=&\frac14\partial_\mu\epsilon_{np}\varphi V C_1\gamma^{mnp}\varphi\nn\\
&=&\frac14\partial_\mu\epsilon_{np}\varphi V C_2\gamma^q\varphi\epsilon_q{^{mnp}}\nn\\
&=&-\frac12\partial_\mu\tilde\epsilon^m{_n}\varphi V C_2\gamma^n\varphi.
\ea
Here $\gamma^{mnp}$ is the totally antisymmetrized product of three Dirac matrices
\be\label{31B}
\gamma^{mnp}=\gamma^{[m}\gamma^n\gamma^{p]}=\epsilon_q{^{mnp}}\bar\gamma\gamma^q,
\ee
and 
\be\label{31Aa}
\tilde\epsilon _{qm}=\frac12\epsilon_{qm}{^{np}}\epsilon_{np}.
\ee
If we choose $C_2$ instead of $C_1$ for the definition of the vierbein, the transformation \eqref{31A} applies with the role of $C_1$ and $C_2$ interchanged. (In a general formulation one may replace $C_1\to C, ~C_2\to C\bar\gamma$.) 

Local $SO(4,{\mathbbm C})$ symmetry of an action involving the vierbein bilinears requires that the inhomogeneous term vanishes when multiplied with the other factors in $S$. For our setting with twelve spinors we have already seen that this happens automatically if the action is invariant under global $SO(4,{\mathbbm C})$ transformations. A similar discussion applies for the Minkowski vierbein bilinear $\tilde E^m_\mu$ if we replace $\gamma^m\to\gamma^m_M$, $\Sigma^{np}\to\Sigma^{np}_M$, $\epsilon_{mn}\to\epsilon^{(M)}_{mn}$ and use $\eta_{mn}$ or $\eta^{mn}$ for raising and lowering indices. Furthermore, the transformation property \eqref{31A} remains also unaffected if we consider $(\tilde E^m_\mu)^{ab}$ instead of $\tilde E^m_\mu$. 

\section*{APPENDIX D: PROOF OF ABSENCE OF COSMOLOGICAL CONSTANT INVARIANT}
\renewcommand{\theequation}{D.\arabic{equation}}
\setcounter{equation}{0}
\label{PROOFS OF ABSENCE}

In this appendix we show that a cosmological constant invariant \eqref{A18} is not compatible with the symmetries of the action. This holds for an arbitrary choice of the flavor matrix $V^{ab}$ in eq. \eqref{BB1}. It is therefore not possible to reorder the spinors in eq. \eqref{30F} such that the expression \eqref{A18} is obtained.

The Lorentz invariance of $\det (\tilde E^m_\mu)$ is related to the existence of an invariant tensor $K$ which is defined as
\ba\label{31A1}
&&K^{a_1\dots a_4b_1\dots b_4}_{\alpha_1\dots\alpha_4\beta_1\dots \beta_4}=\\
&&\quad\hat S(\eta_1\dots\eta_4)\bar K_{\alpha_1\dots\alpha_4\beta_1\dots\beta_4}
V^{a_1b_1}\dots  V^{a_4b_4},\nn
\ea
with 
\ba\label{31B1}
\bar K_{\alpha_1\dots\beta_4}=\frac{1}{24}
\epsilon_{m_1m_2m_3m_4}(C\gamma^{m_1})_{\alpha_1\beta_1}\dots
(C\gamma^{m_4})_{\alpha_4\beta_4}.\nn\\
\ea
Here $C=C_1$ or $C=C_2$ and $\hat S(\eta_1\dots\eta_4)$ denotes total symmetrization over the four double indices $\eta_i=(\beta_i,b_i)$. 

Concerning the determinant of $\tilde E^m_{2,\mu}$ we observe that for antisymmetric $V^{ab}=(\tau_2)^{ab}=-V^{ba}$ the product $\bar K_{\alpha_1\dots\beta_4}V^{a_1b_1}\dots V^{a_4b_4}$ is antisymmetric with respect to the exchange of index pairs $\epsilon_j=(\alpha_j,a_j)\leftrightarrow\eta_j=(\beta_j,b_j)$. It is also symmetric with respect to $(a_j,b_j)\leftrightarrow(a_k,b_k)$ and antisymmetric with respect to $(\alpha_j,\beta_j)\leftrightarrow(\alpha_k,\beta_k)$. In consequence, the symmetrized tensor $K$ in eq. \eqref{31A1} is totally antisymmetric in the first four double-indices $(\epsilon_1,\dots,\epsilon_4)$. Similar symmetry considerations apply for other choices of $C$ or $V$. 

For arbitrary $C$ the action \eqref{A18} may be written in the form \eqref{A1} with 
\be\label{TB}
J_{\epsilon_1\dots\epsilon_8\eta_1\dots\eta_4}=\hat A(\epsilon_1\dots\epsilon_8)
\big[K_{\epsilon_1\dots\epsilon_4\eta_1\dots\eta_4}W_{\epsilon_5\epsilon_6\epsilon_7\epsilon_8}
\big],
\ee
where  $W_{\epsilon_5\epsilon_6\epsilon_7\epsilon_8}$ is an appropriate invariant tensor and $\hat A(\epsilon_1\dots\epsilon_8)$ denotes total antisymmetrization over the indices $\epsilon_1\dots\epsilon_8$. The action \eqref{A18} involves six Weyl spinors $\varphi_+$ and six Weyl spinors $\varphi_-$ and should be invariant under $SO(4,{\mathbbm C})$. Since an action with these properties is unique, one concludes that the expression \eqref{A18} either vanishes or is proportional to the action \eqref{30F}.

We next show that no invariant 
\be\label{55A}
W=W_{\epsilon_5\epsilon_6\epsilon_7\epsilon_8}
\varphi_{\epsilon_5}\varphi_{\epsilon_6}\varphi_{\epsilon_7}\varphi_{\epsilon_8}
\ee
exists which is compatible with the symmetries of the action.  Since $\tilde E$ involves four Weyl spinors $\varphi_+$ and four Weyl spinors $\varphi_-$ we infer that $W$ has to involve two Weyl spinors $\varphi_+$ and $\varphi_-$ each. The invariance under Lorentz transformations requires then $W$ to be of the form 
\be\label{49A}
W=b_{kl}H^+_kH^-_l.
\ee
The symmetry $\varphi_+\leftrightarrow\varphi_-$ leaves the action invariant. Since $\tilde E$ is odd under $\varphi_+\leftrightarrow\varphi_-$, also $W$ has to be odd under this transformation. Furthermore, $\varphi_+\leftrightarrow\varphi_-$ or $\varphi\to\gamma^0\varphi$ exchanges $H^+\leftrightarrow -H^-$, and we conclude that $b_{kl}=-b_{lk}$ must be antisymmetric. The discrete flavor symmetry $\varphi^1\leftrightarrow\varphi^2$ leaves $\tilde E$ invariant, (cf. sect. \ref{Symmetries}), and therefore $W$ must be invariant, containing two spinors $\varphi^1$ and two spinors $\varphi^2$. Under the symmetry $\varphi^2\to-\varphi^2$ the action and $\tilde E$ are invariant, such that $W$ must be invariant. This transforms $H^\pm_3\to-H^\pm_3$ such that $b_{k3}=0$ and $b_{3l}=0$ are required. At this stage $b_{kl}$ can only have the nonvanishing component $b_{12}=-b_{21}$. Under the transformation $\varphi^1\leftrightarrow\varphi^2$ one finally finds $H^\pm_1\to -H^\pm_1$ such that $b_{k1}=b_{1l}=0$ and therefore $b_{kl}=0$. This concludes the proof that no invariant $W$ exists which is compatible with the symmetries of the action. In turn, no cosmological constant invariant is compatible with the symmetries. 

\section*{APPENDIX E: FLAVORED GEOMETRICAL COLLECTIVE FIELDS}
\renewcommand{\theequation}{E.\arabic{equation}}
\setcounter{equation}{0}
\label{FLAVORED GEOMETRY}

There are various ways to group an even number of spinors to a bosonic collective field. We have already encountered several such fields, as $D^\pm_{\mu\nu},A^\pm,\tilde e^m_{1(k)\mu},\tilde e^m_{2,\mu},\bar S^m_{(k)},\bar A^m$ or products thereof. As a characteristic feature of our model of spinor gravity we observe an entanglement between gauge transformations and generalized Lorentz transformations. Many of the possible geometrical objects transform non-trivially under both. In this appendix we discuss a few of these ``flavored'' geometrical quantities, as well as some other collective bosonic fields.

An interesting type of bilinears are the fields
\ba\label{c1}
B^{\pm mn}_\mu&=&\varphi^a_\pm C_\pm\Sigma^{mn}_M(\tau_2)^{ab}\partial_\mu\varphi^b_\pm\nn\\
&=&\partial_\mu\varphi^a_\pm C_\pm\Sigma^{mn}_M(\tau_2)^{ab}\varphi^b_\pm.
\ea
(Here we use the fact that the matrices $C_\pm\Sigma^{mn}_M$ are symmetric according to eqs. \eqref{A6H}, \eqref{SD}.) The field $B^{+mn}_\mu$ transforms as a vector under general coordinate transformations and is an antisymmetric tensor under global Lorentz transformations in the representation $(3,1)$. Similarly, the vector $B^{-mn}_\mu$ belongs to the $(1,3)$ representation of the Lorentz group. Both $B^{+mn}_\mu$ and $B^{-mn}_\mu$ are singlets with respect to the $SU(2,{\mathbbm C})_L\times SU(2,{\mathbbm C})_R$ gauge transformations. With respect to local $SO(4,{\mathbbm C})$ transformations the inhomogeneous part reads
\be\label{c2}
\delta_{inh}B^{\pm mn}_\mu=-\frac12\partial_\mu\epsilon^{(M)}_{pq}\varphi^a_\pm C_\pm\Sigma^{mn}_M\Sigma^{pq}_M(\tau_2)^{ab}\varphi^b_\pm.
\ee
Due to the anticommuting property of Grassmann variables only the symmetric part of $C_\pm\Sigma^{mn}_M\Sigma^{pq}_M$ contributes. This can be written in terms of a commutator
\ba\label{c3}
C_\pm\Sigma^{mn}_M\Sigma^{pq}_M+(C_\pm\Sigma^{mn}_M\Sigma^{pq}_M)^T
=C_\pm[\Sigma^{mn}_M,\Sigma^{pq}_M].
\ea
Since $\Sigma^{mn}_M$ are generators of a Lie group their commutator is again a generator, with structure constants $f^{mnpqst}$. 
\be\label{c4}
[\Sigma^{mn}_M,\Sigma^{pq}_M]=f^{mnpqst}\Sigma^{st}_M.
\ee
The fields $B^{\pm mn}_\mu$ show a certain analogy to the gauge fields associated to Lorentz transformations, but the inhomogeneous part of their transformation differs from the standard setting. 

We may also consider scalars (with respect to diffeomorphisms)
\be\label{c5}
M^{\pm mn}=\varphi^a_\pm C_\pm\Sigma^{mn}_M(\tau_2)^{ab}\varphi^b_\pm
\ee
which transform in the $(3,1)$ or $(1,3)$ representation of $SO(4,{\mathbbm C})$. In terms of these scalars the inhomogeneous part of the transformation of $B^{+mn}_\mu$ becomes 
\ba\label{c6}
\delta_{inh}B^{\pm mn}_\mu&=&-\frac14 f^{mnpqst}\partial_\mu
\epsilon^{(M)}_{pq}M^{\pm st}\\
&=&-\frac12(\partial_\mu\epsilon^{(M)m}_pM^{\pm pn}
+\partial_\mu\epsilon^{(M)n}_p M^{\pm mp}).\nn
\ea
This is also found by noting that we can write $B_\mu$ as a derivative of $M$,
\be\label{c7}
B^{\pm mn}_\mu=\frac12\partial_\mu M^{\pm mn},
\ee
with 
\be\label{A.6A}
\delta M^{\pm mn}=-M^{\pm pn}\epsilon^{(M)m}_p-M^{\pm mp}\epsilon^{(M)n}_p.
\ee

A similar setting holds for the fields
\be\label{c9}
A^\pm_{k,\mu}=\frac12\partial_\mu H^\pm_k,
\ee
which show analogies to the gauge fields of the flavor symmetries $SU(2)_L$ and $SU(2)_R$, respectively. Again, the inhomogeneous part of the local gauge transformation differs from the standard transformation. On the other hand, $A^\pm_{k,\mu}$ are singlets with respect to local $SO(4,{\mathbbm C})$ transformations. They transform homogeneously, albeit trivially, with respect to the local generalized Lorentz transformations.

Other bilinears involving one derivative can be constructed as
\ba\label{c10}
P^{\pm mn}_{k,\mu}&=&\varphi^a_\pm C_\pm\Sigma^{mn}_M(\tau_2\tau_k)^{ab}\partial_\mu\varphi^b_\pm\nn\\
&=&-\partial_\mu\varphi^a_\pm C_\pm \Sigma^{mn}_M(\tau_2\tau_k)^{ab}\varphi^b_\pm.
\ea
Under general coordinate and global generalized Lorentz transformations they show the same transformation properties as $B^{\pm mn}_\mu$, but they are now vectors with respect to global gauge transformations $SU(2,{\mathbbm C})_L$ and $SU(2,{\mathbbm C})_R$, respectively. The inhomogeneous piece of the local Lorentz transformations reads
\ba\label{c11}
\delta_{inh}P^{\pm mn}_{k,\mu}&=&-\frac12\partial_\mu\epsilon^{(M)}_{pq}\varphi^a_\pm C_\pm
\Sigma^{mn}_M\Sigma^{pq}_M(\tau_2\tau_k)^{ab}\varphi^b_\pm\nn\\
&=&-\frac14\partial_\mu\epsilon^{(M)}_{pq}\varphi^a_\pm C_+
\{\Sigma^{mn}_M,\Sigma^{pq}_M\}(\tau_2\tau_k)^{ab}\varphi^b_\pm. \nn\\
\ea

Second rank tensors with respect to diffeomorphisms can be constructed from two derivatives. We are first interested in antisymmetric tensors. Singlets with respect to Lorentz transformations can be found as
\be\label{c12}
Q_{kl,\mu\nu}=\partial_\mu H^+_k\partial_\nu H^-_l-\partial_\nu H^+_k\partial_\mu H^-_l.
\ee
Since $Q_{\mu\nu}$ transforms homogeneously under $SO(4,{\mathbbm C})$, other tensors can be constructed by multiplying with terms not containing derivatives, as 
\be\label{c13}
R^{\pm mn}_{kl,\mu\nu}=Q_{kl,\mu\nu}M^{\pm mn}.
\ee
For every given choice of $(k,l)$ this object is antisymmetric in the first and second index pair $(\mu,\nu)$ and $(m,n)$, respectively. It has the same transformation properties as the curvature tensor multiplied with inverse vierbeins in standard geometry, i.e. $R_{\mu\nu\rho\sigma}e^{m\rho}e^{n\sigma}$. Other objects transforming as $R_{\mu\nu}{^{mn}}$ can be obtained by replacing $M^{\pm mn}$ by other combinations of scalars that transform as an antisymmetric tensor under Lorentz transformations.

From tensors as $R_{\mu\nu}{^{mn}}$ one can form invariants with respect to diffeomorphisms and local Lorentz transformations by multiplication and contraction with suitable invariants. For the object defined by eq. \eqref{c13} we may investigate the diffeomorphism and Lorentz invariants
\ba\label{c14}
I_1&=&A^{k_1k_2k_3k_4}_1R^{+m_1m_2}_{k_1k_2,\mu_1\mu_2}
R^{-m_3m_4}_{k_3k_4,\mu_3\mu_4}\nn\\
&&\times \epsilon^{\mu_1\mu_2\mu_3\mu_4}\epsilon_{m_1m_2m_3m_4},
\ea
or 
\ba\label{c15}
I_2&=&A^{k_1k_2k_3k_4}_2R^{+m_1m_2}_{k_1k_2,\mu_1\mu_2}
R^{-m_3m_4}_{k_3k_4,\mu_1\mu_2}
\epsilon^{\mu_1\mu_2\mu_3\mu_4}\nn\\
&&\times(\eta_{m_1m_3}\eta_{m_2m_4}-\eta_{m_1m_4}\eta_{m_2m_3}),
\ea
with suitable flavor structures given by $A^{k_1k_2k_3k_4}_{1,2}$. Both $I_2$ and $I_2$ involve six Weyl spinors $\varphi_+$ and six Weyl spinors $\varphi_-$. 

It is an interesting question if the action \eqref{30F} can be written in terms of $I_1$ or $I_2$. For this purpose we classify the behavior of $Q_{kl,\mu\nu}$ and $M^{+mn}$ under various discrete symmetry transformations. For $\varphi\to \bar\gamma\varphi$ both $Q$ and $M^\pm$ are invariant. For the flavor reflections $\varphi^a\to \tau^{ab}_3\varphi^b$ and $\varphi^a\to \tau^{ab}_1\varphi^b$ one finds that both $M^+$ and $M^-$ change sign. For the first transformation $Q_{kl}$ changes sign if precisely one of the indices $k,l$ equals three. For the second one the sign flip of $Q_{kl}$ occurs if one index equals one. Thus both $I_1$ and $I_2$ are invariant if for the nonvanishing values of $A^{k_1\dots k_4}_{1,2}$ every index $k=1,2,3$ occurs an even number of times, e.g. $A^{1111}$ or $A^{2233}$ etc.. For $\varphi\to\gamma^0\varphi$ we find $Q_{kl,\mu\nu}\to -Q_{lk,\mu\nu}$, while
\be\label{c15a}
M^{+kl}\leftrightarrow -M^{-kl}~,~M^{+0k}\leftrightarrow M^{-0k}.
\ee
As a result, the combination
\be\label{c16}
J_1=M^{+m_1m_2}M^{-m_3m_4}\epsilon_{m_1m_2m_3m_4}
\ee
is odd under $\varphi\to \gamma^0\varphi$, while 
\be\label{c17}
J_2=M^{+m_1m_2}M^{-m_3m_4}
(\eta_{m_1m_3}\eta_{m_2m_4}-\eta_{m_1m_4}\eta_{m_2m_3})
\ee
is even. If we require $I_1$ and $I_2$ to be invariant under $\varphi\to \gamma^0\varphi$ one needs
\be\label{c18}
A^{k_2k_1k_4k_3}_1=-A^{k_1k_2k_3k_4}_1~,~A^{k_2k_1k_4k_3}_2=A^{k_1k_2k_3k_4}_2.
\ee
This allows combinations of the type $A^{2323}_1=-A^{3232}_1$ etc., while for $A_2$ one has many simple possibilities, as $A_2=\delta^{k_1k_3}\delta^{k_2k_4}$ or $A_2=\delta^{k_1k_2}\delta^{k_3k_4}$. 

Since for six Weyl spinors $\varphi_+$ and six spinors $\varphi_-$ only one independent invariant under diffeomorphisms and generalized Lorentz transformations exists we conclude $I_1=c_1I_0,I_2=c_2I_0$, with $I_0=DA^{(8)}$, up to a total derivative. Here the coefficients $c_i$ depend on the choice of $A^{k_1\dots k_4}_i$ and may vanish. If we require invariance under the chiral  $SU(2,{\mathbbm C})_L\times SU(2,{\mathbbm C})_R$ gauge transformations only
\be\label{c19}
A^{k_1k_2k_3k_4}=\delta^{k_1k_3}\delta^{k_2k_4}
\ee
is allowed since $Q_{kl}$ belongs to the $(3,3)$ representation of this group. Then $I_2$ remains as the only candidate. With the choice \eqref{c19} one also finds that $I_2$ is odd under the transformation $\varphi^2_-\to -\varphi^2_-$, represented by $\varphi^2\to\bar\gamma\varphi^2$ or $\varphi^a_-\to(\tau_3)^{ab}\varphi^b_-$. Under this transformation $M^-$ is odd and $M^+$ even, such that both $J_1$ and $J_2$ change sign. The only component of $H^\pm_k$ that changes sign is $H^-_3$. The combination 
\ba\label{c20}
K_i&=&Q_{k_1k_2,\mu_1\mu_2}Q_{k_3k_4,\mu_3\mu_4}\epsilon^{\mu_1\mu_2\mu_3\mu_4}
A^{k_1k_2k_3k_4}_i,\nn\\
I_{1,2}&=&K_{1,2}J_{1,2}
\ea
is invariant for the choice \eqref{c19}. An action $\sim\int_yI_2$ is therefore invariant under the same parity and time reversal operations as $\int_yI_0$. We conclude $c_1=0$, while $c_2$ may differ from zero such that the action may be represented by a suitable contraction of a product of two ``curvature tensors''. 

What is remarkable in the discussion of geometric fields is the close linkage between Lorentz and flavor structures. In usual geometry these two structures are separated - geometrical objects as the vierbein carry no flavor indices. The observation that this is different in spinor gravity may perhaps be less surprising if we remember the $SO(8,{\mathbbm C})$ symmetry of an extended action which unifies the flavor gauge symmetries and the generalized Lorentz symmetries.

\section*{APPENDIX F: ACTION IN TERMS OF SPIN CONNECTION BILINEARS}
\renewcommand{\theequation}{F.\arabic{equation}}
\setcounter{equation}{0}
\label{ACTION IN TERMS OF}

In this appendix we express the action \eqref{30F} in terms of the spin connection bilinears \eqref{82AB}. On the level of fermions we can perform reorderings of Grassmann variables, relating different bosonic fields to each other. The invariant $D$ in eq. \eqref{30D}, \eqref{32A}, 
\ba\label{b3}
D&=&\epsilon^{\mu_1\mu_2\mu_3\mu_4}(\tau_2)_{\alpha_1\beta_1}(\tau_2)^{a_1b_1}
(\tau_2)_{\alpha_2\beta_2}(\tau_2)^{a_2b_2}\nn\\
&\times&\partial_{\mu_1}\varphi^{a_1}_{+\alpha_1}\partial_{\mu_2}\varphi^{b_1}_{+\beta_1}\partial_{\mu_3}\varphi^{a_2}_{-\alpha_2}\partial_{\mu_4}
\varphi^{b_2}_{-\beta_2},
\ea
can be written as a sum of products of two bilinears different from eq. \eqref{32A}, by using the identity 
\ba\label{b4}
(\tau_2)_{\alpha_1\beta_1}(\tau_2)_{\alpha_2\beta_2}&=&\frac12(\tau_2)_{\alpha_1\alpha_2}(\tau_2)_{\beta_1\beta_2}\nn\\
&&-\frac12(\tau_2\tau_k)_{\alpha_1\alpha_2}(\tau_2\tau_k)_{\beta_1\beta_2}.
\ea
This yields
\ba\label{b5}
D&=&\frac14\epsilon^{\mu_1\mu_2\mu_3\mu_4}\big\{(\partial_{\mu_1}\varphi_+C_+\otimes\tau_2\partial_{\mu_2}\varphi_-)\nn\\
&&\times (\partial_{\mu_3}\varphi_+C_+\otimes\tau_2\partial_{\mu_4}\varphi_-)\nn\\
&&-(\partial\m \varphi_+C_+\tau_k\otimes\tau_2\partial_{\mu_2}\varphi_-)
(\partial_{\mu_3}\varphi_+C_+\tau_k\otimes\tau_2\partial_{\mu_4}\varphi_-)\nn\\
&&-(\partial_{\mu_1}\varphi_+C_+\otimes\tau_2\tau_l\partial_{\mu_2}\varphi_-)
(\partial_{\mu_3}\varphi_+C_+\otimes\tau_2\tau_l\partial_{\mu_4}\varphi_-)\nn\\
&&+(\partial\m\varphi_+C_+\tau_k\otimes\tau_2\tau_l\partial_{\mu_2}\varphi_-)\nn\\
&&\times(\partial_{\mu_3}\varphi_+C_+\tau_k\otimes\tau_2\tau_l\partial_{\mu_4}\varphi_-)\nn\\
&=&\frac14\epsilon^{\mu_1\mu_2\mu_3\mu_4}\eta_{mn}\big\{(\partial\m\varphi C_+\gamma^m_M\otimes\tau_2\partial_{\mu_2}\varphi)\nn\\
&&\times(\partial_{\mu_3}\varphi C_+\gamma^n_M\otimes \tau_2\partial_{\mu_4}\varphi)\nn\\
&&-(\partial\m\varphi C_+\gamma^m_M\otimes\tau_2\tau_l\partial_{\mu_2}\varphi)\nn\\
&&\times(\partial_{\mu_3}\varphi C_+\gamma^n_M\otimes\tau_2\tau_l\partial_{\mu_4}\varphi)\big\}.
\ea
Since all contributions $\sim \partial\m\partial_{\mu_2}X$ vanish due to the contraction with the $\epsilon$-tensor, one finds 
\ba\label{b6}
D=\frac{1}{16}\epsilon^{\mu_1\mu_2\mu_3\mu_4}\eta_{mn}\big\{\bar\Omega^m_{2,\mu_1\mu_2}\bar\Omega^n_{2,\mu_3\mu_4}\nn\\
-\bar\Omega^m_{1(k)\mu_1\mu_2}\bar\Omega^n_{1(k)\mu_3\mu_4}\big\},
\ea
with 
\ba\label{b7}
\bar\Omega^m_{2,\mu_1\mu_2}&=&-\frac12(\partial\m\bar E^m_{2,\mu_2}-\partial_{\mu_2}\bar E^m_{2,\mu_1}),\nn\\
\bar\Omega^m_{1(k)\mu_1\mu_2}&=&-\frac12(\partial\m\bar E^m_{1(k)\mu_2}-\partial_{\mu_2}\bar E^m_{1(k)\mu_2}). 
\ea

The bilinears $\bar\Omega^m_{2,\mu\nu}$ and $\bar\Omega^m_{1(k)\mu\nu}$ transform as antisymmetric second rank tensors under diffeomorphisms, and they are vectors with respect to global generalized Lorentz transformations. With respect to local Lorentz transformations they can be associated to the spin connection. Indeed, for a standard vierbein $e^m_\mu$ in geometry one may define
\be\label{b8}
\Omega_{\mu\nu}{^p}=-\frac12(\partial_\mu e_\nu{^p}-\partial_\nu e_\mu{^p}),
\ee
which is related to the spin connection $\omega_{\mu mn}$ by suitable multiplications with inverse vierbeins $e_m{^\mu}$, i.e.
\ba\label{b9}
\Omega_{mnp}&=&e_m{^\mu}e_n{^\nu}\eta_{pq}\Omega_{\mu\nu}{^q},\nn\\
\omega_{\mu np}&=&-e_\mu{^m}(\Omega_{mnp}-\Omega_{mpn}-\Omega_{npm}).
\ea

Under local Lorentz transformation $\Omega_{\mu\nu}{^p}$ acquires an inhomogeneous piece
\ba\label{b10}
\delta\Omega_{\mu\nu}{^p}&=&-\Omega_{\mu\nu}{^n}\epsilon^{(M)p}_n\nn\\
&&+\frac12(e_\nu{^m}\partial_\mu\epsilon^{(M)p}_m-e_\mu{^m}\partial_\nu\epsilon^{(M)p}_m).
\ea
This is almost the transformation property of $\bar\Omega_{\mu\nu}{^p}$, e.g.
\ba\label{b11}
\delta\bar\Omega^p_{2,\mu\nu}&=&-\bar\Omega^n_{2,\mu\nu}\epsilon^{(M)p}_n\nn\\
&&+\frac12(\bar E^m_{2,\nu}\partial_\mu\epsilon^{(M)p}_m-\bar E^m_{2,\mu}\partial_\nu\epsilon^{(M)p}_m)\nn\\
&&+\frac14(\partial_\mu\tilde\epsilon^{(M)p}_n\partial_\nu\bar A^n-\partial_\nu\tilde \epsilon^{(M)p}_n\partial_\mu\bar A^n),
\ea
and similar for $\bar\Omega_{1(l)}$, with $\bar E_2$ replaced by $\bar E_{1(l)}$ and $\bar A$ by $\bar S_{(l)}$. The last piece vanishes for $\Delta\to 0$ if we consider the transformation property of 
\be\label{b11a}
\tilde \Omega^p_{2,\mu\nu}=\Delta\bar\Omega^p_{2,\mu\nu}
\ee
and replace $\bar E^m_{2,\nu}$ by $\tilde e^m_{2,\nu}$. In this limit $\tilde\Omega^p_{2,\mu\nu}$ has indeed the same transformation property \eqref{b10} as $\Omega_{\mu\nu}{^p}$ in standard geometry. 

We have now collected all pieces that we need for an expression of the action \eqref{30F} in terms of the vierbein bilinears and scalars and derivatives thereof. Using eqs. \eqref{32C}, \eqref{b2}, \eqref{b6} we infer eq. \eqref{82AA}

\section*{APPENDIX G: SYMMETRIES OF THE LATTICE ACTION AND LATTICE DERIVATIVES}
\renewcommand{\theequation}{G.\arabic{equation}}
\setcounter{equation}{0}
\label{SYMMETRIES OF THE LATTICE ACTION}

In this appendix we provide some details of our lattice formulation of spinor gravity.

\medskip\noindent
{\bf 1. Lattice symmetries and action}

For the construction of the lattice action we want to implement the same behavior under $\pi/2$ rotations and reflections as for the continuum action. Our starting point is the Lagrangian
\be\label{V3}
{\cal L}(\tilde y)=s\big\{{\cal F}_+^{1,2,8,7}(\tilde y){\cal F}^{3,4,6,5}_-(\tilde y)\big\}
\ee
with
\ba\label{V4}
&&{\cal F}^{abcd}_\pm=\frac{1}{24}\epsilon^{klm}\big[\h^k_\pm(\tilde x_a)\h^l_\pm(\tilde x_b)\h^m_\pm(\tilde x_c)\nn\\
&&\quad +\h^k_\pm(\tilde x_b)\h^l_\pm(\tilde x_c)\h^m_\pm(\tilde x_d)+\h^k_\pm(\tilde x_c)\h^l_\pm(\tilde x_d)\h^m_\pm(\tilde x_a)\nn\\
&&\quad +\h^k_\pm(\tilde x_d)\h^l_\pm(\tilde x_a)\h^m_\pm(\tilde x_b)\big].
\ea
The symbol $s$ denotes a symmetrization that will be discussed below. 

We observe that ${\cal F}^{1,2,8,7}_+$ is invariant under rotations by $\pi/2$ in the $z^0-z^1$-plane, corresponding to $\tilde x_1\to\tilde x_2~,~\tilde x_2\to\tilde x_8~,~\tilde x_8\to\tilde x_7~,~\tilde x_7\to\tilde x_1$. These rotations exchange the four terms cyclically in eq. \eqref{V4}, and we observe 
\be\label{V5}
{\cal F}^{abcd}_\pm={\cal F}^{bcda}_\pm={\cal F}^{cdab}_\pm={\cal F}^{dabc}_\pm.
\ee
A reflection $\tilde z^0\to -\tilde z^0$ exchanges $\tilde x_1\leftrightarrow \tilde x_8$, while all other $\tilde x_j$ remain invariant. (It also exchanges the positions of the cells by $\tilde y^0\to-\tilde y^0$. Since the action involves a sum over all positions $\tilde y$, it is sufficient to discuss rotations and reflections for the cell at $\tilde y=0$.) The reflection $\tilde x_1\leftrightarrow \tilde x_8$ maps 
\be\label{V6}
{\cal F}^{1,2,8,7}_+\to {\cal F}^{8,2,1,7}_+=-{\cal F}^{1,2,8,7}_+,
\ee
such that ${\cal L}(\tilde y)$ changes sign. Indeed, ${\cal F}^{abcd}_\pm$ is antisymmetric under the exchange of the two indices $a$ and $c$. This amounts to an exchange $k\leftrightarrow m$ for the first and third factor in eq. \eqref{V4}, whereas the second and fourth factor are mapped into each other, together with $k\leftrightarrow m$. The exchange $k\leftrightarrow m$ yields a minus sign due to the total antisymmetry of $\epsilon^{klm}$. 

Similarly, one finds antisymmetry in the second and fourth index of $\f$,
\be\label{VV1}
\f_\pm^{cbad}=\f^{adcb}_\pm =-\f^{abcd}_\pm,
\ee
implying that $\f^{1,2,8,7}_+$ is odd under the reflection $\tilde z^1\to -\tilde z^1$ which exchanges $\tilde x_2\leftrightarrow \tilde x_7$. Furthermore, we may consider a reflection on a diagonal in the $z^0-z^1$-plane, which exchanges simultaneously $\tilde x_1\leftrightarrow\tilde x_7$ and $\tilde x_2\leftrightarrow\tilde x_8$, resulting in $\f^{1,2,8,7}\to \f^{7,8,2,1}=-\f^{2,8,7,1}=-\f^{1,2,8,7}$. The same holds for the other diagonal reflection, $\tilde x_1\leftrightarrow\tilde x_2~,~\tilde x_7\leftrightarrow\tilde x_8$. Since $\f^{3,4,6,5}_-$ is invariant under reflections and rotations in the $z^0-z^1$-plane we conclude that ${\cal L}(y=0)$ and therefore also the action \eqref{L1} are invariant under $\pi/2$-rotations in the $z^0-z^1$-plane, while the action changes sign under the reflections $\tilde z^0\to -\tilde z^0~,~\tilde z^1\to -\tilde z^1~,~\tilde z^0\leftrightarrow \tilde z^1$ and $\tilde z^0\leftrightarrow -\tilde z^1$. These are the required symmetry properties for the continuum action. The same transformation properties hold for rotations and reflections in the $z^2-z^3$-plane. Now $\f^{1,2,8,7}_+$ is invariant, while $\f^{3,4,6,5}_-$ is even under $\pi/2$-rotations and odd under reflections.

The particular hyperlink $\f^{1,2,8,7}_+\f^{3,4,6,5}_-$ is only one out of several possibilities to place the factors $\tilde {\cal H}_+$ and $\tilde{\cal H}_-$ on two planes that are orthogonal to each other. For example, we could equally start with 
\be\label{VV2}
{\cal L}(y)=s\{\f^{1,3,8,6}_+\f^{7,4,2,5}_-\}.
\ee
Now the factors $\tilde {\cal H}_+$ all occur in the $z^0-z^2$-plane, while the factors $\tilde {\cal H}_-$ are placed in the $z^1-z^3$-plane. The hyperlink $\f^{1,3,8,6}_+\f^{7,4,2,5}_-$ can be obtained from $\f^{1,2,8,7}_+\f^{3,4,6,5}_-$ by a $\pi/2$-rotation in the $z^1-z^2$-plane, $\tilde x_2\to \tilde x_3~,~\tilde x_3\to\tilde x_7~,~\tilde x_7\to \tilde x_6~,~\tilde x_6\to \tilde x_2$. The symmetrization $s$ in eqs. \eqref{V3} and \eqref{VV2} sums over all six possibilities to place the factors $\tilde {\cal H}_+$ on the possible planes spanned by two coordinates $\tilde z^\mu$, e.g. $(0,1),(0,2)\dots (2,3)$. The signs of the hyperlinks are thereby chosen such that the six terms can be obtained from each other by $\pi/2$-rotations, similar to the pair in eqs. \eqref{V3}, \eqref{VV2}. (This guarantees that the expressions \eqref{V3} and \eqref{VV2} are indeed equal.) We can write the symmetrization explicitly as
\ba\label{VV3}
{\cal L}(\tilde y)&=&\frac16\big \{\f^{1,2,8,7}_+\f^{3,4,6,5}_-+\f^{1,3,8,6}_+\f^{7,4,2,5}_-\nn\\
&&+\f^{1,4,8,5}_+\f^{3,7,6,2}_-+(\f_+\leftrightarrow \f_-)\big \}.
\ea
Here the third term obtains from the first term by a $\pi/2$-rotation in the $z^1-z^3$ plane, $\tilde x_2\to \tilde x_4$ , $\tilde x_4\to\tilde x_7,~\tilde x_7\to \tilde x_5~,~\tilde x_5\to\tilde x_2$, while the second term is invariant under this rotation. 

As a result, ${\cal L}(\tilde y)$ is invariant under $\pi/2$-rotations in all six planes spanned by two coordinates $\tilde z^\mu$. It is also odd under all four reflections of a single coordinate, $\tilde z^\mu\to -\tilde z^\mu$. For a ``diagonal reflection as $\tilde z^1\leftrightarrow \tilde z^2$ corresponding to $\tilde x_2\leftrightarrow \tilde x_3~,~\tilde x_6\leftrightarrow \tilde x_7$ we observe $R\{\f^{1,2,8,7}_+\f^{3,4,6,5}_-\}=\f^{1,3,8,6}_+\f^{2,4,7,5}_-=-\f^{1,3,8,6}\f^{7,4,2,5}$, such that the sum of the first two terms in eq. \eqref{VV3} changes sign. The third term is odd itself, and the three remaining terms obtained by exchanging $\varphi_+\leftrightarrow\varphi_-$ show the same transformation properties as the first three terms. Thus ${\cal L}(\tilde y)$ in eq. \eqref{VV3} is odd under this reflection, and the same holds for all twelve diagonal reflections of the type $\tilde z^\mu\leftrightarrow \tilde z^\nu$ or $\tilde z^\mu\leftrightarrow-\tilde z^\nu$. The discretized action \eqref{L1} shares with the continuum action the transformation properties with respect to $\pi/2$-rotations in all $z^\mu-z^\nu$-planes, as well as reflections of single $\tilde z^\mu$ or diagonal reflections. Further details about the lattice formulation can be found in appendix H. 

Since the four points $\tilde x_1,\tilde x_2,\tilde x_8$ and $\tilde x_7$ are all in the $z^0-z^1$-plane of the cell we may switch notation and denote 
\be\label{VV17}
\f^\pm_{01}=-\f^\pm_{10}=\f^{1,2,8,7}_\pm.
\ee
Similarly, one defines
\ba\label{VV18}
\f^\pm_{02}&=&-\f^\pm_{20}=\f_\pm^{1,3,8,6},\nn\\
\f^\pm_{03}&=&-\f^\pm_{30}=\f^{1,4,8,5}_\pm,
\ea
and 
\ba\label{VV19}
\f^\pm_{12}&=&-\f^\pm_{21}=\f^{3,7,6,2}_\pm,\nn\\
\f^\pm_{13}&=&-\f^\pm_{31}=\f^{2,4,7,5},\nn\\
\f^\pm_{23}&=&-\f^\pm_{32}=\f^{3,4,6,5}_\pm.
\ea
In this notation we find the intuitive expression 
\be\label{VV21}
{\cal L}(\tilde y)=\frac{1}{24}\epsilon^{\mu_1\mu_2\mu_3\mu_4}\f^+_{\mu_1\mu_2}\f^-_{\mu_3\mu_4}.
\ee
This structure is already very similar to the continuum action \eqref{32F}. Writing
\ba\label{VV12}
\f^{1,2,8,7}_+&=&\frac{1}{24}\epsilon^{klm}\Big\{\tilde \h^k_+(\tilde x_1)\tilde \h^m_+(\tilde x_8)\big[\tilde \h^l_+(\tilde x_2)
-\tilde \h^l_+(\tilde x_7)\big]\nn\\
&&+\tilde \h^k_+(\tilde x_2)\tilde \h^m_+(\tilde x_7)\big[\tilde \h^l_+(\tilde x_8)-
\tilde \h^l_+(x_1)\big]\Big\},\\
&&\hspace{-1.5cm}=\frac{1}{48}\epsilon^{klm}\big(\tilde \h_+^k(\x1)+\tilde \h_+^k(\x2)+\tilde \h_+^k(\x7)
+\tilde \h_+^k(\x8)\big)\nn\\
&\times&\big(\tilde\h_+^l(\x8)-\tilde \h_+^l(\x1)\big)\big(\tilde\h_+^m(\x7)-\tilde \h_+^m(\x2)\big)\nn
\ea
an intuitive expression for $\f^\pm_{\mu\nu}$ is given by
\ba\label{131AA}
&&\f^\pm_{\mu\nu}(\tilde y)=\frac{1}{48}\epsilon^{klm}\\
&&\big\{\tilde\h^k_\pm(\tilde y+v_\mu)+\tilde \h^k_\pm(\tilde y-v_\mu)+\tilde \h^k_\pm(\tilde y+v_\nu)
+\tilde\h^k_\pm(\tilde y-v_\nu)\big\}\nn\\
&&\times[\tilde\h^l_\pm(\tilde y+v_\mu)-\tilde \h^l_\pm(\tilde y-v_\mu)]
[\tilde\h^m_\pm(\tilde y+v_\nu)-\tilde \h^m_\pm(\tilde y-v_\nu)].\nn
\ea
Here we employ unit lattice vectors $v_\mu$ whose components obey $(v_\mu)^\nu=\delta^\nu_\mu$. (The eight points in the cell are $\tilde y\pm v_\mu$.) The expression \eqref{131AA} agrees with eq. \eqref{145A2}. 

Finally, we note that the three components $\tilde \h^k_+$ in eq. \eqref{L4} transform as a three-component vector with respect to global $SU(2,{\mathbbm C})_F$ gauge transformations. Thus the contraction \eqref{V4} with the invariant tensor $\epsilon^{klm}$ yields a $SU(2,{\mathbbm C})_F$-singlet, and $\f^{abcd}_\pm$ is invariant under global $SU(2,{\mathbbm C})_F$ transformations. Since $\f^+_{\mu\nu}$ and $\f^-_{\mu\nu}$ are separately invariant the global gauge symmetry extends to $SU(2,{\mathbbm C})_L\times SU(2,{\mathbbm C}_R$. Again, the discretized action \eqref{V3} shares this symmetry with the continuum action. The issue of local gauge transformations will be discussed below.

\medskip\noindent
{\bf 2. Lattice derivatives}

Lattice derivatives in the $z^\mu$-directions are defined, with $\hat\partial_\mu\widehat{=}\partial/\partial z^\mu$, as
\ba\label{VV4}
\hat\partial_0\varphi(\tilde y)&=&\frac{1}{2\Delta}\big(\varphi(\tilde x_8)-\varphi(\tilde x_1)\big),\nn\\
\hat\partial_1\varphi(\tilde y)&=&\frac{1}{2\Delta}\big(\varphi(\tilde x_7)-\varphi(\tilde x_2)\big),\nn\\
\hat\partial_2\varphi(\tilde y)&=&\frac{1}{2\Delta}\big(\varphi(\tilde x_6)-\varphi(\tilde x_3)\big),\nn\\
\hat\partial_3\varphi(\tilde y)&=&\frac{1}{2\Delta}\big(\varphi(\tilde x_5)-\varphi(\tilde x_4)\big). 
\ea
Here we have suppressed the spinor and flavor indices of $\varphi^a_\alpha$, and $\tilde x_j$ stands for $\tilde x_j(\tilde y)$. Note that we associate the lattice derivatives with positions $\tilde y$ on the dual lattice. To each position $\tilde y$ of a cell we can also associate ``average spinors''
\ba\label{VV5}
\bar\varphi_0(\tilde y)&=&\frac12\big(\varphi(\tilde x_1)+\varphi(\tilde x_8)\big)~,~\bar\varphi_1
(\tilde y)=\frac12\big(\varphi(\tilde x_2)+\varphi(\tilde x_7)\big),\nn\\
\bar\varphi_2(\tilde y)&=&\frac12\big(\varphi(\tilde x_3)+\varphi(\tilde x_6)\big)~,~\bar\varphi_3
(\tilde y)=\frac12\big(\varphi(\tilde x_4)+\varphi(\tilde x_5)\big),\nn\\
\ea
and we write for each cell $\tilde y$
\ba\label{VV6}
\varphi(\tilde x_1)&=&\bar\varphi_0-\Delta\hat\partial_0\varphi~,~\varphi(\tilde x_8)=\bar\varphi_0+\Delta\hat\partial_0\varphi,
\nn\\
\varphi(\tilde x_2)&=&\bar\varphi_1-\Delta\hat\partial_1\varphi~,~\varphi(\tilde x_7)=\bar\varphi_1+\Delta\hat\partial_1\varphi,\nn\\
\varphi(\tilde x_3)&=&\bar\varphi_2-\Delta\hat\partial_2\varphi~,~\varphi(\tilde x_6)=\bar\varphi_2+\Delta\hat\partial_2\varphi,\nn\\
\varphi(\tilde x_4)&=&\bar\varphi_3-\Delta\hat\partial_3\varphi~,~\varphi(\tilde x_5)=\bar\varphi_3+\Delta\hat\partial_3\varphi.
\ea

We next express ${\cal L}(\tilde y)$ in terms of the averages $\bar\varphi$ and lattice derivatives $\hat\partial_\mu\varphi$. We write
\ba\label{VV7}
\tilde{\cal H}^k_\pm(\tilde x_j)=\sigma^\mu_j\bar{\cal H}^k_{\pm\mu}(\tilde y)+2\Delta V^\mu_j
{\cal D}^k_{\pm\mu}(\tilde y)+\Delta^2\sigma^\mu_j{\cal G}^k_{\pm\mu}(\tilde y),\nn\\
\ea
where $\bar{\cal H}_\mu(\tilde y)$ obtains from $\tilde {\cal H}(\tilde x_j)$ by the replacement $\varphi(\tilde x_j)\to\bar\varphi_\mu(\tilde y)$, with $\sigma^\mu_j=(V^\mu_j)^2$ or $\sigma^0_1=\sigma^0_8=\sigma^1_2=\sigma^1_7=\sigma^2_3=\sigma^2_6=\sigma^3_4=\sigma^3_5=1$ and $\sigma^\mu_j=0$ otherwise. We also define (no sum over $\mu$ here and all quantities evaluated at $\tilde y$)
\be\label{W8} 
\D^k_{\pm\mu}=(\bar\varphi_\mu)^a_\alpha(C_\pm)_{\alpha\beta}(\tau_2\tau_k)^{ab}\hat\partial_\mu\varphi^b_\beta
\ee
and 
\be\label{VV9}
\g^k_{\pm\mu}=\hat\partial_\mu\varphi^a_\alpha(C_\pm)_{\alpha\beta}(\tau_2\tau_k)^{ab}\hat\partial_\mu\varphi^b_\beta.
\ee

Expanding $\f^{1,2,8,7}_+$ in powers of $\Delta$ one finds that all contributions $\sim\bar{\cal H}^3$ vanish since there are an equal number of terms with positive and negative signs in the sum \eqref{V4}. For this purpose we can use expressions of the type 
\ba\label{VV10}
\tilde \h^k_\pm(\tilde x_1)\tilde \h^m_\pm(\tilde x_8)&=&\hat \h^k_{\pm 0}\hat \h^m_{\pm 0}+2\Delta
(\hat \h^k_{\pm 0}\D^m_{\pm 0}-\hat \h^m_{\pm 0} \D^k_{\pm 0})\nn\\
&&-4\Delta^2(\D^k_{\pm 0}\D^m_{\pm 0}+\D^m_{\pm 0}\D^k_{\pm 0}),
\ea
where 
\be\label{VV11}
\hat\h^k_{\pm \mu}=\bar \h^k_{\pm \mu}+\Delta^2 \g^k_{\pm\mu}.
\ee
The contraction with $\epsilon^{klm}$ implies that only the term linear in $\Delta$ contributes in the expression \eqref{VV10}. Employing 
\ba\label{VV13}
\tilde \h^k_\pm(\tilde x_1)-\tilde \h^k_\pm(\tilde x_8)=-4\Delta \D^k_{\pm 0},\nn\\
\tilde \h^k_\pm(\tilde x_2)-\tilde \h^k_\pm(\tilde x_7)=-4\Delta \D^k_{\pm 1},
\ea
one obtains
\ba\label{VV14}
\f^{1,2,8,7}_+=\frac{2\Delta^2}{3}\epsilon^{klm}\big\{\hat\h^k_{+1} \D^m_{+1} \D^l_{+0}-\hat \h^k_{+0}\D^m_{+0}\D^l_{+1}\}.
\ea
We finally define 
\ba\label{VV15}
\h^k_{\pm ab}&=&\frac12(\hat \h^k_{\pm a}+\hat\h^k_{\pm b}),\nn\\
{\cal M}^k_{\pm ab}&=&\frac12(\hat\h^k_{\pm a}-\hat\h^k_{\pm b}),
\ea
and note that ${\cal M}_{ab}$ drops out due to the contraction with $\epsilon^{klm}$. This yields a simple final result 
\be\label{VV16}
\f^{1,2,8,7}_+=\frac{2\Delta^2}{3}\epsilon^{klm}\h^k_{+01}
(\D^l_{+0}\D^m_{+1}-\D^l_{+1}\D^m_{+0}).
\ee

The relations similar to \eqref{VV16} can then be summarized as
\be\label{VV20}
\f^\pm_{\mu\nu}=\frac{2\Delta^2}{3}\epsilon^{klm}\h^k_{\pm\mu\nu}
(\D^l_{\pm\mu}\D^m_{\pm\nu}-\D^l_{\pm\nu}\D^m_{\pm \mu}).
\ee

\medskip\noindent
{\bf 3. Continuum limit}

The continuum limit obtains formally as $\Delta\to 0$ at fixed $y^\mu$. In this limit we can replace $\bar\varphi_\mu(\tilde y)$ by $\varphi(y)$. The lattice derivative $\hat\partial_\mu$ becomes the continuum derivative $\partial_\mu=\partial/\partial y^\mu$. This results in 
\be\label{VV22}
\D^k_{\pm \mu}(\tilde y)\to \varphi^a_\alpha (y)(C_\pm)_{\alpha\beta}(\tau_2\tau_k)^{ab}\partial_\mu\varphi^b_\beta(y),
\ee
and 
\be\label{VV23}
\h^k_{\pm\mu\nu}(\tilde y)\to\varphi^a_\alpha(y)(C_\pm)_{\alpha\beta}(\tau_2\tau_k)^{ab}\varphi^b_\beta(y)=H^\pm_k (y).
\ee
In terms of the scalar bilinears $H^\pm_k$ the continuum limit \eqref{VV22} reads
\be\label{151A}
\D^k_{\pm\mu}(\tilde y)\to\frac12\partial_\mu H^\pm_k(y),
\ee
such that 
\be\label{151B}
\f^\pm_{\mu\nu}=\frac{\Delta^2}{3}\epsilon^{klm}H^\pm_k\partial_\mu H^\pm_l\partial_\nu H^\pm_m.
\ee

We next want to establish the connection to the form \eqref{32F} of the continuum action, thereby also relating $\tilde\alpha$ to $\alpha$. For this purpose we compare the continuum limit of $\f^\pm_{\mu\nu}$ in eq. \eqref{VV20} with $F^\pm_{\mu\nu}$ in eq. \eqref{32E}. It is useful to employ the matrices
\be\label{173A}
\tilde \tau_k=\tau_2\tau_k~,~\tilde \tau_0=\tau_2,
\ee
where $\tilde\tau^T_k=\tilde \tau_k$ and $\tilde\tau^T_0=-\tilde\tau_0$. With the help of the identities
\ba\label{VV24}
&&(C_+)_{\alpha_1\beta_1}(C_+)_{\alpha_2\beta_2}=\frac12(C_+)_{\alpha_1\alpha_2}(C_+)_{\beta_1\beta_2}\nn\\
&&\qquad \qquad \qquad ~  -\frac12(C_+\tau_k)_{\alpha_1\alpha_2}(C_+\tau_k)_{\beta_1\beta_2},
\ea
and 
\ba\label{VV25}
&&\epsilon^{klm}(\tilde \tau_l)^{a_1b_1}(\tilde \tau_m)^{a_2b_2}=\nn\\
&&\qquad  -i\big[(\tilde\tau_k)^{a_1a_2}(\tilde\tau_0)^{b_1b_2}+(\tilde\tau_0)^{a_1a_2}(\tilde\tau_k)^{b_1b_2}\big],
\ea
we infer (cf. eq. \eqref{32B})
\ba\label{VV26}
&&\hspace{-1.0cm}\epsilon^{klm}\D^l_{+\mu}\D^m_{+\nu}=-\frac i2 H_k^+(y)D^+_{\mu\nu}\nn\\
&&\quad -\frac i2\varphi^{a_1}_{\alpha_1}(C_+\tau_n)_{\alpha_1\alpha_2}
(\tilde \tau_0)^{a_1a_2}\varphi^{a_2}_{\alpha_2}\nn\\
&&\quad\times \partial_\mu\varphi^{b_1}_{\beta_1}
(C_+\tau_n)_{\beta_1\beta_2}(\tilde \tau_k)^{b_1b_2}\partial_\nu\varphi^{b_2}_{\beta_2}.
\ea

We next investigate the product
\be\label{VV27}
H_k^+H_k^+=A^{a_1a_2a_3a_4}_{\alpha_1\alpha_2\alpha_3\alpha_4}
\varphi^{a_1}_{+\alpha_1}
\varphi^{a_2}_{+\alpha_2}
\varphi^{a_3}_{+\alpha_3}
\varphi^{a_4}_{+\alpha_4}
\ee
with $\alpha_1\dots\alpha_4$ taking values $1,2$ and
\ba\label{VV28}
A^{a_1a_2a_3a_4}_{\alpha_1\alpha_2\alpha_3\alpha_4}&=&(C_+)_{\alpha_1\alpha_2}
(C_+)_{\alpha_3\alpha_4}(\tilde\tau_k)^{a_1a_2}
(\tilde\tau_k)^{a_3a_4}\\
&=&\epsilon_{\alpha_1\alpha_2}\epsilon_{\alpha_3\alpha_4}
(\epsilon^{a_1a_4}\epsilon^{a_3a_2}+\epsilon^{a_1a_3}\epsilon^{a_4a_2}),\nn
\ea
and $\epsilon$ the two-index antisymmetric tensor $\epsilon_{12}=-\epsilon_{21}=1$. Insertion into eq. \eqref{VV27} yields (similar for $H^-$)
\be\label{VV29}
H^\pm_k H^\pm_k=-24A^\pm.
\ee
This term appears when we multiply the first term in eq. \eqref{VV26} with $H_k^+$. 

For the second term the multiplication with $H_k^+$ yields an expression involving four spinors without derivatives
\ba\label{VV30}
&&\varphi^{a_1}_{\alpha_1}(C_+\tau_n)_{\alpha_1\alpha_2}(\tilde\tau_0)^{a_1a_2}\varphi^{a_2}_{\alpha_2}\varphi^{a_3}_{\alpha_3}
(C_+)_{\alpha_3\alpha_4}(\tilde\tau_k)^{a_3a_4}\varphi^{a_4}_{\alpha_4}\nn\\
&&\qquad \qquad =(B_{nk})^{a_1a_2a_3a_4}_{\alpha_1\alpha_2\alpha_3\alpha_4}
\varphi^{a_1}_{\alpha_1}
\varphi^{a_2}_{\alpha_2}
\varphi^{a_3}_{\alpha_3}
\varphi^{a_4}_{\alpha_4}.
\ea
Under Lorentz transformations this expression transforms as $(3,1)$. Since a product of four spinors $\varphi_+$ must be a Lorentz singlet we expect the expression \eqref{VV30} to vanish. This can be verified by explicit computation of the totally antisymmetric part of
\be\label{VV31}
(B_{nk})^{a_1a_2a_3a_4}_{\alpha_1\alpha_2\alpha_3\alpha_4}=
(\tilde\tau_n)_{\alpha_1\alpha_2}(\tilde\tau_0)_{\alpha_3\alpha_4}(\tilde\tau_0)^{a_1a_2}
(\tilde\tau_k)^{a_3a_4}.
\ee
Using the identity
\ba\label{VV32}
&&(\tilde\tau_0)_{\alpha_1\alpha_2}(\tilde\tau_k)_{\alpha_3\alpha_4}=\frac12\big[(\tilde\tau_0)_{\alpha_1\alpha_4}
(\tilde\tau_k)_{\alpha_2\alpha_3}\\
&&\quad -(\tilde\tau_k)_{\alpha_1\alpha_4}
(\tilde\tau_0)_{\alpha_2\alpha_3}\big]+\frac i2\epsilon_{klm}(\tilde\tau_l)_{\alpha_1\alpha_4}(\tilde\tau_m)_{\alpha_2\alpha_3}\nn
\ea
one finds
\ba\label{VV33}
X&=&(B_{nk})^{a_1a_2a_3a_4}_{\alpha_1\alpha_2\alpha_3\alpha_4}\\
&&-\frac14(B_{nk})^{a_1a_4a_3a_2}_{\alpha_1\alpha_4\alpha_3\alpha_2}
-\frac14(B_{nk})^{a_3a_2a_1a_4}_{\alpha_3\alpha_2\alpha_1\alpha_4},\nn
\ea
where $X$ is symmetric under the exchange of at least one index pair $\eta_i=(\alpha_i,a_i)$ and $\eta_j=(\alpha_j,a_j)$. The totally antisymmetric part of $X_{\eta_1\eta_2\eta_3\eta_4}$ vanishes, while the totally antisymmetric part of the left hand side of eq. \eqref{VV33} yields $(3/2)$ times the totally antisymmetric part of $B_{\eta_1\eta_2\eta_3\eta_4}$. Thus the totally antisymmetric part of $B$ vanishes indeed, and the second expression in eq. \eqref{VV26} does not contribute. 

In the continuum limit we arrive at the simple result 
\be\label{VV34}
\f^+_{\mu\nu}=16i\Delta^2 F^+_{\mu\nu}.
\ee
The discussion for $\f^-_{\mu\nu}$ proceeds in parallel, where we observe an additional minus sign of the first term in eq. \eqref{VV26} which arises from the definition \eqref{W3},
\be\label{VV35}
\f^-_{\mu\nu}=-16i\Delta^2F^-_{\mu\nu}
\ee
This yields the continuum limit of ${\cal L}(y)$, 
\be\label{VV31}
{\cal L}(y)\to\frac{32}{3}\Delta^4\epsilon^{\mu_1\mu_2\mu_3\mu_4}
F^+_{\mu_1\mu_2}F^-_{\mu_3\mu_4}.
\ee
Again the factor $\Delta^4$ cancels in $\Sigma_{\tilde y}{\cal L}(\tilde y)$, the action is independent of the lattice distance $\Delta$, and we arrive at eq. \eqref{VV38}.

\medskip\noindent
{\bf 4. Gauge symmetry}

The continuum action is invariant under local $SU(2,{\mathbbm C})_F$ gauge transformations. For the discretized action \eqref{L1}, \eqref{V3} we have so far only established the invariance under global $SU(2,{\mathbbm C})_F$ gauge transformations. The question arises how $\f^{abcd}_\pm$ in eq. \eqref{V4} transforms under local gauge transformations.

Under a local gauge transformation the spinors transform infinitesimally according to eq. \eqref{S1}, which implies that the bilinear $\tilde \h^k_\pm$ transforms as a three-component vector 
\be\label{W1a}
\delta\tilde\h^k_\pm(\tilde x_j)=-\epsilon^{klm}\tilde\alpha_l(\tilde x_j)\tilde \h^m_\pm(\tilde x_j).
\ee
In turn, the product of three $\h$-factors at three different positions, contracted with $\epsilon^{klm}$, transforms as
\ba\label{W2a}
&&\delta\big\{\epsilon^{klm}\tilde\h^k(\tilde x_a)\tilde \h^l(\tilde x_b)\tilde \h^m(\tilde x_c)\big \}=A_H(\tilde x_a,\tilde x_b,\tilde x_c)\\
&&=\tilde\alpha_k(\tilde x_a)\tilde\h^l(\tilde x_a)\big\{\tilde\h^l(\tilde x_b)\tilde\h^k(\tilde x_c)-\tilde\h^l(\tilde x_c)
\tilde \h^k(\tilde x_b)\big\}\nn\\
&&+\tilde\alpha_k(\tilde x_b)\tilde\h^l(\tilde x_b)\big\{ \tilde\h^l(\tilde x_c)
\tilde\h^k(\tilde x_a)-\tilde \h^l(\tilde x_a)\tilde \h^k(\tilde x_c)\big\}\nn\\
&&+\tilde \alpha_k(\tilde x_c)\tilde\h^l(\tilde x_c)\big\{
\tilde\h^l(\tilde x_a)\tilde \h^k(\tilde x_b)-\tilde\h^l(\tilde x_b)\tilde\h^k(\tilde x_a)\big\}.\nn\\
\ea
In general, this variation does not vanish and the lattice action is not invariant under local gauge transformations.

The r.h.s. of eq. \eqref{W2a} vanishes for global gauge transformations where $\tilde\alpha_k(\x{j_1})=\tilde\alpha_k(\x{j_2})$. This is seen most easily in the form 
\ba\label{W3}
A_H(\x a,\x b,\x c)&=&\big (\tilde\alpha_k(\x a)-\tilde\alpha_k(\x b)\big)\tilde\h^l(\x a)\tilde\h^l(\x b)\tilde \h^k(\x c)\nn\\
&+&\big(\tilde\alpha_k(\x b)-\tilde\alpha_k(\x c)\big)\tilde\h^l(\x b)\tilde\h^l(\x c)\tilde \h^k(\x a)\nn\\
&+&\big(\tilde\alpha_k(\x c)-\tilde\alpha_k(\x a)\big)\tilde \h^l(\x c)\tilde \h^l(\x a)\tilde \h^k(\x b).\nn\\
\ea

The invariance under global $SU(2, {\mathbbm C})_F$ transformations extends to $\f^+_{\mu\nu}$ and therefore to the action. The global invariance also holds for chiral $SU(2,{\mathbbm C})_L$ transformations where only the Weyl spinors $\varphi_+$ are transformed, while $\varphi_-$ is left invariant, and similarly for $SU(2,{\mathbbm C})_R$. 

\section*{APPENDIX H: LATTICE GEOMETRY}
\renewcommand{\theequation}{H.\arabic{equation}}
\setcounter{equation}{0}
\label{Lattice geometry}

In this appendix we investigate further aspects of the lattice formulation. The lattice coordinates $\tilde z$ make no distinction between time and space coordinates. So far we have associated one of the $\tilde z^\mu$ coordinates, e.g. $\tilde z^0$, with a time coordinate, which is appropriate if the metric takes the expectation value $g_{\mu\nu}=diag(-1,1,1,1)$. Even if a Minkowski-signature is selected by the ground state or cosmological solution, it is a priori not clear how the ``equal time-hyperface'' is embedded in the lattice geometry. The time direction could also correspond to a ``diagonal direction'' in the space of lattice coordinates $\tilde z^\mu$. For this reason we reformulate in this appendix our lattice also in different lattice coordinates. 

\medskip\noindent
{\bf 1. Diagonal coordinates}

Let us consider in the continuum new coordinates $u^\mu=(t,u_1,u_2,u_3)$ that are expressed in terms of the coordinates $z^\mu$ used in the main text by
\ba\label{B1}
z^0&=&\frac12(t+u_1+u_2+u_3),\nn\\
z^1&=&\frac12(t-u_1+u_2-u_3),\nn\\
z^2&=&\frac12(t+u_1-u_2-u_3),\nn\\
z^3&=&\frac12(t-u_1-u_2+u_3).
\ea
In a matrix notation this reads
\be\label{B2}
z^\mu=R^\mu{_\nu}u^\nu
\ee
with an orthogonal matrix
\be\label{B3}
R=\frac12
\left(\begin{array}{rrrrrrr}
1&,&1&,&1&,&1\\
1&,&-1&,&1&,&-1\\
1&,&1&,&-1&,&-1\\
1&,&-1&,&-1&,&1
\end{array}\right)
\ee
obeying
\be\label{B4}
R^2=1~,~R=R^T~,~\det R=1.
\ee
This implies
\be\label{B5}
u^\mu=R^\mu{_\nu}z^\nu~,~\delta_{\mu\nu}z^\mu z^\nu=\delta_{\mu\nu}u^\mu u^\nu
\ee
and 
\be\label{B6}
\int d^4u=\int d^4 z.
\ee
Defining $\partial_\mu=\partial/\partial z^\mu$ and $\tilde\partial_\mu=\partial/\partial u^\mu$ yields
\be\label{B7}
\partial_\mu=\tilde\partial_\nu R^\nu{_\mu}~,~\tilde\partial_\mu=\partial_\nu R^\nu{_\mu}.
\ee
Since $\epsilon^{\mu\nu\rho\sigma}$ is invariant under rotations the continuum action \eqref{30F} has the same form in the new and old coordinate system.

We next introduce integer lattice coordinates $\tilde u^\mu$ that are obtained from $\tilde z^u$ by
\be\label{B.A}
\tilde u^\mu=R^\mu{_\nu}\tilde z^\nu+\left(\frac12,\frac12,\frac12,\frac12\right).
\ee
(We have performed in addition to the rotation an additive shift such that $\tilde u^\mu$ are integers with $\Sigma_\mu \tilde u^\mu$ even. With this shift the points $\tilde u=(\tilde u^0,\tilde u^1,\tilde u^2,\tilde u^3)$ are on the same even sublattice as the cell positions $\tilde y$. For a regular lattice the continuum coordinates are $u^\mu=\Delta\tilde u^\mu)$. 

\begin{figure}[htb]
\begin{center}
\includegraphics[width=0.45\textwidth]{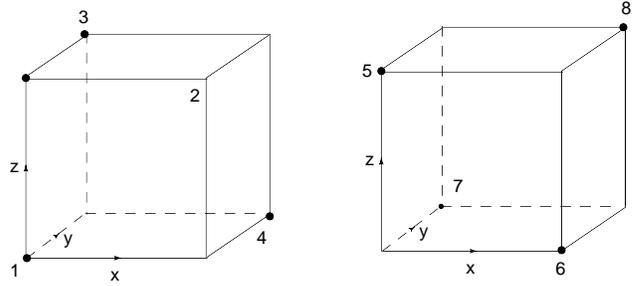}
\caption{Points of the fundamental lattice belonging to a cell. The time coordinate of the right cube is one unit higher than the one of the left cube.}
\label{fig1}
\end{center}
\end{figure}

To each $\tilde y$ we associate again the eight positions $\tilde x_j(\tilde y)$ that belong to the cell at $\tilde y$, $j=1\dots 8$. They can be written as
\be\label{L2}
\tilde x^\mu_j(\tilde y)=\tilde y^\mu+\frac12(v^\mu_j+1),
\ee
with 
\ba\label{L3}
\begin{array}{lllllll}
v_1&=&(-1,-1,-1,-1)&,&v_2&=&(-1,1,-1,1),\\
v_3&=&(-1,-1,1,1)&,&v_4&=&(-1,1,1,-1),\\
v_5&=&(1,-1,-1,1)&,&v_6&=&(1,1,-1,-1),\\
v_7&=&(1,-1,1,-1)&,&v_8&=&(1,1,1,1).
\end{array}
\ea
(The positions in the cell obey for each $j$ the constraint $\prod_\mu v^\mu=1$.) The points $\tilde x_j$ are indicated in fig. \ref{fig1}. Here the time coordinate of the points on the right cube is one unit higher than the one of the points on the left cube. Two neighboring hypercubes have two common points on the fundamental lattice.

We have chosen the $\tilde u$-coordinates such that for $\tilde y=(0,0,0,0)$ the first four cell points $\x1,\x2,\x3,\x4$ all have a ``time coordinate'' $\tilde t=\tilde u^0=0$, and the other four $\x5,\x6,\x7,\x8$ are at $\tilde t=1$. The same situation holds for all cells, with time coordinate of $\x5,\dots \x8$ one unit higher than for $\x1,\dots \x4$. Our definition of lattice derivatives \eqref{VV4} involves for all $\hat\partial_\mu$ one point in the cell among $\x1,\dots\x4$, and another one among $\x5,\dots\x8$. For the $\tilde u$ coordinates these are a type of ``light-cone derivatives''. In the following we discuss a few lattice properties in the $\tilde u$-coordinates. 

\medskip\noindent
{\bf 2. Reflections}

The continuum action \eqref{A1} changes  sign under the reflection  of one coordinate, for example $\varphi^a_\alpha(u^0,\vec u)\to\varphi^a_\alpha(-u^0,\vec u)$. More generally, it changes sign under reflections at arbitrary $3$-dimensional hyperplanes. We want to preserve this property for the lattice action for suitable hyperplanes consistent with the lattice symmetries. As an example, we consider the exchange of the coordinates $u^1$ and $u^2$, $\varphi(u^0,u^1,u^2,u^3)\to \varphi(u^0,u^2,u^1,u^3)$. This is realized by the combination of an exchange of the locations of the cells, $(\tilde y^0,\tilde y^1,\tilde y^2,\tilde y^3)\to(\tilde y^0,\tilde y^2,\tilde y^1,\tilde y^3)$, together with a reflection inside each cell (fixed $\tilde y$), whereby 
\be\label{L10}
\tilde x_2(\tilde y)\leftrightarrow \tilde x_3(\tilde y)~,~\tilde x_6(\tilde y)\leftrightarrow\tilde x_7(\tilde y),
\ee
while the positions $\tilde x_1,\tilde x_4, \tilde x_5$ and $\tilde x_8$ remain invariant. In other words, we reflect in eq. \eqref{L2} $\tilde y^1\leftrightarrow \tilde y^2~,~v^1_j\leftrightarrow v^2_j$, whereby the transformation of $v$ can also be expressed as $v_2\leftrightarrow v_3~,~v_6\leftrightarrow v_7$ with $v_{1,4,5,8}$ invariant. As an example, a term that is odd under this reflection is realized by the combination $\tilde C(\tilde x_1,\tilde x_4,\tilde x_8,\tilde x_3,\tilde x_7,\tilde x_6)-\tilde C(\tilde x_1,\tilde x_4,\tilde x_8,\tilde x_2,\tilde x_6,\tilde x_7)$. 

We list the twelve lattice reflections of this type in table 1 by specifying which positions are exchanged within a given cell. (Positions not listed are invariant within a given cell.) We also indicate by shorthands which 

\bigskip
\noindent
\small 
\begin{tabular}{|p{4mm}|p{27mm}|p{7mm}||p{5mm}|p{27mm}|p{9mm}|}
\hline 
$R_1$&$\tilde x_2\leftrightarrow\tilde x_5~,~\tilde x_4\leftrightarrow\tilde x_7$&$(0,1)$&$R_7$&
$\tilde x_1\leftrightarrow\tilde x_6~,~\tilde x_3\leftrightarrow\tilde x_8$&$(0,1)_-~$ \\ \hline
$R_2$&$\tilde x_3\leftrightarrow \tilde x_5~,~\tilde x_4\leftrightarrow \tilde x_6$&$(0,2)$&$R_8$&
$\tilde x_1\leftrightarrow \tilde x_7~,~\tilde x_2\leftrightarrow \tilde x_8$&$(0,2)_-$\\ \hline
$R_3$&$\tilde x_2\leftrightarrow\tilde x_6~,~\tilde x_3\leftrightarrow\tilde x_7$&$(0,3)$&$R_9$&
$\tilde x_1\leftrightarrow\tilde x_5~,~\tilde x_4\leftrightarrow\tilde x_8$&$(0,3)_-$\\ \hline
$R_4$&$\tilde x_2\leftrightarrow\tilde x_3~,~\tilde x_6\leftrightarrow\tilde x_7$&$(1,2)$&$R_{10}$&
$\tilde x_1\leftrightarrow\tilde x_4~,~\tilde x_5\leftrightarrow\tilde x_8$&$(1,2)_-$\\ \hline
$R_5$&$\tilde x_2\leftrightarrow\tilde x_4~,~\tilde x_5\leftrightarrow\tilde x_7$&$(2,3)$&$R_{11}$&
$\tilde x_1\leftrightarrow\tilde x_3~,~\tilde x_6\leftrightarrow\tilde x_8$&$(2,3)_-$\\ \hline 
$R_6$&$\tilde x_3\leftrightarrow \tilde x_4~,~\tilde x_5\leftrightarrow\tilde x_6$&$(1,3)$&$R_{12}$&
$\tilde x_1\leftrightarrow\tilde x_2~,~\tilde x_7\leftrightarrow\tilde x_8$&$(1,3)_-$\\ \hline 
\end{tabular}

\normalsize
\medskip
\noindent
table 1: lattice reflections

\medskip\noindent
coordinates are exchanged. For example, $R_4$ with the shorthands $(1,2)$ corresponds to $u^1\leftrightarrow u^2$, or to an exchange of the components $v^1_j\leftrightarrow v^2_j$. Similarly, $R_{10}$ with shorthand $(1,2)_-$ denotes the exchange of the components $v^1_j\leftrightarrow -v^2_j$, together with $\tilde y^1\leftrightarrow -\tilde y^2$. For the coordinates of the lattice points this corresponds to the reflection $\tilde u^1\leftrightarrow 1-\tilde u^2$. For the action to be odd under these reflections it is sufficient that ${\cal L}\big(\tilde y=(0,0,0,0)\big)$ changes sign under each reflection. For appropriate boundary conditions the sum over cells in the action \eqref{L1} contains for every $\tilde y$ also the reflected $\tilde y$. Then a change of sign of ${\cal L}(\tilde y)$ under the exchange of positions specified in table 1 implies a change of sign of the action. Since the transformations in table 1 only refer to an exchange of positions within a cell, but not to the location of the cell, we can investigate an arbitrary $\tilde y$, in particular $\tilde y=(0,0,0,0)$. 

Each of the reflections $R_1\dots R_{12}$ leaves four positions in the cell invariant. They are on opposite ends of two maximal diagonals. For example, both $R_1$ and $R_5$ leave the points $\tilde x_1,\tilde x_3,\tilde x_6$ and $\tilde x_8$ invariant, and the corresponding maximal diagonals join the points $(\tilde x_1,\tilde x_8)$ and $(\tilde x_3,\tilde x_6)$, respectively. These four points define a two-dimensional plane since the vector from $\tilde x_1$ to $\tilde x_3$ is the same as the one from $\tilde x_6$ to $\tilde x_8$. The three dimensional hyperplane on which the reflection is performed needs the specification of one further direction, and this distinguishes the hyperplanes relevant for $R_1$ and $R_5$. This further direction is fixed by indicating which pairs of points are exchanged by the reflection, e.g. $\tilde x_2\leftrightarrow\tilde x_5~,~\tilde x_4\leftrightarrow\tilde x_7$ for $R_1$, and $\tilde x_2\leftrightarrow\tilde x_4~,~\tilde x_5\leftrightarrow\tilde x_7$ for $R_5$. We observe that an exchange of pairs at the ends of a maximal diagonal, i.e. $\tilde x_2\leftrightarrow\tilde x_7$ and $\tilde x_4\leftrightarrow\tilde x_5$, cannot be realized by a reflection. (The vectors from $\tilde x_2$ to $\tilde x_7$, and from $\tilde x_4$ to $\tilde x_5$, are not parallel.) These properties are shared by all reflections $R_s,s=1\dots 12$:  the pairs of exchanged points are never ends of maximal diagonals within a hypercube.

The list of twelve reflections $R_s$ in table 1 is complete for this type - there are no more reflections that map two pairs of points of the cell (shown in fig. 1) into each other. This type of reflections corresponds to the diagonal reflections in the $\tilde z$ coordinates, e.g. of the type $\tilde z^0\leftrightarrow \tilde z^1$. (There are four further reflections changing the sign of one coordinate $\tilde z^u$. These exchange only two points within a cell, e.g. $\x1$ and $\x8$, leaving all others invariant.) The lattice action should change sign whenever one of the reflections is applied. For the example $R_1$ the lattice action should change sign if we replace in each hyperlink $\varphi ^a_\alpha (\tilde x _2)\to \varphi^a_\alpha(\tilde x_5)~,~\varphi^a_\alpha(\tilde x_5)\to\varphi^a_\alpha(\tilde x_2)~,~\varphi^a_\alpha(\tilde x_4)\to  \varphi^a_\alpha(\tilde x_7)$ and $\varphi^a_\alpha(\tilde x_7)\to\varphi^a_\alpha(\hat x_4)$. 

None of the reflections $R_s$ leaves the hyperlink $\tilde C(\tilde x_1,\tilde x_4,\tilde x_8,\tilde x_3,\tilde x_7,\tilde x_6)$ invariant. This is required since otherwise the properly symmetrized lattice action would vanish. For all $R_s$ except $R_1,R_7,R_{11}$ either $\tilde x_2$ or $\tilde x_5$ appear as an ``occupied position'' of the reflected hyperlink, which differs therefore from the original hyperlink for which the positions $\tilde x_2$ and $\tilde x_5$ are ``empty''. The reflection $R_1$ exchanges the positions $\tilde x_4$ and $\x7$ of the hyperlink. The hyperlink $\tilde C(\x1,\x7,\x8,\x3,\x4,\x6)$ differs from the original link since the position $\x7$ is now occupied by a Weyl spinor $\psi_-$, in contrast to $\psi_+$ for the original link. Also the reflection $R_7$ and $R_{11}$ exchange different types of Weyl spinors. Actually, the requirement of the lattice action being odd under all reflections $R_s$ restricts the possible choice of hyperlinks severely. For example, the hyperlink $\tilde C(\x1,\x4,\x3,\x8,\x6,\x7)$ is invariant under the reflection $R_1$. We recall at this place that hyperlinks are uniquely determined by the occupied positions $\{j_+\}$ and $\{j_-\}$. Different link orderings can correspond to identical hyperlinks.

\medskip\noindent
{\bf 3. Rotations}

We may also consider combinations of two reflections. The lattice action should remain invariant under such transformations. As an example, we may consider the combinations of the reflections $R_1$ and $R_2$, 
\ba\label{L11}
R_2R_1:&&\tilde x_2\to \tilde x_5~,~\tilde x_5\to \tilde x_3~,~\tilde x_3\to \tilde x_2,\nn\\
&&\tilde x_4\to\tilde x_6~,~\tilde x_6\to\tilde x_7~,~\tilde x_7\to \tilde x_4,
\ea
and 
\ba\label{L12}
R_1R_2:&&\x1\to \x3~,~\x3\to \x5~,~\x5\to \x2,\nn\\
&&\x4\to \x7~,~\x7\to \x6~,~\x6\to \x4.
\ea
The order matters, reflections do not commute. We note $(R_2R_1)^2=R_1R_2$ and $(R_2R_1)^3=1$. Indeed, $R_2R_1$ is a rotation of $2\pi/3$ around an axis which is the maximal diagonal linking the invariant points $\tilde x_1$ and $\tilde x_8$, with $R_1R_2$ a rotation around the same axis with opposite direction. All combinations of two reflections with a common pair of invariant points are similar $2\pi/3$-rotations around suitable axes. There are a total of eight independent transformations of this type, corresponding to the four maximal diagonals $(1,8),(2,7),(3,6)$ and $(4,5)$, with two rotation directions for each axis. The lattice action should be invariant with respect to these eight discrete rotations that we denote by $T_r,r=1\dots 8$. 

The reflections that do not share common invariant points commute, e.g.
\ba\label{L13}
R_1R_7=R_7R_1&:&\tilde x_1\leftrightarrow\tilde x_6~,~\tilde x_2\leftrightarrow\tilde x_5~,~\tilde x_3\leftrightarrow \tilde x_8,\nn\\
&&\tilde x_4\leftrightarrow \tilde x_7,\nn\\
R_4R_{10}=R_{10}R_4&:&\tilde x_1\leftrightarrow \tilde x_4~,~\tilde x_2\leftrightarrow \tilde x_3~,~\tilde x_5\leftrightarrow \tilde x_8,\nn\\
&&\tilde x_6\leftrightarrow \tilde x_7.
\ea
Their multiplication combines to rotations by $\pi$ around suitable axes. For example, $R_4R_{10}$ is a rotation around the $u^3$-axis through the center of the hypercube, with $u^0$ kept fixed. There are a total of twelve $\pi$-rotations of this type, corresponding to the products of reflections $R_6 R_{12},R_2R_{12},R_6R_8,R_2R_8,R_3R_{10},R_4R_{10},R_3R_9,$ $R_4R_9,R_1 R_7,R_1R_{11}R_5R_7,R_5R_{11}$. Together, we have 20 independent discrete transformations which can be obtained by products of two reflections $R_{s_1}R_{s_2}$. We can further form products of three or more reflections. All products with an odd number of reflections change the sign of the lattice action, whereas products with an even number of reflections leave the action invariant. The products of the twelve basic reflections form a large non-abelian discrete group. 

\medskip\noindent
{\bf 4. Diagonal sublattices}

One may perform a rotation of the coordinates $u^\mu$ by defining new coordinates $s^\mu$ as
\ba\label{D1}
s^0&=&\frac{1}{\sqrt{2}}(u^0+u^2)~,~s^1=\frac{1}{\sqrt{2}}(u^1+u^3),\nn\\
s^2&=&\frac{1}{\sqrt{2}}(u^0-u^2)~,~s^3=\frac{1}{\sqrt{2}}(u^1-u^3).
\ea
The rotation matrix $R$, $s^\mu=\tilde R^{\mu}{_\nu}u^\nu$, obeys
\be\label{D2}
\tilde R=\frac{1}{\sqrt{2}}\left(\begin{array}{ccccccc}
1&,&0&,&1&,&0\\
0&,&1&,&0&,&1\\
1&,&0&,&-1&,&0\\
0&,&1&,&0&,&-1
\end{array}\right)~,~\tilde R^T=\tilde R~,~\tilde R^2=1.
\ee

Similarly, we may introduce
\be\label{D3}
\tilde s^\mu=\frac{1}{\sqrt{2}}\tilde R^{\mu}{_\nu}\tilde u^\nu.
\ee
Integer values of $\tilde s^\mu$ define a hypercubic lattice with lattice distance $\sqrt{2}\Delta$. This is a sublattice of the fundamental lattice that we may denote by ${\cal L}^{(1)}$. Indeed, for $\tilde u^\mu=\sqrt{2}\tilde R^{\mu}{_\nu}\tilde s^\nu$ one finds an even value for $\Sigma_\mu\tilde u^\mu=2(\tilde s^0+\tilde s^1)$. The number of lattice points of ${\cal L}^{(1)}$ amounts to one half of the points of the fundamental lattice. If all lattice coordinates $\tilde s^\mu$ are half-integer one finds again that $\tilde u$ belongs to the fundamental lattice. The half-integer values of $\tilde s^\mu$ define a second hypercubic lattice ${\cal L}^{(2)}$, again with lattice distance $\sqrt{2}\Delta$. It obtains from ${\cal L}^{(1)}$ by a translational shift by a vector $\Delta\tilde s^\mu=\frac12(1,1,1,1)$ or $\Delta\tilde u^\mu=(1,1,0,0)$. We call ${\cal L}^{(1)}$ and ${\cal L}^{(2)}$ the diagonal sublattices. The sum of ${\cal L}^{(1)}
$ and ${\cal L}^{(2)}$ constitutes the fundamental lattice. 

For the cell at $\tilde y=0$ we list in table $2$ first the four positions on the diagonal sublattice ${\cal L}^{(1)}$, and subsequently those on ${\cal L}^{(2)}$. We list the coordinates $\tilde z,\tilde u$ and $\tilde s$ for each position in the cell.

\medskip
\begin{center}
\begin{tabular}{|c|c|c|c|}
\hline
&$\tilde z$&$\tilde u$&$\tilde s$\\ \hline
$\x1$&$(-1,0,0,0)$&$(0,0,0,0)$&$(0,0,0,0)$\\
$\x2$&$(0,-1,0,0)$&$(0,1,0,1)$&$(0,1,0,0)$\\
$\x7$&$(0,1,0,0)$&$(1,0,1,0)$&$(1,0,0,0)$\\
$\x8$&$(1,0,0,0)$&$(1,1,1,1)$&$(1,1,0,0)$\\ \hline
$\x3$&$(0,0,-1,0)$&$(0,0,1,1)$&$\left(\frac12,\frac12,-\frac12,-\frac12\right)$\\
$\x4$&$(0,0,0,-1)$&$(0,1,1,0)$&$\left(\frac12,\frac12,-\frac12,\frac12\right)$\\
$\x5$&$(0,0,0,1)$&$(1,0,0,1)$&$\left(\frac12,\frac12,\frac12,-\frac12\right)$\\
$\x6$&$(0,0,1,0)$&$(1,1,0,0)$&$\left(\frac12,\frac12,\frac12,\frac12\right)$\\ \hline
\end{tabular}

\medskip
\noindent
table 2: Positions $\x j(\tilde y)$ in the hypercube $\tilde y=0$ for \\
different lattice coordinates.~~~~~~~~
\end{center}

\noindent
For the hypercube at $\tilde y=(1,0,1,0)$ one adds to $\tilde s$ the vector $(1,0,0,0)$, such that $\x{1,2,7,8}$ belong again to ${\cal L}^{(1)}$, and $\x{3,4,5,6}$ are on ${\cal L}^{(2)}$. On the other hand, for the hypercube located at $\tilde y=(1,1,0,0)$ the $\tilde s$-coordinates are shifted by a vector $(\frac12,\frac12,\frac12,\frac12)$ as compared to the values shown in table 2. Now $\x{1,2,7,8}$ belong to the sublattice ${\cal L}^{(2)}$ and $\x{3,4,5,6}$ to ${\cal L}^{(1)}$. The two common points for the hypercubes at $\tilde y=(1,1,0,0)$ and $\tilde y=(0,0,0,0)$ are
\ba\label{D4}
\x1(1,1,0,0)&=&\x6(0,0,0,0),\nn\\
\x5(1,1,0,0)&=&\x8(0,0,0,0),
\ea
consistent with the assignment to sublattices.

The reflection $R_2$ acts in a simple way within the diagonal lattice: it simply reverses the coordinate $\tilde s^2$, i.e. $R_2(\tilde s^0,\tilde s^1,\tilde s^2,\tilde s^3)=(\tilde s^0,\tilde s^1,-\tilde s^2,\tilde s^3)$. Similarly, the reflection $R_8$ describes a reflection of the diagonal coordinate $\tilde s^0$, now at the hyperplane $\tilde s^0=\frac12$, such that $R_8(\s0,\s1,\s2,\s3)=(1-\s0,\s1,\s2,\s3)$. In an analogous way $R_6$ describes a reflection of $\s3$ at the hyperplane $\s3=0,R_6(\s0,\s1,\s2,\s3)=(\s0,\s1,\s2,-\s3$), while $R_{12}$ reflects $\s1$ at the hyperplane $\s1=\frac12$, i.e. $R_{12}(\s0,\s1,\s2,\s3)=(\s0,1-\s1,\s2,\s3)$. We conclude that the four reflections $(R_2,R_6,R_8,R_{12})$ describe the reversal of the $s$-coordinates.

For the hypercube at $\tilde y=0$ the local $SO(4,{\mathbbm C})$ invariant $\tilde C_+(\x1,\x7,\x8)$ in eq. \eqref{LLA3} involves only Weyl spinors $\varphi_+$ which are all situated on the sublattice ${\cal L}^{(1)}$, while $\tilde C_-(\x3,\x4,\x6)$ involves Weyl spinors $\varphi_-$ with positions on ${\cal L}^{(2)}$. The reflections $R_2$ and $R_6$ act only on the points of ${\cal L}^{(2)}$ within this hypercube, while the points $\x1,\x2,\x7$ and $\x8$ are left invariant. The symmetrized expression 
\ba\label{D5}
\bar C_-&=&\frac14\big\{\tilde C_-(\x3,\x4,\x6)-\tilde C_-(\x4,\x5,\x6)\nn\\
&&+\tilde C_-(\x3,\x5,\x6)-\tilde C_-(\x3,\x4,\x5)\big\}
\ea
changes sign under the reflections $R_2$ and $R_6$ and is invariant under $R_8$ and $R_{12}$. Similarly, the symmetrized combination 
\ba\label{D6}
\bar C_+&=&\frac14\big\{\tilde C_+(\x1\,\x7,\x8)-\tilde C_+(\x2,\x7,\x8)\nn\\
&&+\tilde C_+(\x1,\x2,\x8)-\tilde C_+(\x1,\x2,\x7)\big\}
\ea
is odd under $R_8$ and $R_{12}$ and even under $R_2$ and $R_6$. We conclude that the combination $\bar C_+\bar C_-$ changes sign under all four reflections $R_2,R_6,R_8$ and $R_{12}$. We may therefore use
\be\label{D6a}
{\cal L}(y)\sim \check s [\bar C_+\bar C_-],
\ee
where $\check s$ denotes the remaining necessary symmetrization which guarantees the correct transformations with respect to the remaining eight reflections. 

\medskip\noindent
{\bf 5. Alternative lattice derivatives}

One may investigate alternative choices of lattice derivatives by suitable linear combinations of spinors on sites belonging to the cell. As an example, we can define the four derivatives in the diagonal directions
\ba\label{E1}
\check\partial_0\varphi_\eta(\tilde y)&=&\frac{1}{2\sqrt{2}\Delta}
\big \{\varphi_\eta (\x7)-\varphi_\eta(\x1)+\varphi_\eta(\x8)-\varphi_\eta(\x2)\big\},\nn\\
\check\partial_1\varphi_\eta(\tilde y)&=&\frac{1}{2\sqrt{2}\Delta}\big\{\varphi_\eta(\x2)-\varphi_\eta(\x1)+\varphi_\eta(\x8)-\varphi_\eta(\x7)\big\}\nn\\
\check\partial_2\varphi_\eta(\tilde y)&=&\frac{1}{2\sqrt{2}\Delta}\big\{\varphi_\eta(\x5)-\varphi_\eta(\x3)+\varphi_\eta(\x6)-\varphi_\eta(\x4)\big\},\nn\\
\check\partial_3\varphi_\eta(\tilde y)&=&\frac{1}{2\sqrt{2}\Delta}\big\{\varphi_\eta(\x4)-\varphi_\eta(\x3)+\varphi_\eta(\x6)
-\varphi_\eta(\x5)\big\}\nn\\
\ea
Here $\check\partial_\mu$ stands for $\partial/\partial s^\mu$. All four lattice derivatives are odd with respect to one of the four reflections $R_2,R_6,R_8,R_{12}$, namely
\ba\label{E2}
R_8(\check\partial_0\varphi)&=&-\check\partial_0\varphi~,~R_{12}(\check\partial_1\varphi)=-\check\partial_1\varphi,\nn\\
R_2(\check\partial_2\varphi)&=&-\check\partial_2\varphi~,~R_6(\check\partial_3\varphi)=-\check\partial_3\varphi.
\ea
They are invariant under the remaining three reflections. 

We can also introduce mixed second derivatives as
\be\label{E3}
\check\partial_0\check\partial_1\varphi=\check\partial_1\check\partial_0\varphi=
\frac{1}{2\Delta^2}\big\{\varphi(\tilde x_1)-\varphi(\x2)+\varphi(\x8)-\varphi(\x7)\big\}.
\ee
This expression changes sign under each of the two reflections $R_8$ and $R_{12}$ and is invariant under $R_2$ and $R_6$. Similarly, the mixed derivative 
\be\label{E4}
\check\partial_2\check\partial_3\varphi=\check\partial_3\check\partial_2\varphi=\frac{1}{2\Delta^2}
\big\{ \varphi (\x3)-\varphi(\x4)-\varphi(\x5)+\varphi(\x6)\big\}
\ee
changes sign under $R_2$ and $R_6$ and is invariant under $R_8$ and $R_{12}$. The ``average value'' of $\varphi$ within the hypercube
\be\label{E5}
\varphi_\eta(\tilde y)=\frac18\sum^8_{j=1}\varphi_\eta (\tilde x_j)
\ee
is invariant under all reflections. Finally, the combination
\ba\label{E6}
E_\eta(\tilde y)&=&\frac18\big\{\varphi_\eta(\x1)+\varphi_\eta(\x2)+\varphi_\eta(\x7)+\varphi_\eta(\x8)\nn\\
&-&\varphi_\eta(\x3)-\varphi_\eta(\x4)-\varphi_\eta (\x5)-\varphi_\eta(\x6)\big\}
\ea
is invariant under $R_2,R_6,R_8,R_{12}$, while it changes sign under the $\pi$-rotations $R_1R_7,R_3R_9,R_4R_{10}$ and $R_5R_{11}$. 

We can combine the expressions \eqref{E1}-\eqref{E6} in order to express the variables within a given hypercube in terms of the lattice derivatives at the location of this hypercube,
\ba\label{E7}  
\varphi(\x1)&=&\varphi+E-\frac{\Delta}{\sqrt{2}}\hat\partial_0\varphi-\frac{\Delta}{\sqrt{2}}\check\partial_1\varphi
+\frac{\Delta^2}{2}\check\partial_0\check\partial_1\varphi,\nn\\
\varphi(\x2)&=&\varphi+E-\frac{\Delta}{\sqrt{2}}\check\partial_0\varphi+\frac{\Delta}{\sqrt{2}}\check\partial_1\varphi
-\frac{\Delta^2}{2}\check\partial_0\check\partial_1\varphi,\nn\\
\varphi(\x7)&=&\varphi+E+\frac{\Delta}{\sqrt{2}}\check\partial_0\varphi-\frac{\Delta}{\sqrt{2}}\check\partial_1\varphi-
\frac{\Delta^2}{2}\check\partial_0\check\partial_1\varphi,\nn\\
\varphi(\x8)&=&\varphi+E+\frac{\Delta}{\sqrt{2}}\check\partial_0\varphi+\frac{\Delta}{\sqrt{2}}\check\partial_1\varphi+
\frac{\Delta^2}{2}\check\partial_0\check\partial_1\varphi.\nn\\
\ea
and similar for the remaining spinors which involve the lattice derivatives $\check\partial_2$ and $\check\partial_3$. (Here we suppress spinor indices and the position $\tilde y$ of the hypercube.)

\section*{APPENDIX I: RELATION BETWEEN LINKS AND LATTICE VIERBEIN BILINEARS}
\renewcommand{\theequation}{I.\arabic{equation}}
\setcounter{equation}{0}
\label{RELATION BETWEEN LINKS}

The discretized version of the vierbein bilinears is in close correspondence with the links. It can be obtained by contraction of the links with Dirac matrices and some internal matrix $U^{ab}$. We observe the identities
\ba\label{182A}
&&(\gamma^m_M)_{\beta\alpha}U^{ba}
L^{(+-)ab}_{(k)\alpha\beta}(\x{j_1},\x{j_2})=\nn\\
&&\qquad -\varphi^b_-(\x{j_2})C_-\gamma^m_M\varphi^a_+(\x{j_1})V^{ba},
\ea
and
\ba\label{VB1}
&&(\gamma^m_M)_{\beta\alpha}U^{ba}L^{(-+)ab}_{(k)\alpha\beta}(\x{j_1},\x{j_2})=\nn\\
&&\qquad -\varphi^b_+(\x{j_2})C_+\gamma^m_M\varphi^a_-(\x{j_1})V^{ba},
\ea
with 
\be\label{VB2}
V^{ba}=(\tau_2\tau_kU)^{ba}.
\ee
(Dirac indices are suppressed on the r.h.s. of eqs. \eqref{182A}, \eqref{VB1}.) For symmetric $V^{ba}=V^{ab}$ we consider 
\ba\label{VB3}
E^m_1(\x{j_2},\x{j_1})&=&-(\gamma^m_M)_{\beta\alpha} U^{ba}
\big[L^{(+-)ab}_{(k)\alpha\beta}(\x{j_1},\x{j_2})\nn\\
&&+L^{(-+)ab}_{(k)\alpha\beta}(\x{j_1},\x{j_2})\big]\nn\\
&=&\varphi^a(\x{j_2})C_1\gamma^m_M V^{ab}\varphi^b(\x{j_1}),
\ea
while for antisymmetric  $V^{ba}=-V^{ab}$ we take 
\ba\label{VB4}
E^m_2(\x{j_2},\x{j_1})&=&(\gamma^m_M)_{\alpha\beta}U^{ba}\big[L^{(+-)ab}_{(k)\alpha\beta}(\x{j_1},\x{j_2})\nn\\
&&-L^{(-+)}_{(k)\alpha\beta}(\x{j_1},\x{j_2})\big]\nn\\
&=&\varphi^a(\x{j_2})C_2\gamma^m_M V^{ab}\varphi^b(\x{j_1}).
\ea

The continuum limit of $E_1$ and $E_2$ corresponds to the vierbein (or linear combinations of vierbeins) with the same $V^{ab}$. Indeed, the matrices $C_1\gamma^m\otimes V$ or $C_2\gamma^m\otimes V$ for $E_1$ or $E_2$, respectively, are symmetric. Defining similarly to eq. \eqref{VV6}
\be\label{VB5}
\varphi(\x j)=\bar\varphi(\tilde y)+\Delta V^\mu_j\hat\partial_\mu\varphi(\tilde y)+0(\Delta^2),
\ee
the contribution $\sim \bar\varphi$ vanishes due to the anticommuting properties of the Grassmann variables. The leading terms in $E_{1,2}$ are therefore 
\ba\label{VB6}
E^m_i(\x{j_2},\x{j_1})&=&\Delta(V^\mu_{j_1}-V^\mu_{j_2})\bar\varphi^a(\tilde y)C_i\gamma^m_M V^{ab}
\hat\partial_\mu\varphi^b(\tilde y)\nn\\
&\to&\Delta(V^\mu_{j_1}-V^\mu_{j_2})\tilde E^m_\mu.
\ea

We want to express the links in terms of vierbein bilinears. This is, in principle, simple algebra that we perform here in several steps. We first introduce the bilinears $(m=0\dots 3)$
\ba\label{VB7}
Q^{m,ab}_{+-}(\x{j_1},\x{j_2})&=&\varphi^a_{+,\alpha}(\x{j_1})(C_+\tau_m)_{\alpha\beta}\varphi^b_{-,\beta}(\x{j_2}),\nn\\
Q^{m,ab}_{-+}(\x{j_1},\x{j_2})&=&\varphi^a_{-,\alpha}(\x{j_1})(C_-\tau_m)_{\alpha\beta}\varphi^b_{+,\beta}(\x{j_2}),
\ea
with $\tau_0=1_2$. Using the identity
\be\label{VB8}
(\tau_m)_{\alpha\beta}(\tau_m)_{\gamma\delta}=2\delta_{\alpha\delta}\delta_{\beta\gamma},
\ee
one finds
\ba\label{VB9}
L^{(+-)ab}_{(k)\alpha\beta}(\x{j_1},\x{j_2})=-\frac12 Q^{m,ca}_{-+}(\x{j_2},\x{j_1})(\tau_m)_{\alpha\beta}(\tilde\tau_k)^{cb},\nn\\
\ea
and a similar equation for $L^{-+}$ with $+$ and $-$ exchanged. We also define the linear combinations
\ba\label{VB10b}
R^{0,ab}_1&=&i(Q^{0,ab}_{+-}+Q^{0,ab}_{-+}),\nn\\
R^{k,ab}_1&=&-i(Q^{k,ab}_{+-}-Q^{k,ab}_{-+}),\nn\\
R^{0,ab}_2&=&i(Q^{0,ab}_{+-}-Q^{0,ab}_{-+}),\nn\\
R^{k,ab}_2&=&-i(Q^{k,ab}_{+-}+Q^{k,ab}_{-+}),
\ea
such that 
\be\label{VB10c}
R^{m,ab}_{1,2}(\x{j_1},\x{j_2})=\varphi^a_\alpha(\x{j_1})(C_{1,2}\gamma^m_M)_{\alpha\beta}\varphi^b_\beta(\x{j_2}).
\ee
We next consider the symmetric and antisymmetric parts of $R$ separately and define 
\ba\label{VB10}
E^{m,ab}_1&=&\frac12(R^{m,ab}_1+R^{m,ba}_1),\nn\\
A^{m,ab}&=&\frac12(R_1^{m,ab}-R_1^{m,ba}).
\ea
In the continuum limit \eqref{VB5} one obtains 
\ba\label{EA}
E\ma_1(\x{j_1},\x{j_2})&=&\frac12\Delta(V^{\mu}_{j_2}-V^\mu_{j_1})\big[(\tilde E^m_{1,\mu})^{ab}+(\tilde E^m_{1,\mu})^{ba}\big]\nn\\
A\ma(\x{j_1},\x{j_2})&=&\varphi^aC_1\gamma^m_M\varphi^b.
\ea
Similarly, we define 
\ba\label{EB}
S\ma&=&\frac12(R\ma_2+R^{m,ba}_2),\nn\\
E\ma_2&=&\frac12(R\ma_2-R^{m,ba}_2),
\ea
with continuum limit 
\ba\label{EC}
S\ma(\x{j_1},\x{j_2})&=&\varphi^aC_2\gamma^m_M\varphi^b,\\
E\ma_2(\x{j_1},\x{j_2})&=&\frac12\Delta(V^\mu_{j_2}-V^\mu_{j_1})\big[(\tilde E^m_{2,\mu})^{ab}-
(\tilde E^m_{2,\mu})^{ba}\big].\nn
\ea

Insertion of these relations yields for the link bilinears the relation \eqref{ED}, with 
\be\label{EE}
(V^m_{\pm k})^{ab}(\x{j_1},\x{j_2})=(\tilde \tau_k)^{ac}
(E^m_1\mp E^m_2+A^m\mp S^m)^{cb}
(\x{j_1},\x{j_2}).
\ee

We finally use $SU(2)_V$ representations similar to eqs. \eqref{31A1a}-\eqref{57D} for the ``lattice vierbein bilinears''
\ba\label{L10a}
&&\bar E^m_{1(k)}(\tilde x_{j_1},\x{j_2})=E^{m,ab}_1(\x{j_1},\x{j_2})(\tilde\tau_k)^{ab}\nn\\
&&\qquad= \varphi(\x{j_1})C_1\gamma^m_M\otimes\tilde\tau_k\varphi(\x{j_2}),\nn\\
&&\bar E^m_2(\x{j_1},\x{j_2})=E^{m,ab}_2(\x{j_1},\x{j_2})(\tilde\tau_0)^{ab}\nn\\
&&\qquad=\varphi(\x{j_1})C_2\gamma^m_M\otimes\tilde\tau_0\varphi(\x{j_2}),\nn\\
&&\bar S^m_{(k)}(\x{j_1},\x{j_2})=S^{m,ab}(\x{j_1},\x{j_2})(\tilde\tau_k)^{ab}\nn\\
&&\qquad=\varphi(\x{j_1})C_2\gamma^m_M\otimes\tilde\tau_k\varphi(\x{j_2}),\nn\\
&&\bar A^m(\x{j_1},\x{j_2})=A^{m,ab}(\x{j_1},\x{j_2})(\tilde\tau_0)^{ab}\nn\\
&&\qquad=\varphi(\x{j_1})C_1\gamma^m_M\otimes \tilde\tau_0\varphi(\x{j_2}).
\ea
The quantities $E^{m,ab}_{1,2}, S^{m,ab}$ and $A^{m,ab}$ can be expressed in terms of those bilinears
\ba\label{EO}
E^{m,ab}_1(\x{j_1},\x{j_2})&=&\frac12\bar E^m_{1(k)}(\x{j_1},\x{j_2})(\tau_k\tau_2)^{ab}\nn\\
E^{m,ab}_2(\x{j_1},\x{j_2})&=&-\frac12\bar E^m_2(\x{j_1},\x{j_2})(\tau_2)^{ab}\nn\\
S^{m,ab}(\x{j_1},\x{j_2})&=&\frac12\bar S^m_{(k)}(\x{j_1},\x{j_2})(\tau_k\tau_2)^{ab}\nn\\
A^{m,ab}(\x{j_1},\x{j_2})&=&-\frac12\bar A^m(\x{j_1},\x{j_2})(\tau_2)^{ab}.
\ea
Insertion into eq. \eqref{EE} yields the relation \eqref{EP}.

\section*{APPENDIX J: LATTICE ACTION IN TERMS OF VIERBEIN BILINEARS}
\renewcommand{\theequation}{J.\arabic{equation}}
\setcounter{equation}{0}
\label{Lattice action in terms of vierbein}

One can use the expression \eqref{L8A} in order to write the lattice action as a sum of terms involving six powers of lattice vierbeins \eqref{207A1} or scalars \eqref{A21a}. This employs the relations \eqref{L8Aa}, \eqref{ED} and \eqref{EP}. In this appendix we briefly sketch the necessary algebra. One has to evaluate the hyperlink 
\ba\label{EF}
&&\tilde {\cal C}(\x1,\x3,\x2,\x4,\x8,\x6)=\frac{1}{36}\frac{1}{4096}
\epsilon^{klj}\epsilon^{pqr}Y_{m_1m_2\dots m_6}\nn\\
&\times&\tr \big\{ V^{m_6}_{-r}(\x1,\x6)V^{m_5}_{+j}(\x6,\x8)V^{m_4}_{-q}
(\x8,\x4)V^{m_3}_{+l}(\x4,\x2)\nn\\
&\times&V^{m_2}_{-p}(\x2,\x3)V^{m_1}_{+k}(\x3,\x1)\big\},
\ea
where 
\be\label{EG}
Y_{m_1\dots m_6}=\tr (\tau_{m_1}\hat\tau_{m_2}\tau_{m_3}
\hat\tau_{m_4}\tau_{m_5}\hat\tau_{m_6}).
\ee
The trace in eq. \eqref{EF} is over flavor indices, with $V^m_{\pm k}$ interpreted as $2\times 2$ matrices in flavor space and matrix multiplication implied. We recall the definitions $\tau_0=-\hat\tau_0=1,\hat\tau_k=\tau_k$. 

The first task is an evaluation of $Y_{m_1\dots m_6}$. We first consider those components of $Y$ for which an even number of indices equals zero, as
\ba\label{EH}
Y_{000000}&=&-2,\nn\\
Y_{0000kl}&=&-Y_{000k0l}=2\delta_{kl}.
\ea
We can use the identities
\ba\label{EI}
\tau_0\hat\tau_0&=&-1~,~\tau_0\hat\tau_k=-\hat\tau_0\tau_k=\tau_k,\nn\\
\tau_k\hat\tau_l&=&\tau_k\tau_l=\delta_{kl}+i\epsilon_{klm}\tau_m,
\ea
and 
\ba\label{EJ}
\tau_i\tau_j\tau_k\tau_l&=&\delta_{ij}\delta_{kl}+i\epsilon_{ijm}\tau_m\delta_{kl}
+i\epsilon_{klm}\tau_m\delta_{ij}\nn\\
&&-\epsilon_{ijn}\epsilon_{klm}\tau_n\tau_m
\ea
in order to establish
\be\label{EK}
Y_{00ijkl}=2(\delta_{ik}\delta_{jl}-\delta_{il}\delta_{jk}-\delta_{ij}\delta_{kl})
\ee
and 
\ba\label{EL}
Y_{ijklpq}&=&Y_{jklpqi}=2(\delta_{ij}\delta_{kl}\delta_{pq}\nn\\
&&+\delta _{ij}\delta_{kq}\delta_{lp}-\delta_{ij}\delta_{kp}\delta_{lq}\nn\\
&&+\delta_{iq}\delta_{jp}\delta_{kl}-\delta_{ip}\delta_{jq}\delta_{kl}\nn\\
&&+\delta_{il}\delta_{jk}\delta_{pq}-\delta_{ik}\delta_{jl}\delta_{pq}\nn\\
&&+\delta_{ik}\delta_{jp}\delta_{lq}-\delta_{ik}\delta_{jq}\delta_{lp}\nn\\
&&+\delta_{il}\delta_{jq}\delta_{kp}-\delta_{il}\delta_{jp}\delta_{kq}\nn\\
&&+\delta_{ip}\delta_{jl}\delta_{kq}-\delta_{ip}\delta_{jk}\delta_{lq}\nn\\
&&+\delta_{iq}\delta_{jk}\delta_{lp}-\delta_{iq}\delta_{jl}\delta_{kp})
\ea
The value of $Y$ for other positions of the index $0$ is easily obtained by keeping track of the minus sign if $\tau_0$ in eq. \eqref{EG} is replaced by $\hat\tau_0$ and vice versa. We observe the cyclic property for index pairs
\ba\label{EM}
Y_{m_1m_2m_3m_4m_5m_6}&=&
Y_{m_3m_4m_5m_6m_1m_2}\nn\\
&=&Y_{m_5m_6m_1m_2m_3m_4},
\ea
which is a direct consequence of the definition by the trace \eqref{EG}. 

We denote by $Y^+$ the part with an even number of indices $0$, and $Y^-$ the part with an odd number,
\be\label{ELA}
Y_{m_1\dots m_6}=Y^+_{m_1\dots m_6}+Y^-_{m_1\dots m_6}.
\ee
Since $Y$ must be invariant under $SO(1,3)$-Lorentz transformations it is suggestive to generalize eq. \eqref{EL} to 
\ba\label{EN}
&&Y^+_{m_1m_2m_3m_4m_5m_6}=2(\eta_{m_1m_2}\eta_{m_3m_4}\eta_{m_5m_6}\nn\\
&&+\eta_{m_1m_2}\eta_{m_3m_6}\eta_{m_4m_5}-\eta_{m_1m_2}\eta_{m_3m_5}\eta_{m_4m_6}\nn\\
&&+\eta_{m_1m_6}\eta_{m_2m_5}\eta_{m_3m_4}-\eta_{m_1m_5}\eta_{m_2m_6}\eta_{m_3m_4}\nn\\
&&+\eta_{m_1m_4}\eta_{m_2m_3}\eta_{m_5m_6}-\eta_{m_1m_3}\eta_{m_2m_4}\eta_{m_5m_6}\nn\\
&&+\eta_{m_1m_3}\eta_{m_2m_5}\eta_{m_4m_6}-\eta_{m_1m_3}\eta_{m_2m_6}\eta_{m_4m_5}\nn\\
&&+\eta_{m_1m_4}\eta_{m_2m_6}\eta_{m_3m_5}-\eta_{m_1m_4}\eta_{m_2m_5}\eta_{m_3m_6}\nn\\
&&+\eta_{m_1m_5}\eta_{m_2m_4}\eta_{m_3m_6}-\eta_{m_1m_5}\eta_{m_2m_3}\eta_{m_4m_6}\nn\\
&&+\eta_{m_1m_6}\eta_{m_2m_3}\eta_{m_4m_5}-\eta_{m_1m_6}\eta_{m_2m_4}\eta_{m_3m_5}).
\ea
Indeed, it is straightforward to verify the relations \eqref{EH}, \eqref{EK}, and \eqref{EL}. This can easily be extended to other positions of the index $0$. One finds the relations
\ba\label{ENA}
Y^+_{m_1m_2m_3m_4m_5m_6}&=&Y^+_{m_2m_3m_4m_5m_6m_1}\nn\\
&=&Y^+_{m_6m_5m_4m_3m_2m_1}.
\ea

What remains is the part $Y^-$ with an odd number of indices $0$, as
\ba\label{ENB}
Y_{00000k}&=&0,\nn\\
Y_{000ijk}&=&-\tr(\tau_i\tau_j\tau_k)=-2i\epsilon_{ijk}.
\ea
The contribution with five indices different from zero reads
\ba\label{J12A}
Y_{0ijklm}&=&\tr(\tau_i\tau_j\tau_k\tau_l\tau_m)\\
&=&2i(\delta_{ij}\epsilon_{klm}+\delta_{kl}\epsilon_{ijm}+\delta_{im}\epsilon_{jkl}
-\delta_{jm}\epsilon_{ikl}).\nn
\ea
The values for other positions of the index $0$ are obtained easily.

For a systematic expression the next step expresses $V$ in terms of $\bar E,\bar A, \bar S$. For this purpose we write 
\ba\label{EQ}
(V^m_{-k})^{ab}&=&\frac12\big[\tau_2(v^{0,m}_{-k}\tau_0+v^{j,m}_{-k}\tau_j)\tau_2\big ]^{ab}\nn\\
(V^m_{+k})^{ab}&=&\frac12\big[\tau_2(v^{j,m}_{+k}\hat\tau_0+v^{j,m}_{+k}\hat\tau_j)\tau_2\big]^{ab},
\ea
with 
\ba\label{ER}
v^{0,m}_{-k}&=&\bar E^m_{1(k)}+\bar S^m_{(k)},\nn\\
v^{0,m}_{+k}&=&-\bar E^m_{1(k)}+\bar S^m_{(k)},\nn\\
v^{j,m}_{-k}&=&-(\bar E^m_2+\bar A^m)\delta_{jk}+i\epsilon_{klj}(\bar E^m_{1(l)}+\bar S^m_{(l)}),\nn\\
v^{j,m}_{+k}&=&(\bar E^m_2-\bar A^m)\delta_{jk}+i\epsilon_{klj}(\bar E^m_{1(l)}-\bar S^m_{(l)}).
\ea
The $\tau_2$-factors drop out in the trace in eq. \eqref{EF} and we recognize the structure \eqref{EG}, such that with $s_i=0,1,2,3$ one obtains 
\ba\label{ES}
&&\tilde{\cal C}(\x1,\x3,\x2,\x4,\x8,\x6)=\frac{1}{36}\frac{1}{2^{18}}
\epsilon^{k_1k_3k_5}\epsilon^{k_2k_4k_6}\nn\\
&&\times Y_{m_6m_5m_4m_3m_2m_1}
Y_{s_1s_2s_3s_4s_5s_6}\nn\\
&&\times v^{s_1,m_1}_{-k_1}(\x1,\x6)v^{s_2,m_2}_{+k_2}(\x6,\x8)
v^{s_3,m_3}_{-k_3}(\x8,\x4)\nn\\
&&\times v^{s_4,m_4}_{+k_4}(\x4,\x2)v^{s_5,m_5}_{-k_5}(\x2,\x3)v^{s_6,m_6}_{+k_6}(\x3,\x1).
\ea
Eq. \eqref{ES} expresses the hyperlink ${\cal C}$ and therefore the lattice action \eqref{L8A} in terms of the lattice vierbeins $\bar E^m_{1(k)},\bar E^m_2$ and $\bar S^m_{(k)},\bar A^m$.

The combinatorics of eq. \eqref{ES} is still rather involved. Insight in the structure can be gained by studying special cases. We consider the case where only $\bar E^m_2$ differs from zero, such that $v^{0,m}_{\pm k}=0,v^{j,m}_{\pm k}=\bar E^m_2\delta_{jk}$. Using 
\be\label{ET}
\epsilon^{k_1k_3k_5}\epsilon^{k_2k_4k_6}Y_{k_1k_2k_3k_4k_5k_6}=-24
\ee
yields
\ba\label{EU}
&&{\cal C}(\x1,\x3,\x2,\x4,\x8,\x6)=\frac16\frac{1}{2^{16}}Y_{m_6\dots m_1}\nn\\
&&\quad\times\bar E^{m_1}_2(\x1,\x6)\bar E^{m_2}_2(\x6,\x8)\bar E^{m_3}_2(\x8,\x4)\bar E^{m_4}_2
(\x4,\x2)\nn\\
&&\quad\times\bar E^{m_5}_2(\x2,\x3)\bar E^{m_6}_2(\x3,\x1).
\ea
The contribution from $Y^+$ yields a Lorentz-invariant expression where index pairs $(m_a,m_b)$ are contracted by $\eta_{m_am_b}$. 

In the continuum limit we may consider first the leading order relations
\be\label{EV}
\bar E^m_2(\x{j_1},\x{j_2})=(V^\mu_{j_2}-V^\mu_{j_1})\tilde e^m_{2,\mu}(y),
\ee
with
\be\label{A24}
\tilde e^m_{2,\mu}(y)=\Delta\varphi^a(y)C_2\gamma^m_M(\tau_2)^{ab}\partial_\mu\varphi^b(y).
\ee

We also can use
\ba\label{EX}
&&\bar E^m_2(\x{j_1},\x{j_2})\bar E^n_2(\x{j_3},\x{j_4})\eta_{mn}\nn\\
&&\qquad=(V^\mu_{j_2}-V^\mu_{j_1})(V^\nu_{j_4}-V^\nu_{j_3})\tilde g_{2,\mu\nu},
\ea
with 
\be\label{EY}
\tilde g_{2,\mu\nu}=\tilde e^m_{2,\mu}\tilde e^n_{2,\nu}\eta_{mn}.
\ee
Insertion into eq. \eqref{EU} yields
\be\label{EZ}
{\cal C}(\x1,\x3,\x2,\x4,\x8,\x6)=\frac16\frac{1}{2^{16}}\hat Y_{m_6\dots m_1},
\ee
where $\hat Y_{m_6\dots m_1}$ obtains from $Y^+_{m_6\dots m_1}$ by replacing in eq. \eqref{EN} $\eta_{m_am_b}\to\tilde g_{2,\mu\nu}w^\mu_{m_a}w^\nu_{m_b}$, with 
\ba\label{FA}
w_{m_1}&=&V_6-V_1=(1,0,1,0),\nn\\
w_{m_2}&=&V_8-V_6=(1,0,-1,0),\nn\\
w_{m_3}&=&V_4-V_8=(-1,0,0,-1),\nn\\
w_{m_4}&=&V_2-V_4=(0,-1,0,1),\nn\\
w_{m_5}&=&V_3-V_2=(0,1,-1,0),\nn\\
w_{m_6}&=&V_1-V_3=(-1,0,1,0).
\ea
A second contribution arises from $Y^-$. No diffeomorphism invariant exists which involves six powers of the vierbein 
$\tilde e^m_{2,\mu}$. We therefore expect that for the action the sum of all contributions of this type vanishes. 

For the particular term where only $E^m_2$ contributes we can, of course, also use directly eq. \eqref{EP}, 
\be\label{J.24}
(V^m_{\pm k})^{ab}=\pm\frac12\bar E^m_2(\tau_2\tau_k\tau_2)^{ab}.
\ee
Eqs. \eqref{L8Aa}, \eqref{ED} then yield
\ba\label{J.25}
&&\tilde{\cal C}(\x1,\x3,\x2,\x4,\x8,\x6)=-\frac{1}{36}\frac{1}{2^{18}}Y_{m_1\dots m_6}\nn\\
&\times&\epsilon^{klm}\epsilon^{qpr} \tr (\tau_r\tau_m\tau_q\tau_l\tau_p\tau_k)\nn\\
&\times&\bar E^{m_1}_2(\x3,\x1)\bar E^{m_2}_2(\x2,\x3)\bar E^{m_3}_2(\x4,\x2)\nn\\
&\times&\bar E^{m_4}_2(\x8,\x4)\bar E^{m_5}_2(\x6,\x8)\bar E^{m_6}_2(\x1,\x6),
\ea
which agrees with \eqref{EU}.

\section*{APPENDIX K: LATTICE DIFFEOMORPHISM INVARIANT ACTION IN FOUR DIMENSIONS}
\renewcommand{\theequation}{K.\arabic{equation}}
\setcounter{equation}{0}
\label{LATTICE DIFFEOMORPHISM INVAIRANT ACTION}

In this appendix we discuss the general structure of a lattice diffeomorphism invariant action \eqref{DL15} in four dimensions. We relate this to the lattice action \eqref{145A1}.

In our model of spinor gravity we have six fields $\h_A$ which corresponds to the spinor bilinears $\tilde\h^+_k,\tilde\h^-_k$, i.e. $A=(k_+,k_-)$. A lattice diffeomorphism invariant action can be written as
\ba\label{DL17}
{\cal L}(y)&=&c_BB^{A_1\dots A_6}{\cal H}_{A_1}\y{\cal H}_{A_2}\y\big)\\
&\times&({\cal H}_{A_3}(\x8)-{\cal H}_{A_3}(\x1)\big)
\big({\cal H}_{A_4}(\x7)-{\cal H}_{A_4}(\x2)\big)\nn\\
&\times&\big({\cal H}_{A_5}(\x6)-{\cal H}_{A_5}(\x3)\big)
\big({\cal H}_{A_6}(\x5)
-{\cal H}_{A_6}(\x4)\big),\nn
\ea
with $B^{A_1\dots A_6}$ totally antisymmetric in the last four indices $A_3,A_4,A_5,A_6$ and ${\cal H}_{A_1}\y,{\cal H}_{A_2}\y$ average fields in the cell
\be\label{DL18}
{\cal H}_A\y=\frac18\sum^{8}_{j=1}{\cal H}(\x j).
\ee
The use of $\tilde \h^\pm_k$ guarantees local generalized Lorentz symmetry and $SU(2,{\mathbbm C})_F$ flavor symmetry follows for suitable $B$. For our order of the indices $A=1,2,3$ corresponds to the index $k$ of $\h^+_k,k_+=1,2,3$, while $A=4,5,6$ corresponds to $k_-=1,2,3$ for $\h^-_k$. For an action with six Weyl spinors $\varphi_+$ and six Weyl spinors $\varphi_-$ three of the indices of $B$ must be of the type $k_+$, and the other three of the type $k_-$. The chiral flavor symmetry $SU(2,{\mathbbm C})_L\times SU(2,{\mathbbm C})_R$ is realized if $B$ is totally antisymmetric in the three indices $(k_+,l_+,m_+)$, as well as in the three indices $(k_-,l_-,m_-)$. 

We specify $B^{A_1A_2A_3A_4A_5A_6}$ by the following properties: (1) $B$ vanishes whenever two indices are equal, (2) $B$ vanishes if the first two indices $A_1,A_2$ are both of the type $k_+$ or both of the type $k_-$, (3) $B$ vanishes whenever the last four indices $A_3,A_4,A_5,A_6$ do not have two indices of the type $k_+$, and the other two of the type $k_-$, (4) $B$ is totally antisymmetric in the last four indices, (5) $B$ is symmetric in the first two indices, (6) $B$ is specified by its value when $A_1,A_3,A_4$ are of the type $k_+$, 
\be\label{DL19}
B^{k_+k_-l_+m_+l_-m_-}=\epsilon^{k_+l_+m_+}\epsilon^{k_-l_-m_-}.
\ee
Property (4) fixes then the value of all other combinations of the last four indices $B^{k_+k_-A_3A_4A_5A_6}$ consistent with the constraint (3), i.e.
\ba\label{DL20}
B^{k_+k_-l_+l_-m_+m_-}&=&-\epsilon^{k_+l_+m_+}
\epsilon^{k_-l_-m_-},\nn\\
B^{k_+k_-l_-m_-l_+m_+}&=&\epsilon^{k_+l_+m_+}
\epsilon^{k_-l_-m_-}.
\ea
Finally, the condition (5),
\be\label{DL21}
B^{k_-k_+A_3A_4A_5A_6}=B^{k_+k_-A_3A_4A_5A_6},
\ee
fixes the remaining nonzero values of $B$. Our conditions, which are not all independent, determine $B$ completely.

The symmetries of lattice reflections and rotations become apparent if we use the expression \eqref{DL16} in terms of lattice derivatives
\be\label{DL22}
{\cal L}\y=\frac{c_B}{3}V\y B^{A_1\dots A_6}{\cal H}_{A_1}{\cal H}_{A_2}
\epsilon^{\mu_1\dots\mu_4}\hat\partial_{\mu_1}
{\cal H}_{A_3}\dots\hat\partial_{\mu_4}{\cal H}_{A_6}.
\ee
The cell volume $V\y$ is invariant under rotations and reflections. For $\bl={\cal L}/V$ we can use the particular coordinates $x^\mu=\tilde z^\mu\Delta$ such that lattice rotations and reflections are easily realized by the contraction of derivatives with $\epsilon^{\mu_1\dots \mu_4}$. The average fields ${\cal H}\y$ are invariant under lattice reflections and rotations.

For a suitable choice of the constant $c_B$ the action \eqref{DL17}, with $B$ determined by eqs. \eqref{DL19}, \eqref{DL20}, equals the leading term in the lattice action \eqref{VV3}. For a comparison with the lattice action \eqref{VV3} we employ relations of the type
\ba\label{DL23}
&&\frac12\big(\tilde{\cal H}^k_+(\x1)\tilde {\cal H}^m_+(\x8)-\tilde {\cal H}^m_+(\x1)\tilde{\cal H}^k_+(\x8)\big)\nn\\
&&=-\frac14\Big\{(\tilde{\cal H}^k_+(\x1)+\tilde {\cal H}^k_+(\x8)\big)\big(\tilde \h^m_+(\x1)-
\tilde \h^m_+(\x8)\big)\nn\\
&&\qquad\qquad -(k\leftrightarrow m)\big\}
\ea
in order to write
\ba\label{DL24}
{\cal F}^{1,2,8,7}_+&=&\frac{1}{12}\epsilon^{klm}\tilde\h^k_{+,1278}\nn\\
&\times&\big(\tilde\h^l_+(\x8)-\tilde \h^l_+(\x1)\big)
(\tilde \h^m_+(\x7)-\tilde\h^m_+(\x2)\big),\nn
\ea
employing shorthands
\be\label{DL25}
\tilde\h^k_{\pm,abcd}=\frac14\big\{\tilde\h^k_+(\x a)+\tilde\h^k_+(\x b)+\tilde \h^k_+(\x c)+\tilde\h^k_+(\x d)\big\}.
\ee
One finds for the action \eqref{VV3} 
\ba\label{DL27a}
{\cal L}\y&=&\frac16\frac{1}{12^2}\epsilon^{k_+l_+m_+}\epsilon^{k_-l_-m_-}\nn\\
&\times&\big\{\tilde \h^{k_+}_{+,1278}\tilde \h^{k_-}_{-,3456}D_{l_+m_+l_-m_-}\nn\\
&&-\tilde\h^{k_+}_{+,1368}\tilde\h^{k_-}_{-,2457} D_{l_+l_-m_+m_-}\nn\\
&&-\tilde\h^{k_+}_{+,1458}\tilde\h^{k_-}_{-,2367}D_{l_+m_-l_-m_+}\nn\\
&&+\tilde\h^{k_+}_{+,3456}\tilde\h^{k_-}_{-,1278} D_{l_-m_-l_+m_+}\nn\\
&&-\tilde\h^{k_+}_{+,2457}\tilde\h^{k_-}_{-,1368}D_{l_-l_+m_-m_+}\nn\\
&&-\tilde\h^{k_+}_{+,2367}\tilde\h^{k_-}_{-,1458} D_{l_-m_+l_+m_-}\big\},
\ea
with 
\ba\label{DL27}
&&D_{l_+m_+l_-m_-}=\\
&&\big(\tilde\h^{l_+}_+(\x8)-\tilde\h^{l_+}_+(\x1)\big)
\big(\tilde\h^{m_+}_+(\x7)-\tilde\h^{m_+}_+(\x2)\big)\nn\\
&&\times \big(\tilde\h^{l_-}_-(\x6)-\tilde\h^{l_-}_-(\x3)\big)
\big(\tilde\h^{m_-}_-(\x5)-\tilde\h^{m_-}_-(\x4)\big),\nn
\ea
while $D_{l_+m_-l_-m_+}$ obtains from $D_{l_+m_+l_-m_-}$ by exchanging $\h^{m_+}_+$ and $\h^{m_-}_-$, etc.. (The ordering of indices in $D$ is such that the first refers to $\x8-\x1$, the second to $\x7-\x2$, the third and fourth to $\x6-\x3$ and $\x5-\x4$.  For an index $l_+$ the $\h$-factor is $\h^{l_+}_+$, and for $l_-$ one uses $\h^{l_-}_-$.) 

We next express $\h_{abcd}$ in eq. \eqref{DL25} by the cell average $\tilde\h\y$ defined in eq. \eqref{DL18}, and
\be\label{DL28}
\tilde{\cal G}_{abcd,efgh}=\frac12(\h_{abcd}-\h_{efgh}),
\ee
which results in 
\ba\label{DL29}
&&\tilde\h^{k_+}_{+,1278}\tilde\h^{k_-}_{-,3456}=\tilde\h^{k_+}_+\y\tilde\h^{k_-}_-\y\nn\\
&&\quad-\tilde\h^{k_+}_+\y\tilde{\cal G}^{k_-}_{-,1278,3456}\y
+\tilde{\cal G}^{k_+}_{+,1278,3456}\y\tilde\h^k_-\y\nn\\
&&\quad-\tilde{\cal G}^{k_+}_{+,1278,3456}\y
\tilde{\cal G}^{k_-}_{-,1278,3456}\y.\nn\\
\ea
We concentrate first on the first term on the r.h.s. of eq. \eqref{DL29} which only contains $\tilde \h\y$. This term is the same for all six terms in eq. \eqref{DL27a} and one obtains 
\ba\label{DL30}
{\cal L}^{(1)}\y&=&\frac{1}{864}\epsilon^{k_+l_+m_+}\epsilon^{k_-l_-m_-}\nn\\
&\times&\h^{k_+}_+\y\h^{k_-}_-\y D^{(A)}_{l_+l_-m_+m_-},
\ea
with 
\ba\label{DL31}
&&D^{(A)}_{l_+m_+l_-m_-}=\hat A\{ D_{l_+m_+l_-m_-}+D_{l_+l_-m_-m_+}\nn\\
&&\qquad +D_{l_+m_-m_+l_-}+D_{l_-m_-l_+m_+}+D_{l_-l_+m_+m_-}\nn\\
&&\qquad +D_{l_-m_+m_-l_+}\},
\ea
with $\hat A$ denoting antisymmetrization in $l_+\leftrightarrow m_+$ and $l_-\leftrightarrow m_-$. The combination $D^{(A)}$ is totally antisymmetric in all four indices. The action ${\cal L}^{(1)}$ coincides with the lattice diffeomorphism invariant action \eqref{DL17} for $c_B=1/864$. 

In the continuum limit the factors $D$ involve four derivatives and therefore four powers of $\Delta$. The term $\tilde{\cal G}_{1278,3456}$ involves further derivatives and therefore further factors of $\Delta$ as compared to $\tilde \h\y$. The leading term in the continuum limit therefore only retains the lattice diffeomorphism invariant contribution ${\cal L}^{(1)}\y$. This explains the diffeomorphism symmetry of the action in the continuum limit.

\end{document}